\documentclass[twocolumn,tighten]{aastex63}
\usepackage{amssymb,mathtools}
\usepackage{url}

\begin{document}

\newcommand{\afluxa}{450~$\mu$m\ }
\newcommand{\afluxb}{850~$\mu$m\ }
\newcommand{\afluxar}{450~$\mu$m}
\newcommand{\afluxbr}{850~$\mu$m}

\title{A Submillimeter Survey of Faint Galaxies Behind Ten Strong Lensing Clusters}

\author[0000-0002-6319-1575]{L.~L.~Cowie}
\affiliation{Institute for Astronomy, University of Hawaii, 2680 Woodlawn Drive, Honolulu, HI 96822, USA}

\author[0000-0002-3306-1606]{A.~J.~Barger}
\affiliation{Institute for Astronomy, University of Hawaii, 2680 Woodlawn Drive, Honolulu, HI 96822, USA}
\affiliation{Department of Astronomy, University of Wisconsin-Madison, 475 N. Charter Street, Madison, WI 53706, USA}
\affiliation{Department of Physics and Astronomy, University of Hawaii, 2505 Correa Road, Honolulu, HI 96822, USA}

\author{F.~E.~Bauer}
\affiliation{Instituto de Astrof\'isica and Centro de Astroingenier\'ia, Facultad de F\'isica, 
Pontificia Universidad Cat\'olica de Chile, Casilla 306, Santiago 22, Chile}
\affiliation{Millennium Institute of Astrophysics (MAS), Nuncio Monse{\~{n}}or S{\'{o}}tero 
Sanz 100, Providencia, Santiago, Chile} 
\affiliation{Space Science Institute,
4750 Walnut Street, Suite 205, Boulder, Colorado 80301, USA} 

\author[0000-0002-3805-0789]{C.-C.~Chen}
\affiliation{Academia Sinica Institute of Astronomy and Astrophysics,
P.O. Box 23-141, Taipei 10617, Taiwan}

\author[0000-0002-1706-7370]{L.~H.~Jones}
\affiliation{Department of Astronomy, University of Wisconsin-Madison, 475 N. Charter Street, Madison, WI 53706, USA}
\affiliation{Space Telescope Science Institute, 3700 San Martin Drive, Baltimore, MD 21218}

\author{C.~Orquera}
\affiliation{Instituto de Astrof\'isica and Centro de Astroingenier\'ia, Facultad de F\'isica, 
Pontificia Universidad Cat\'olica de Chile, Casilla 306, Santiago 22, Chile}
\affiliation{Millennium Institute of Astrophysics (MAS), Nuncio Monse{\~{n}}or S{\'{o}}tero 
Sanz 100, Providencia, Santiago, Chile} 

\author[0000-0003-3910-6446]{M.~J. Rosenthal}
\affiliation{Department of Astronomy, University of Wisconsin-Madison, 475 N. Charter Street, Madison, WI 53706, USA}

\author[0000-0003-1282-7454]{A.~J.~Taylor}
\affiliation{Department of Astronomy, University of Wisconsin-Madison, 475 N. Charter Street, Madison, WI 53706, USA}

%-----------------------------------------------------------------------------
%   Abstract
%-----------------------------------------------------------------------------
\begin{abstract}
We present deep SCUBA-2 \afluxa and \afluxb imaging of ten strong lensing clusters.
We provide a $>4\sigma$ SCUBA-2 \afluxb catalog of the 404 sources lying within a radius of
$4\farcm5$ from the cluster centers. We also provide catalogs of the $>4.5\sigma$ ALMA 
\afluxb detections in the clusters A370, 
MACSJ1149.5+2223, and MACSJ0717.5+3745 
from our targeted ALMA observations, 
along with catalogs of all other $>4.5\sigma$ ALMA
(mostly 1.2~mm) detections in any of our cluster fields from archival ALMA observations.
For the ALMA detections, we give spectroscopic
or photometric redshifts, where available, from our own Keck observations or from the literature.
We confirm the use of the \afluxa to \afluxb flux ratio for estimating redshifts.
We use lens models to determine magnifications, most of which are in the
1.5--4 range. After supplementing the ALMA cluster sample with Chandra Deep Field (CDF) 
ALMA and SMA samples, we find no evidence for evolution in
the redshift distribution of submillimeter galaxies down to demagnified \afluxb fluxes of 0.5~mJy. 
Given this result, we conclude that our observed trend of increasing F160W to \afluxb flux ratio 
from brighter to fainter demagnified \afluxb flux results from the fainter submillimeter
galaxies having less extinction. However, there is wide spread in this relation, including the
presence of some optical/NIR dark galaxies down to fluxes below 1~mJy.
Finally, with insights from our ALMA analysis, we analyze our SCUBA-2 sample
and present 55 \afluxbr-bright $z>4$ candidates.
\end{abstract}

%-----------------------------------------------------------------------------
\section{Introduction}
%-----------------------------------------------------------------------------
The discovery of the far-infrared (FIR) Extragalactic Background Light (EBL) 
demonstrated that about half of the universe's starlight at UV/optical wavelengths 
is absorbed by dust 
and re-radiated into the FIR (Puget et al.\ 1996; Fixsen et al.\ 1998; Dole et al.\ 2006). 
At high redshifts, observations with single-dish submillimeter telescopes, such as the 
15~m James Clerk Maxwell Telescope (JCMT), can provide direct detections
of the most luminous, dusty, star-forming galaxies 
(Smail, Ivison, \& Blain 1997; Barger et al.\ 1998; Hughes et al.\ 1998;
Eales et al.\ 1999). 
However, at \afluxbr, such surveys become confusion limited at 
$\sim1.6$~mJy (4$\sigma$; Cowie et al.\ 2018), preventing the detection of fainter 
submillimeter galaxies (SMGs) with infrared luminosities $\lesssim10^{12}~L_\odot$, 
or star formation rates (SFRs)
$\lesssim200~M_\odot$~yr$^{-1}$ for a Kroupa (2001) initial mass function (IMF). 
SMGs selected from other ground-based single-dish surveys (e.g., LABOCA,
the South Pole Telescope)
are brighter and hence more extreme starbursts (see, e.g., Hodge et al.\ 2013),
even after taking into account gravitational lensing when present
(see, e.g., Spilker et al.\ 2016).

SMGs selected from blank field, confusion-limited surveys with 
SCUBA-2 (Holland et al.\ 2013) on 
the JCMT are found to be substantially distinct from the extinction-corrected 
UV-selected population (Barger et al.\ 2014; Cowie et al.\ 2017).
However, these SMGs only contain about 20--30\% of the
submillimeter EBL. Fainter SMGs are more common objects that 
contribute the majority of the EBL, but interferometry of any field
or single-dish observations of massive lensing cluster
fields are required to detect them.
Since many of these fainter SMGs may also be selected in UV samples, 
we need to obtain a census of faint SMGs and determine their redshifts and
properties to avoid double-counting and biasing the star formation history.

To this end,
we have been observing massive lensing cluster fields with SCUBA-2 to study
the population of faint SMGs with SFRs comparable to those of the brighter UV-selected
population.
Gravitational lensing by foreground massive galaxy clusters is an
excellent way to detect intrinsically faint SMGs, but it also has the
advantages that lensed sources are magnified at all wavelengths and
lensed images benefit from enhanced spatial resolution.
Number counts in lensed fields are now well established at both \afluxa and \afluxb
(e.g., Chen et al.\ 2013; Hsu et al.\ 2016), 
allowing us to determine the flux levels above which we see 50\% of the submillimeter light.
These flux levels are $\sim3$~mJy at \afluxa and $\sim1$~mJy at \afluxb
for the Fixsen et al.\ (1998) EBL measurements, with
the primary uncertainty being the submillimeter EBL.

Our goal is to generate a uniformly selected
sample of many hundreds of galaxies
in cluster lensing fields that have been intensively
studied at other wavelengths.
We can then optimize the extremely valuable interferometric time on the Atacama Large 
Millimeter/submillimeter Array (ALMA), the Northern 
Extended Millimeter Array (NOEMA), and the Submillimeter Array (SMA)
to measure accurate positions and redshifts for the detected sample.
In combination with the high-quality lensing models, the interferometry  
also allows us to determine the amplifications at the galaxy positions and to 
measure the de-lensed fluxes.
We note that fainter SMGs can also be directly detected with ALMA, but 
the field-of-view is so small---even at millimeter
wavelengths---as to make developing large samples through ALMA 
mosaicking inefficient (see, e.g., Zavala et al.\ 2021, who found 13
sources ($>5\sigma$) at 2~mm over 184~arcmin$^2$).

Our SCUBA-2 program targets 10 massive lensing cluster fields (see
Table~\ref{table1}). We chose these fields to have good
lensing models and a wealth of optical, near-infrared (NIR),
mid-infrared (MIR), FIR, radio, and X-ray
data from the Hubble Space Telescope (HST), the Spitzer Space Telescope,
the Herschel Space Observatory, the Karl G. Jansky Very Large Array, 
the Chandra X-ray Observatory, and ground-based observatories.

Where possible, we chose clusters from the 
HST Frontier Fields (HFF; Lotz et al.\ 2017) program with its 
extraordinarily deep data
and well-defined lensing models from 10 independent teams.
Five of the six HFFs are in our sample,
while the sixth, Abell S1063, is too far south to be observed with the JCMT.
The HFF images have now been enlarged under the
Beyond Ultra-deep Frontier Fields And Legacy Observations 
(BUFFALO) HST Treasury program (Steinhardt et al.\ 2020), providing
a better match in area to the submillimeter observations, though with shallower
images than the HFF regions. We take the BUFFALO images from the
Mikulski Archive for Space Telescopes (MAST).
We mark the clusters with HFF/BUFFALO data in Table~\ref{table1}.
Three other clusters in our survey come from the Cluster Lensing And Supernova Survey
with Hubble (CLASH) HST Treasury program (Postman et al.\ 2012).

The Herschel data listed in Table~\ref{table1} come from 
Oliver et al.\ (2012; Herschel Multi-tiered Extragalactic Survey, or HerMES), 
Smith et al.\ (2010; Local Cluster Substructure Survey or LoCuSS),
Egami et al.\ (2010; Herschel Lensing Survey or HLS), and 
Eales et al.\ (2010, Herschel ATLAS or HATLAS).
We will use the Herschel data in a subsequent 
paper when constructing spectral energy distributions for the sources.
Here we only show some of the Herschel/PACS 100~$\mu$m images
for illustrative purposes.

In Section~2, we provide a description of our SCUBA-2 observations,
data reduction, and sample construction,
together with the final catalog of 404 \afluxb ($>4\sigma$) sources.
In Section~3, we discuss our follow-up with ALMA for the A370,
MACJ0717, and MACSJ1149 clusters, and we catalog the ALMA detected ($>4.5\sigma$)
sources from our work and from the ALMA archive for all the clusters.
In Section~4, we interpret the ALMA cluster sample, supplemented 
with a CDF-S ALMA sample and a CDF-N SMA sample.
In Section~5, we interpret our SCUBA-2 cluster sample with insights from our
ALMA analysis. Finally, in Section~6, we summarize our results.

%---------------------------------------------------------------------
\section{The SCUBA-2 Survey}
\begin{figure}
\includegraphics[width=3.2in,angle=0]{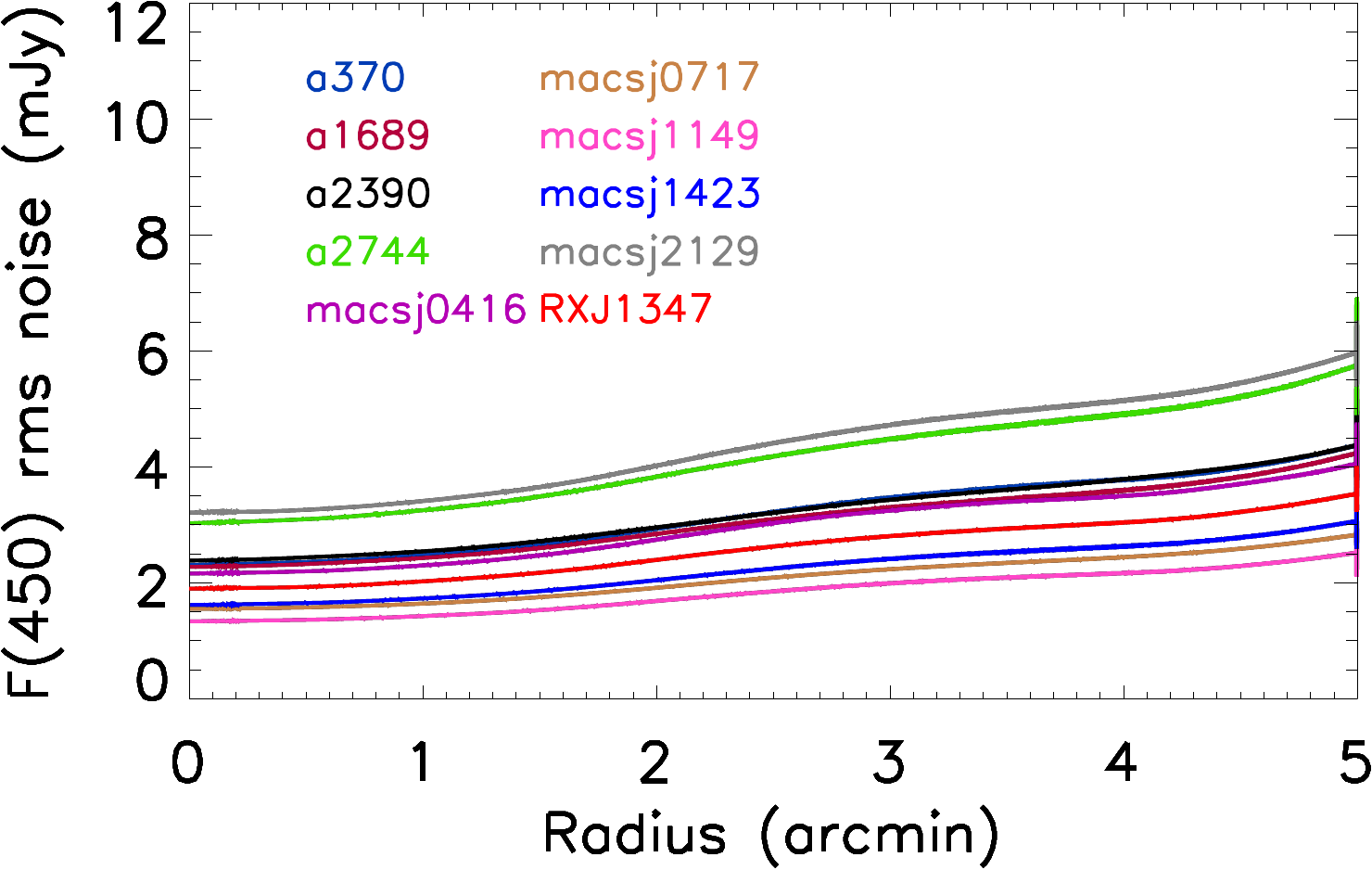}
\caption{
Azimuthally averaged \afluxa rms noise for the 10
clusters vs. radius. Note that the two clusters A2744 (green curve) and MACSJ2129
(gray curve) have shallower observations than the remaining eight clusters
(see Table~\ref{table1}).
\label{450_rad}
}
\end{figure}

%---------------------------------------------------------------------
For our SCUBA-2 observations, we
use two scan patterns: CV DAISY, which has a field
size of $5\farcm5$ radius, and PONG-900, which has a field size of $10\farcm5$
radius. Detailed information
about the SCUBA-2 scan patterns can be found in Holland et al.\ (2013).
In our reductions, we use all available \afluxa and \afluxb data from SCUBA-2 in the 
CADC archive, together with more recent observations from our own programs.
In Table~\ref{table1} and Figure~\ref{450_rad},
we summarize the current status of the SCUBA-2 data.

Chen et al.\ (2013) and Cowie et al.\ (2017) provide
a detailed description of the data reduction, which uses the Dynamic Iterative Map Maker
(DIMM) in the {\sc SMURF} package from the STARLINK software developed
by the Joint Astronomy Centre (Jenness et al.\ 2011; Chapin et al.\ 2013).
We expect nearly all of the galaxies to 
appear as unresolved sources. Thus, we apply
a matched-filter to our maps, which provides a maximum-likelihood
estimate of the source strength for unresolved sources (e.g., Serjeant et al.\ 2003).
Each matched-filter image has a PSF with
a Mexican hat shape and a FWHM corresponding to the telescope resolution.
The FWHM values for the JCMT are $\sim7\farcs5$ at \afluxa and $\sim14''$ at \afluxbr.

%---------------------------------------------------------------------
% FIGURE 1: RMS noise; 450_rad
%---------------------------------------------------------------------
%---------------------------------------------------------------------

%---------------------------------------------------------------------
% TABLE 1
%---------------------------------------------------------------------
\begin{deluxetable*}{ccccccccc}
\tablecaption{Massive Cluster Lensing Fields Observed with SCUBA-2 \label{table1}}
\tablehead{
Field  & R.A. & Decl. &  Central RMS & HFF/ &    Herschel  & 870~$\mu$m  \\
           &       &        &    [450, 850\,$\mu$m] & BUFFALO   &   & Interferometric  \\
           &       &         & (mJy) &         &   &  
           \\ (1) & (2) & (3) & (4) & (5) & (6) & (7) }
\startdata
A370                   &  02 39 53.1 & -01 34 35.0      & [2.31, 0.29] & y &HerMES  & ALMA \\
A1689                  &  13 11 29.0 & -01 20 17.0      & [2.27, 0.32] & n &LoCuSS  &\\
A2390                  &  21 53 36.8 &  \,\,17 41 44.2  & [2.38, 0.28] & n & LoCuSS &\\
A2744                  &  00 14 21.2 &  -30 23 50.1  & [3.03, 0.30] &y &HLS &\\
MACSJ0416.1-2403     &  04 16 08.9 &  -24 04 28.7  & [2.16, 0.31] & y & HLS\\
MACSJ0717.5+3745     &  07 17 34.0 &  \,\,37 44 49.0  & [1.55, 0.22] & y &  HLS  &ALMA \\
MACSJ1149.5+2223     &  11 49 36.3 &  \,\,22 23 58.1  & [1.34, 0.23] & y & HLS  &ALMA\\
MACSJ1423.8+2404     &  14 23 48.3 & \,\,24 04 47.0  & [1.61, 0.23]  & n &  HLS &\\
MACSJ2129.4-0741     & 21 29 26.2  &  -07 41 26.0  & [3.21, 0.40] &  n & HLS   &\\
RXJ1347                &  13 47 31.5 &  -11 44 19.0    & [1.90, 0.32] & n &  HATLAS   & \\
\enddata
\tablecomments{
The central rms noise quoted in Column~(4) is white noise and does not include confusion noise.
We list ALMA in Column (7) for the fields where we have obtained near-complete ALMA
870~$\mu$m targeted follow-up of the SCUBA-2 sources.
}
\end{deluxetable*}
%---------------------------------------------------------------------

%---------------------------------------------------------------------
% FIGURE 2: SCUBA-2 images for MACSJ1149; MACSJ1149_sample
%---------------------------------------------------------------------
\begin{figure*}
\includegraphics[width=7.1in,angle=0]{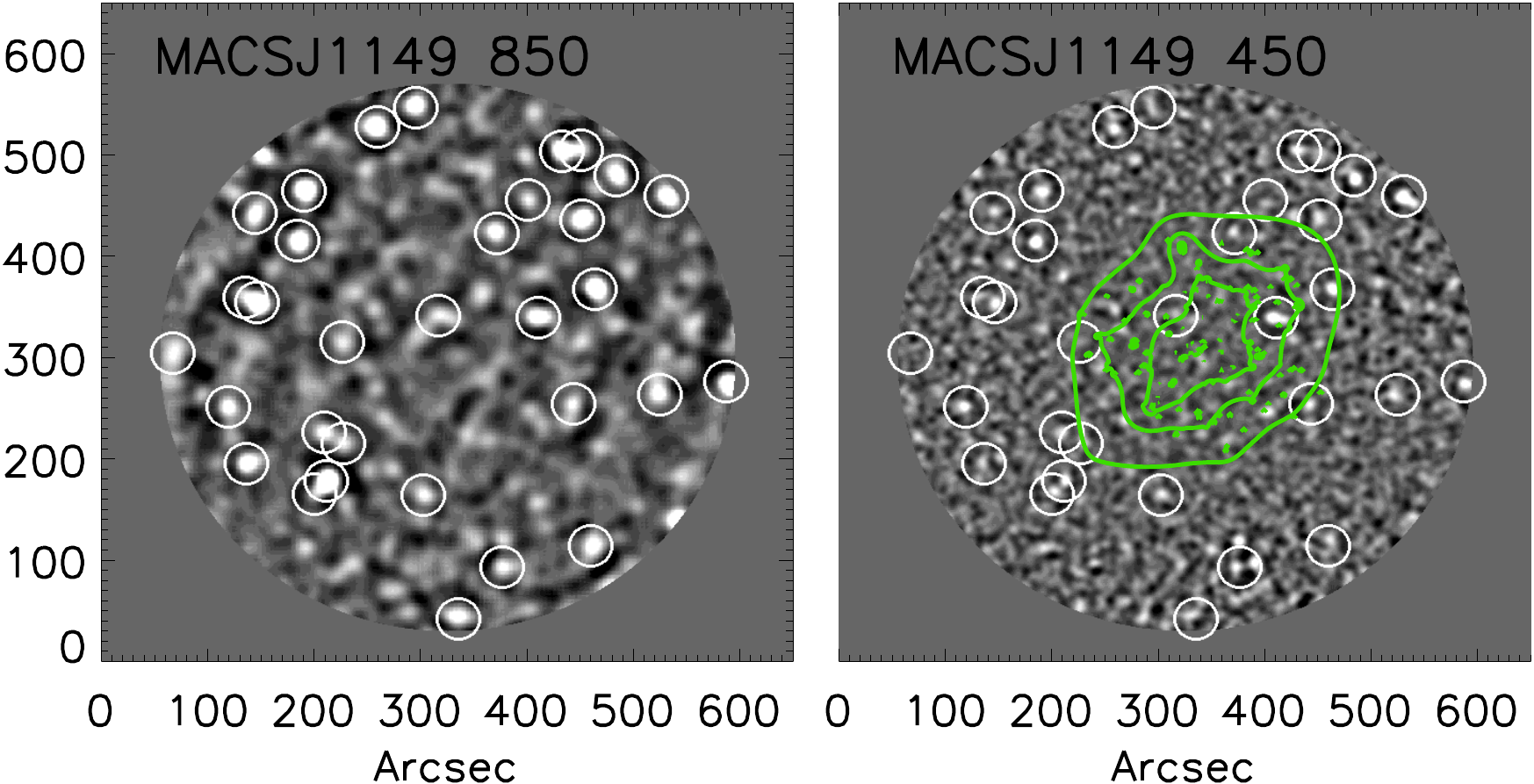}
\caption{
Representative SCUBA-2 {\em (left)} \afluxb and {\em (right)} 
\afluxa images of the cluster MACSJ1149 showing the central 
$4\farcm5$ radius region.
(The images for all 10 clusters are shown in the Appendix.)
The white circles show the $>4\sigma$ \afluxb selected
sample, including a confusion noise of 0.33~mJy.
The green contours on the \afluxa image
show the Zitrin (2021) lens model for $z=2$ at
magnifications (moving inwards) of 1.4, 2, and 4.
\label{macs1149_sample}
}
\end{figure*}
%---------------------------------------------------------------------

We next generate the \afluxb source catalog by identifying the peak signal-to-noise (S/N) pixel,
subtracting this peak pixel and its surrounding areas using the PSF scaled to the
source flux and centered on the value and position of that pixel, and then searching 
for the next S/N peak. We iterate this process until we reach a S/N of 3.5.
The reason for this iterative process is to remove contamination by brighter sources 
before we identify fainter sources and measure their fluxes. 
We then limit the catalog to the sources with a S/N above 4, 
where we include a confusion noise of 0.33~mJy (Cowie et al.\ 2017) added in quadrature.
This choice of S/N provides a robust sample. For example,
after analyzing the negatives of the images in exactly
the same way as the actual images, we find a 4\%
false positive rate, which is consistent with historical estimates of
$\le5$\% (e.g., see Figure~7 of Casey et al.\ 2013 for the field environment and
Section~5.1 of Chen et al.\ 2013a for the cluster environment).

Within a radius of $4\farcm5$
from the cluster centers, we have an \afluxb catalog of 404 sources.
This radius corresponds to the position where the \afluxa
noise is roughly twice the central noise (see Figure~\ref{450_rad}).
At larger radii, the noise rises rapidly.
We measure the \afluxa fluxes (whether positive or negative) and statistical uncertainties 
by searching for the brightest pixel in a $4''$ radius around
each \afluxb source position in the \afluxa map. 
We chose this search radius based on the positional uncertainties
in the \afluxb sample (see, e.g., Cowie et al.\ 2017).
We detect 341 of the 850~$\mu$m sources above the $2\sigma$ level
at \afluxar, and 261 above the $3\sigma$ level.
This procedure produces an upward bias. Based on randomized
samples, we estimate a false positive rate of 14\% at
$2\sigma$ and 6\% at $3\sigma$.

In Figure~\ref{macs1149_sample}, we show representative SCUBA-2
images (of the cluster MACSJ1149) based on just under  60~hours of exposure in 
weather band~1 conditions ($\tau_{225~{\rm GHz}}<0.05$).
We indicate where strongly magnified sources are expected to be by overplotting
green contours on the \afluxa image that show the Zitrin (2021) lens model for 
$z=2$ at magnifications of 1.4, 2, and 4. In general, higher redshifts will have similar 
but somewhat more compact contours.

We show the rest of the \afluxb and \afluxa SCUBA-2 images for the cluster fields in
the Appendix. In Table~\ref{tab2}, we summarize our \afluxb source catalog.
We do not apply any corrections from peak fluxes to total fluxes or
any de-boosting correction. For the \afluxb noise, we use the measured
white noise and a confusion noise of 0.33~mJy (Cowie et al.\ 2017)
added in quadrature.
We retain the brightest cluster galaxy (BCG) sources in Table~\ref{tab2}
[labeled as ``BCG" in Column~(10)], even though they
are associated with cluster cooling flows and are not produced 
by star formation (e.g., Edge et al. 2010).

In the next section, we describe the ALMA observations of
the various cluster fields.
Table~\ref{tab2} provides the target list for this follow up.

%---------------------------------------------------------------------
\section{ALMA Imaging}
%---------------------------------------------------------------------
Our SCUBA-2 sample provides a large number of sources for 
interferometric follow-up. The primary goals of the interferometric work 
are to obtain accurate positions,
to determine the optical/NIR counterparts (if any) to the SMGs, from which
we can then derive photometric redshifts (hereafter, photzs).

The follow-up interferometric continuum imaging is most efficiently carried 
out with ALMA. In Table~\ref{table1}, we mark the fields where we have 
done targeted ALMA observations. This is a nearly complete 870~$\mu$m 
interferometric
follow-up of the SCUBA-2 sources in the central areas of these fields.
With ALMA's subarcsecond resolution, we can obtain accurate positions 
for the SCUBA-2 sources and detect any multiple counterparts 
that are blended into a single source at the single-dish resolution. 
Based on our follow-up observations of SCUBA-2 sources with 
SMA and ALMA (e.g., Chen et al.\ 2013a; 
Cowie et al.\ 2017, 2018), we expect $\sim10$\% of our 
sample to be blended multiples, but there are still few studies of the 
SCUBA-2 multiplicity fraction for our flux range. 
In order to separate multiples, we aim for an rms sensitivity of $\sim0.2$~mJy 
at \afluxbr, which allows for the detection of the brighter component of any
$\sim2:1$ ratio blends that combine to form a single 2~mJy 
source, which is the typical flux of our targets.
(Note that this is the observed flux; the corresponding 
de-lensed fluxes probe the desired 1~mJy range.)

%---------------------------------------------------------------------
\subsection{A370, MACSJ1149, and MACSJ0717}
%---------------------------------------------------------------------
In our program
``An ALMA Survey of Lensed SMGs in the Hubble Frontier Fields"
(ALMA program \#2017.1.00341.S, \#2018.1.00003.S; PI: F.~Bauer),
we observed fields in A370, MACSJ1149, and MACSJ0717. For each cluster,
we observed all objects with SCUBA-2 fluxes above 1.6~mJy lying within
a $2'$ radius from the cluster center, together with a small
number of fainter objects. The observations were made
in band~7 (870~$\mu$m) using the C43-3 array configuration.
We centered the observations on the targeted source positions and used a
spectral set-up configured with four 1.875~GHz spectral windows (using time division mode)
placed around a central frequency of 343.5~GHz in order to match the SCUBA-2
\afluxb observations.
This returned a representative spectral resolution of $\sim28$~km~s$^{-1}$.

Nominal natural-weighted beams of $\approx0\farcs9\times0\farcs5$, 
$0\farcs6\times0\farcs5$, and $0\farcs9\times0\farcs4$ were achieved for our 
observations of A370, MACSJ1149 and MACSJ0717, respectively.
We made images of the targets using the task {\sc clean}. We made dirty
images using natural weighting and a mild {\em uv}-taper to achieve a
synthesized beam of $1\farcs0$ 
for the major axis; the minor axis was still $\sim0.6-0\farcs7$.
Based on the size estimates from Simpson et al.\ (2015), this offers the best trade-off 
between retaining a relatively low
rms and increasing the sensitivity to extended sources in our data; tapering the data
to larger beams means weighting toward shorter baselines, and hence fewer antennas
and lower sensitivity.

We performed an initial source search on the dirty images 
prior to primary beam correction, with the dual purpose of
assessing the rms sensitivity and finding all secure detections.
We produced cleaned images by placing $2''\times2''$ clean boxes around all secure
sources detected with S/N$\geq5$ in the dirty images. We stopped the final cleaning
process after 1000 iterations, 
such that most of the emission associated with the sources was recovered.
We note that this choice does not strongly affect the resulting fluxes or rms; for example, 
opting for only 100 iterations and/or setting the clean threshold to $1\sigma$ results in drops 
of 1--4\% in peak flux and 5--8\% in rms.

We searched each of the $1\farcs0$ cleaned images
for significant sources. We restricted our search to the area contained
within an $8\farcs75$ radius, which corresponds to the ALMA half
power. This also is well matched to the SCUBA-2
FWHM and should contain all sources contributing to
the corresponding SCUBA-2 flux. 

We selected all sources
with a peak flux S/N $>4.5$ (see below). We determined the peak
flux noise using the dispersion in 100 independent
beam positions surrounding the source.
We measured a median central rms noise of 0.24~mJy for the sources
in A370, 0.21~mJy in MACSJ1149, and 0.4~mJy in MACSJ0717.

The total searched area corresponds to 15,000
independent beams. For a Gaussian distribution in the
noise, we expect 0.5 false sources above a S/N cut of 4, and
0.05 false sources above a S/N cut of 4.5. We restrict our final
sample to sources with S/N $>4.5$, which we expect to be highly robust.
We have tested this selection by searching for sources in the negative
of the images. At S/N = 4.5, we find no negative image detections,
which confirms the robustness of the detected sample. 

We find that the ALMA peak fluxes, even in the $1\farcs0$ tapered images,
slightly underestimate the total fluxes. The reason for this is that
the sources are resolved. 
However, $2''$ aperture fluxes well approximate the
total ALMA fluxes. We list these aperture fluxes in Tables~\ref{a370_band7} and 
\ref{macsj1149_band7} for A370 and MACSJ1149, respectively. For the blended 
sources in A370 (numbers 1 and 2), we give the peak flux times 1.3 to approximate 
the total ALMA flux.
We note that the $2''$ aperture fluxes closely 
match the SCUBA-2 fluxes, with a median ratio of SCUBA-2/ALMA
of 1.12. This shows that ALMA is recovering most of the SCUBA-2 flux.

We now describe the ALMA observations for each cluster individually:

%---------------------------------------------------------------------
% FIGURE 3: ALMA images in A370; kohno_image.pro and kohno_other.pro in a370 dir
% final_a370_kohno_image and final_a370_kohno_other
%---------------------------------------------------------------------
\begin{figure*}
\hskip 5.0cm
\centering{\includegraphics[width=7in,angle=0]{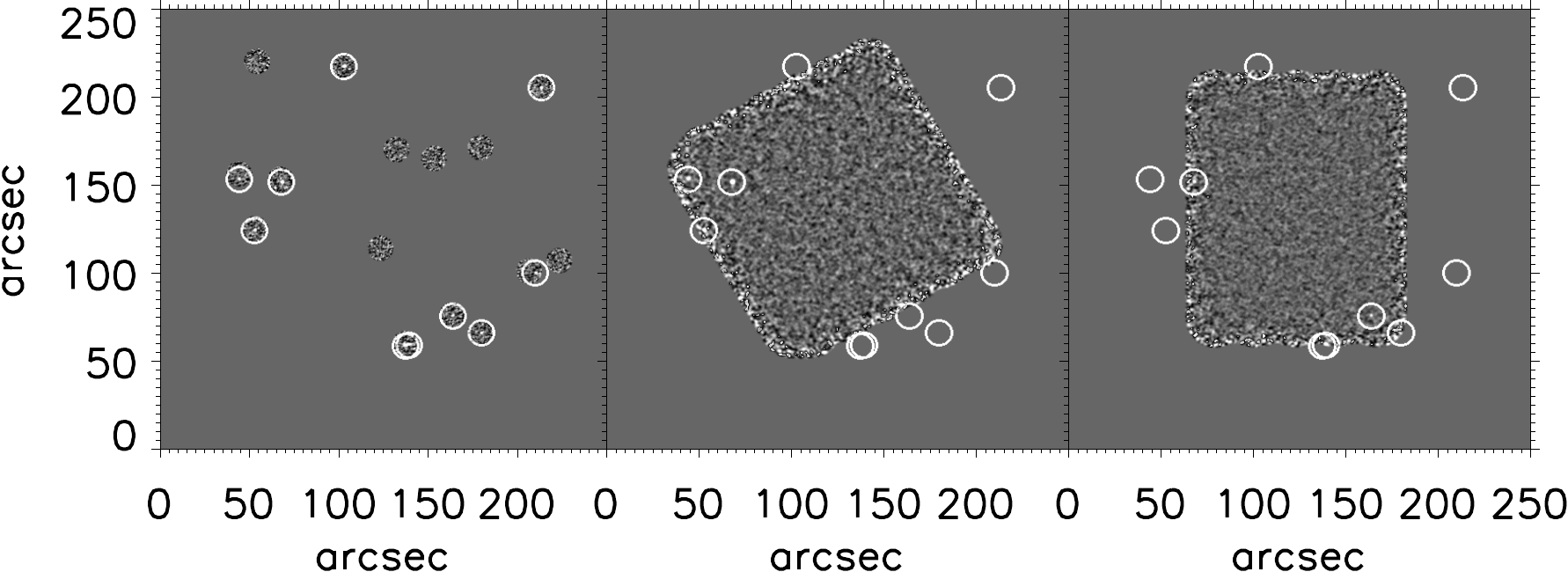}}
\vskip 0.7cm
\centering{\includegraphics[width=7in,angle=0]{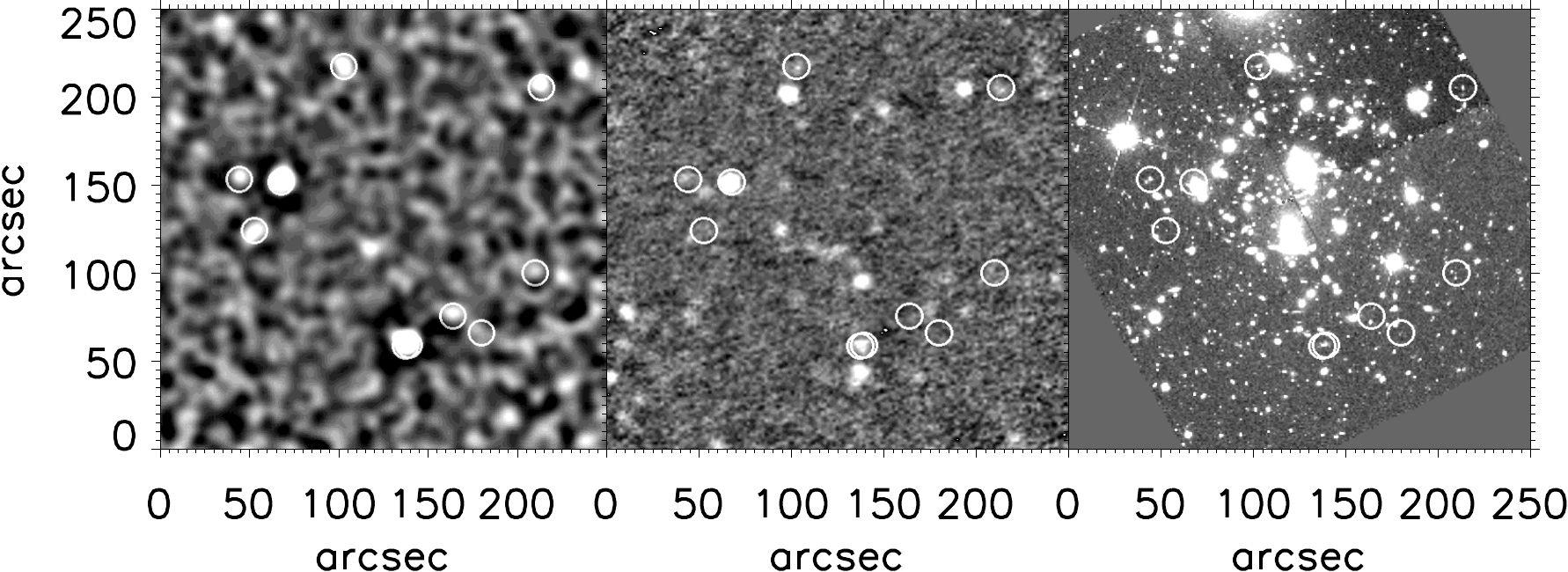}}
\caption{
{\em (Upper three panels)\/} ALMA observations in A370.
The left panel shows the targeted fields at 870~$\mu$m,
each with a radius of $9''$.
The white circles in all panels show our detected sources
from these targeted observations. The circles are of arbitrary size but
roughly match this ALMA field-of-view.
The center panel shows the mosaicked 1.2~mm image
from ALMA programs \#2013.1.00999.S
and \#2015.1.01425.S
(PI:~F.~Bauer; Gonz\'alez-L\'opez et al.\ 2017),
while the right panel shows that from ALMA program \#2018.1.00035.L (PI:~K.~Kohno). 
The mosaicked images are taken from the JVO archive.
There are no additional detected sources in the mosaics.
{\em (Lower three panels)\/} The SCUBA-2 \afluxa image (left),
the HerMES 100~$\mu$m image (center), and the BUFFALO
F160W image.
\label{alma_images}
}
\end{figure*}
%---------------------------------------------------------------------

%---------------------------------------------------------------------
% FIGURE 4: A370 arc; a370_arc
%---------------------------------------------------------------------
\begin{figure}[htb]
\hskip 5.0cm
\centering{\includegraphics[width=2.72in,angle=0]{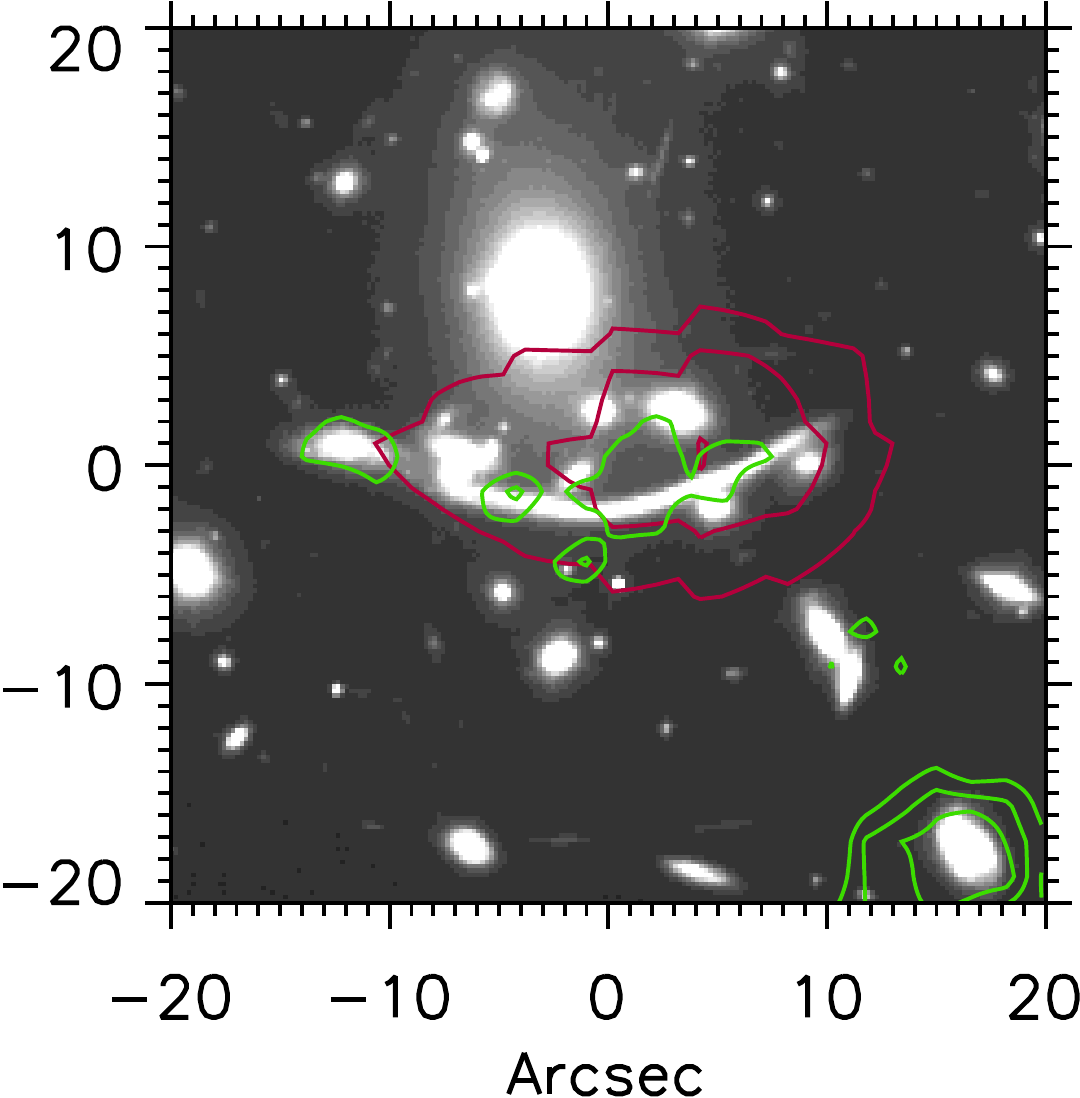}}
\caption{HFF F160W image of the giant arc in A370.
The red contour shows our \afluxb SCUBA-2 observation of the arc, and 
the green contour shows the Herschel/PACS
100~$\mu$m image (Rawle et al.\ 2016).
The image is $40''$ on a side.
\label{a370_arc}
}
\end{figure}
%---------------------------------------------------------------------

%---------------------------------------------------------------------
% FIGURE 5: HST images with ALMA overlays in A370
%---------------------------------------------------------------------
\begin{figure*}
\centering{\includegraphics[width=2.5in,angle=0]{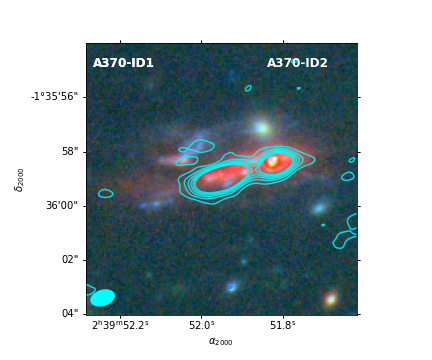}}
\hskip -0.75cm
\centering{\includegraphics[width=2.5in,angle=0]{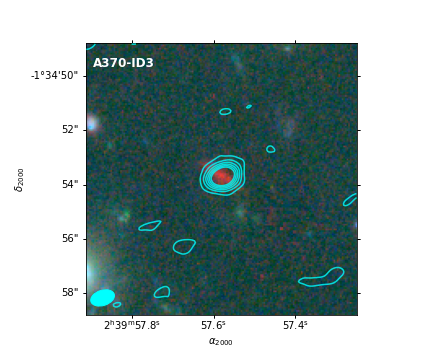}}
\hskip -0.75cm
\centering{\includegraphics[width=2.5in,angle=0]{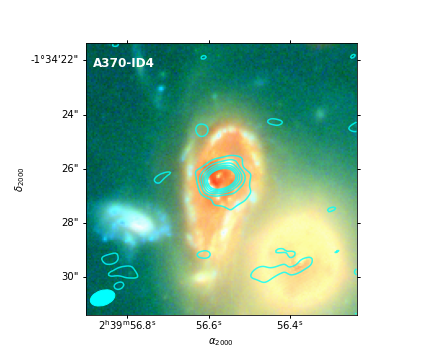}}
\vskip -0.8cm
\centering{\includegraphics[width=2.5in,angle=0]{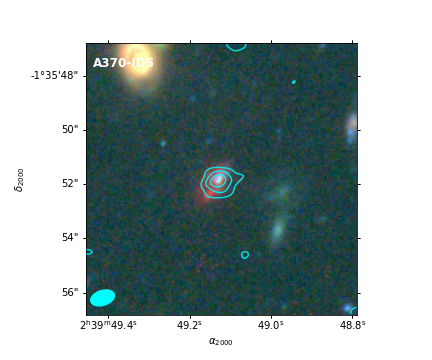}}
\hskip -0.75cm
\centering{\includegraphics[width=2.5in,angle=0]{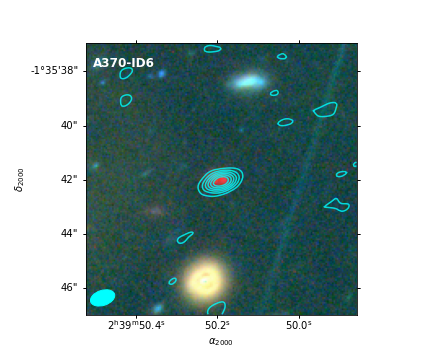}}
\hskip -0.75cm
\centering{\includegraphics[width=2.5in,angle=0]{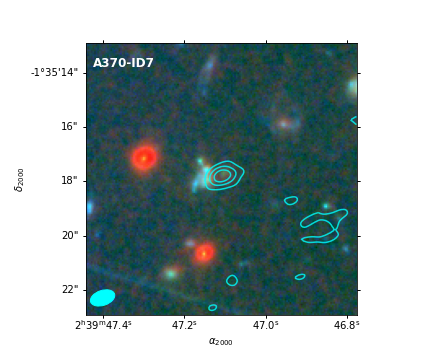}}
\vskip -0.8cm
\centering{\includegraphics[width=2.5in,angle=0]{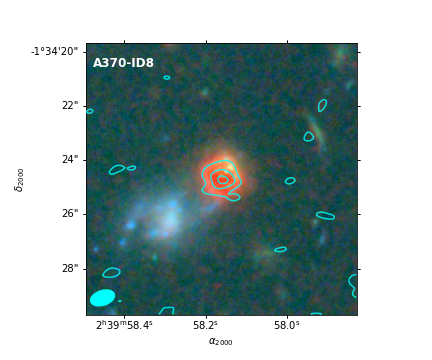}}
\hskip -0.75cm
\centering{\includegraphics[width=2.5in,angle=0]{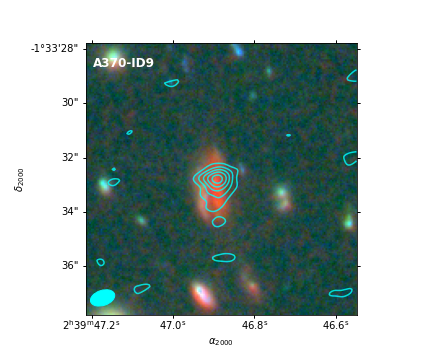}}
\hskip -0.75cm
\centering{\includegraphics[width=2.5in,angle=0]{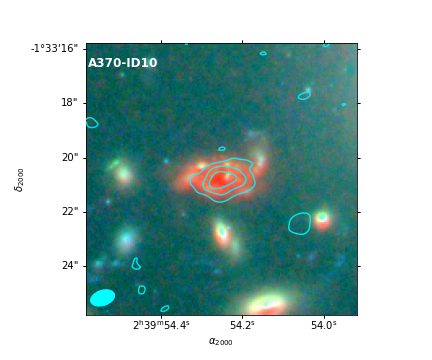}}
\caption{
HST three-color (red: BUFFALO/HFF F160W; green: HFF F814W; blue: HFF F606W) images of the A370 
sources with ALMA emission overlaid (green contours: 2, 4, 6, 8, 10, and $12\sigma$;
beam size is also shown in the lower left corner of each image).
\label{contour_images}
}
\end{figure*}
%---------------------------------------------------------------------

%%%%%%%%
\vskip 0.5cm
{\bf A370:}   
%%%%%%%%
In band~7 (870~$\mu$m), we observed 15 ALMA fields within a $2'$ radius from the cluster center.
These covered all 13 SCUBA-2 sources with measured \afluxb fluxes
in excess of 1.6~mJy in the region and two additional fields. We show
the ALMA images in the upper left panel of Figure~\ref{alma_images}.
We detect 10 sources ($>4.5\sigma$) in the ALMA images.
We summarize the properties of these sources
in Table~\ref{a370_band7}, and we mark their positions with
white circles in all panels of Figure~\ref{alma_images}. 
This includes one double source in the ALMA observations
(a well-known pair at $z=2.8$ studied by
Ivison et al.\ 1998), which blends into a single SCUBA-2 source.
Eight of the ALMA detections are  single sources
(one of which is at $z=1.056$ and was
studied by Barger et al.\ 1999).  In total, 9
of the 13 SCUBA-2 sources are detected with ALMA.

One of the four SCUBA-2 sources that was not detected by ALMA corresponds to
the giant arc at $z=0.724$ (Soucail et al.\ 1988).
This source is extended even in the SCUBA-2 image
(Figure~\ref{a370_arc}) and is seen as an elongated
source in the 100~$\mu$m Herschel/PACS image (Rawle et al.\ 2016).
Although the source is over-resolved and undetected in the ALMA
image, the SCUBA-2 identification is clear.

The three undetected sources are the faintest in the
SCUBA-2 sample with \afluxb fluxes between 1.6~mJy and
1.8~mJy. These sources may be missed if they are extended
or multiple, but they are also the most likely to be spurious
in the \afluxb sample.

We also searched the ALMA archive for additional data. 
Mosaics of the field in band~6 (1.2~mm) have been obtained by Gonz\'alez-L\'opez et al.\ (2017) 
and by ALMA program \#2018.1.00035.L (PI:~K.~Kohno). We show these in the upper
center and upper right panels of Figure~\ref{alma_images}, respectively,
using images taken from the Japanese Virtual Observatory (JVO) archive.
We summarize the three band~6 detected sources (two from Gonz\'alez-L\'opez et al.\
and one from ALMA program \#2018.1.00035.L) in Table~\ref{a370_band6},
but they all overlap with our band~7 sample. The
giant arc is not detected in either of the mosaicked images.

In the lower panels of Figure~\ref{alma_images}, we show the SCUBA-2 \afluxar, 
HerMES 100~$\mu$m, and BUFFALO F160W images, respectively.
In Figure~\ref{contour_images}, we show BUFFALO/HFF three-color images with 
the ALMA emission overlaid.

%---------------------------------------------------------------------
% FIGURE 6: ALMA images in MACSJ1149; kohno_image.pro and kohno_other.pro in macsj1149 dir
% final_macsj1149_kohno_image and final_macsj1149_kohno_other
%---------------------------------------------------------------------
\begin{figure*}
\hskip 5.0cm
\centering{\includegraphics[width=7in,angle=0]{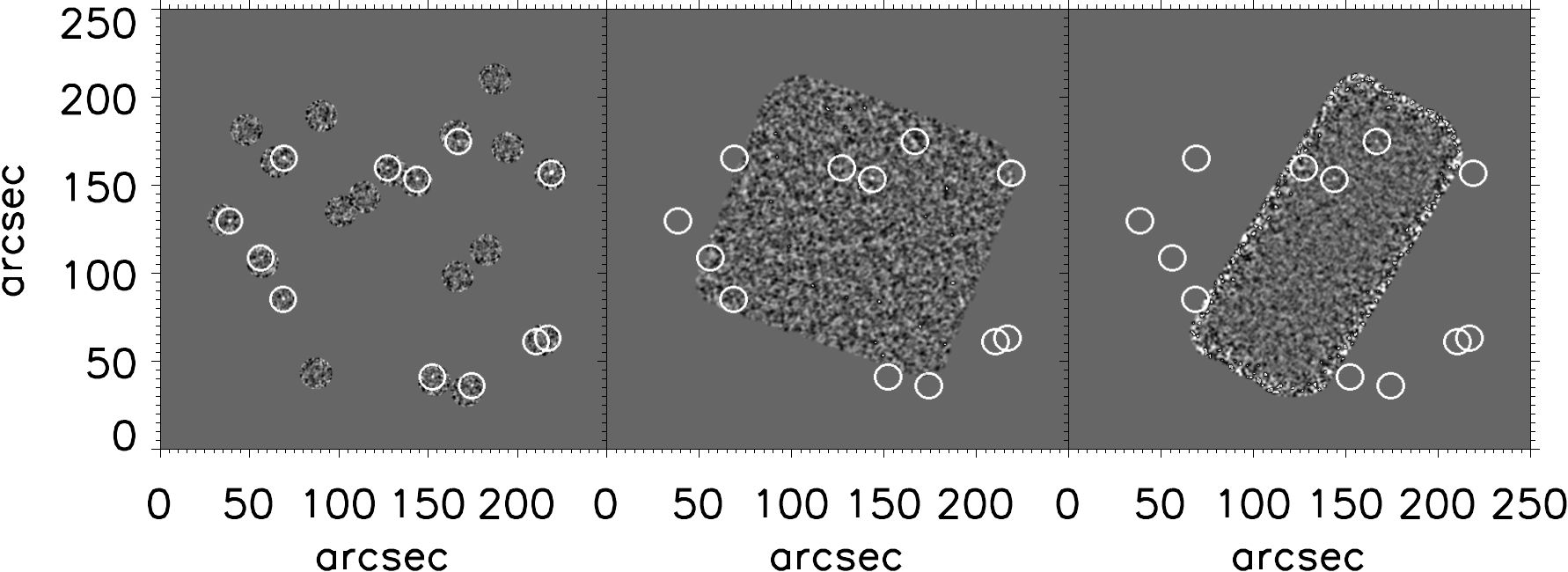}}
\vskip 0.7cm
\centering{\includegraphics[width=7in,angle=0]{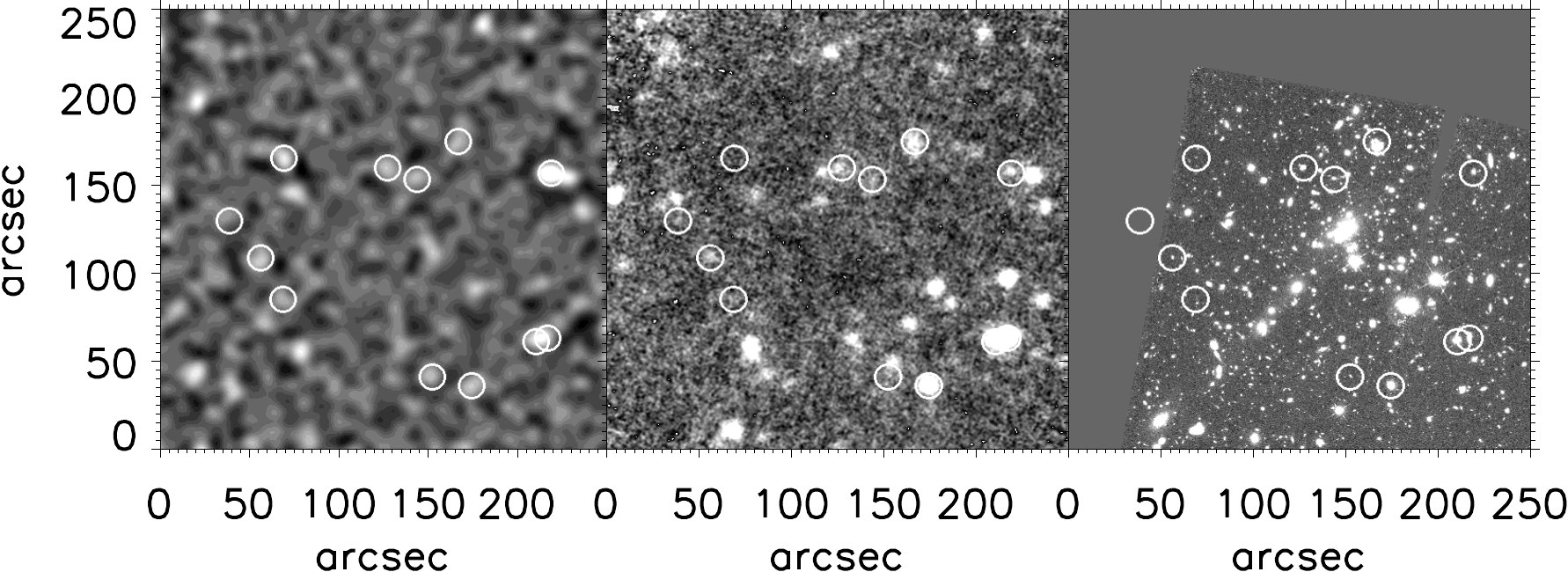}}
\caption{{\em (Upper three panels)\/} 
ALMA observations in MACSJ1149.
The left panel shows our targeted fields at 870~$\mu$m,
each with a radius of $9''$.
The white circles in all panels show our detected sources
from these targeted observations. The circles are of arbitrary size
but roughly match this ALMA field-of-view.
The center panel shows the mosaicked 1.2~mm image
from ALMA programs \#2013.1.00999.S
and \#2015.1.01425.S
(PI:~F.~Bauer; Gonz\'alez-L\'opez et al.\ 2017),
while the right panel shows that from ALMA program \#2018.1.00035.L (PI:~K.~Kohno). 
The mosaicked images are taken from the JVO archive.
There are no additional detected sources in the mosaics.
{\em (Lower three panels)\/} The SCUBA-2 \afluxa image (left),
the HLS 100~$\mu$m image (center), and the BUFFALO HST F160W image.
\label{alma_images2}
}
\end{figure*}
%---------------------------------------------------------------------

%%%%%%%%
\vskip 0.5cm
{\bf MACS\,J1149.5+2223:} 
%%%%%%%%
In band~7 (870~$\mu$m), we observed 22 ALMA fields within a $2'$ radius
from the cluster center. These included 19 of the 20
SCUBA-2 sources with \afluxb fluxes above 1.6~mJy, together
with 3 fainter sources. One SCUBA-2 source with a flux of 1.7~mJy, which
was marginally below our S/N selection threshold at the time of 
setting up the observations, was omitted. 
We show these targeted ALMA observations
in the upper left panel of Figure~\ref{alma_images2}.
We detect 12 sources ($>4.5\sigma$) in the ALMA images. We
summarize the properties of these sources in Table~\ref{macsj1149_band7},
and we mark their positions with white circles in all panels of Figure~\ref{alma_images2}. 
This includes one double source, where the separation of the two ALMA sources
is $6\farcs7$. This is small enough to produce a blended SCUBA-2 source.

ALMA mosaics of the field in band~6 (1.2~mm) were obtained by 
Gonz\'alez-L\'opez et al.\ (2017) and by ALMA program \#2018.1.00035.L (PI:~K.~Kohno).
We show these images, which we took from the JVO archive, in the upper center and 
upper right panels of Figure~\ref{alma_images2}, respectively.
There is one unique $>4.5\sigma$ detected source in each image, and they both 
overlap with our band~7 sample.
Targeted band~6 observations based on an AzTEC image
were also made as part of ALMA program
\#2016.1.00293.S (PI:~A.~Pope), and we summarize these eight detections 
in Table~\ref{macsj1149_band6}. Only two lie within our $2'$ selection radius,
both of which we detected in band~7.

In the lower panels of Figure~\ref{alma_images2}, we show the SCUBA-2 \afluxar, 
HLS 100~$\mu$m, and BUFFALO F160W images, respectively.

%---------------------------------------------------------------------
% FIGURE 7: ALMA images in MACSJ0717; kohno_image.pro and kohno_other.pro in macsj0717 dir
% final_macsj0717_kohno_image and final_macsj0717_kohno_other
%---------------------------------------------------------------------
\begin{figure*}
\hskip 5.0cm
\centering{\includegraphics[width=7in,angle=0]{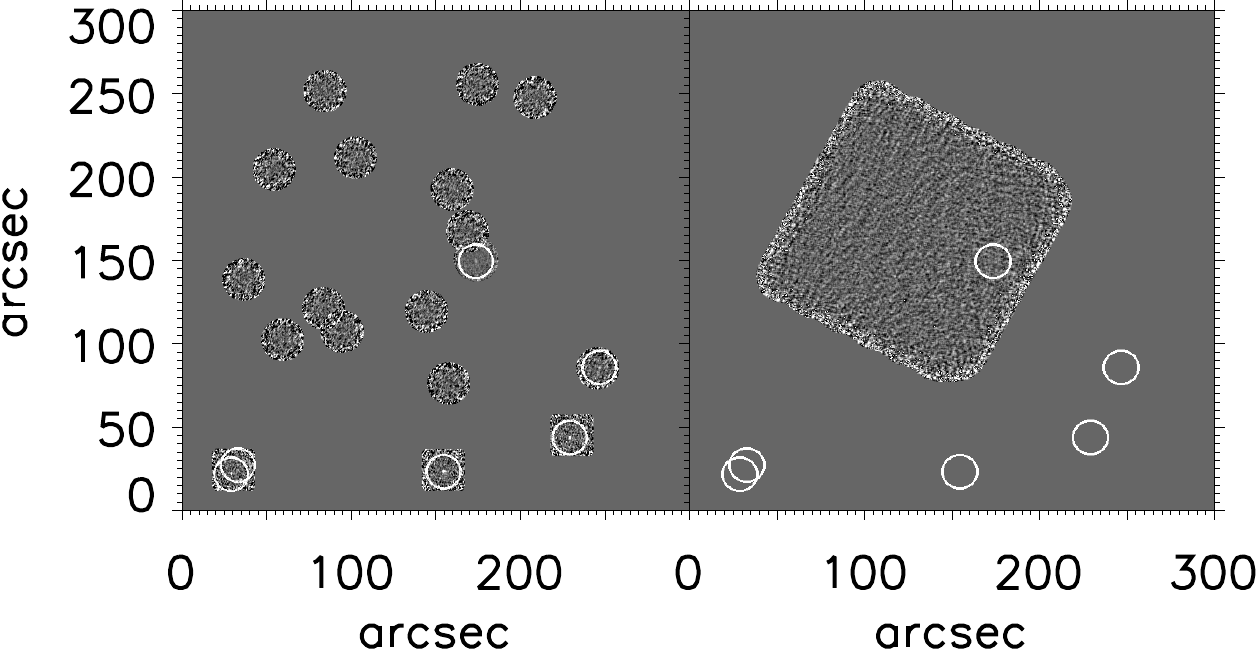}}
\vskip 0.7cm
\centering{\includegraphics[width=7in,angle=0]{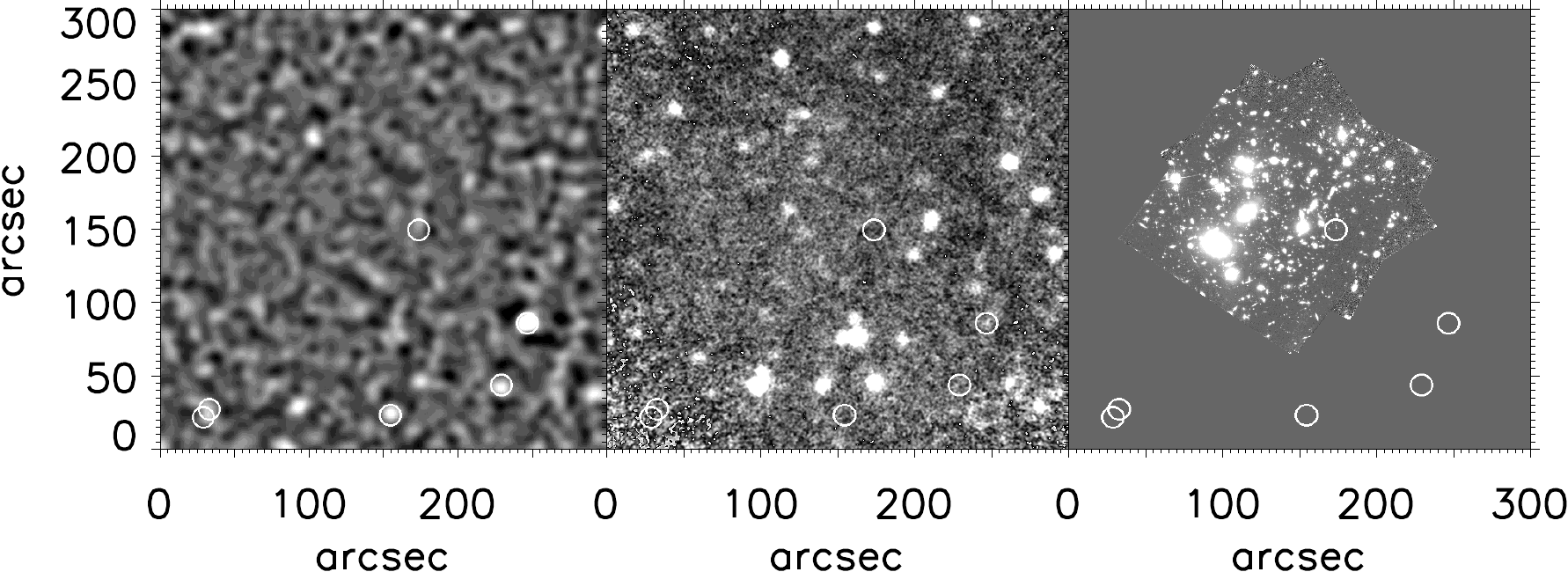}}
\caption{
{\em (Upper two panels)\/} ALMA observations in MACSJ0717.
The left panel shows the targeted fields at 870~$\mu$m from 
our program and from ALMA program \#2017.1.00091.S
(PI:~A.~Pope) as circular fields with a radius of $9''$,
as well as the targeted fields at 1.2~mm
from ALMA program \#2016.1.00293 (PI:~A.~Pope) as
square fields. 
The white circles in all panels show the detected sources in either band 
from these targeted observations.
The circles are of arbitrary size but roughly match this ALMA field-of-view.
The right panel shows the mosaicked
1.2~mm observations from ALMA programs \#2013.1.00999.S
and \#2015.1.01425.S (PI:~F.~Bauer; Gonz\'alez-L\'opez et al.\ 2017).
There are no additional detected sources in the mosaics.
{\em (Lower three panels)\/} The SCUBA-2 \afluxa image (left),
the HLS 100~$\mu$m image (center), and the BUFFALO
HST F160W image.
\label{alma_images3}
}
\end{figure*}
%---------------------------------------------------------------------

%---------------------------------------------------------------------
% FIGURE 8: Lensed images in MACSJ0717
% create_lens and create_lens_450 and create_alma and show_5_2
%---------------------------------------------------------------------
\begin{figure*}
\hskip -0.5cm
\centering{\includegraphics[width=3.5in,angle=0]{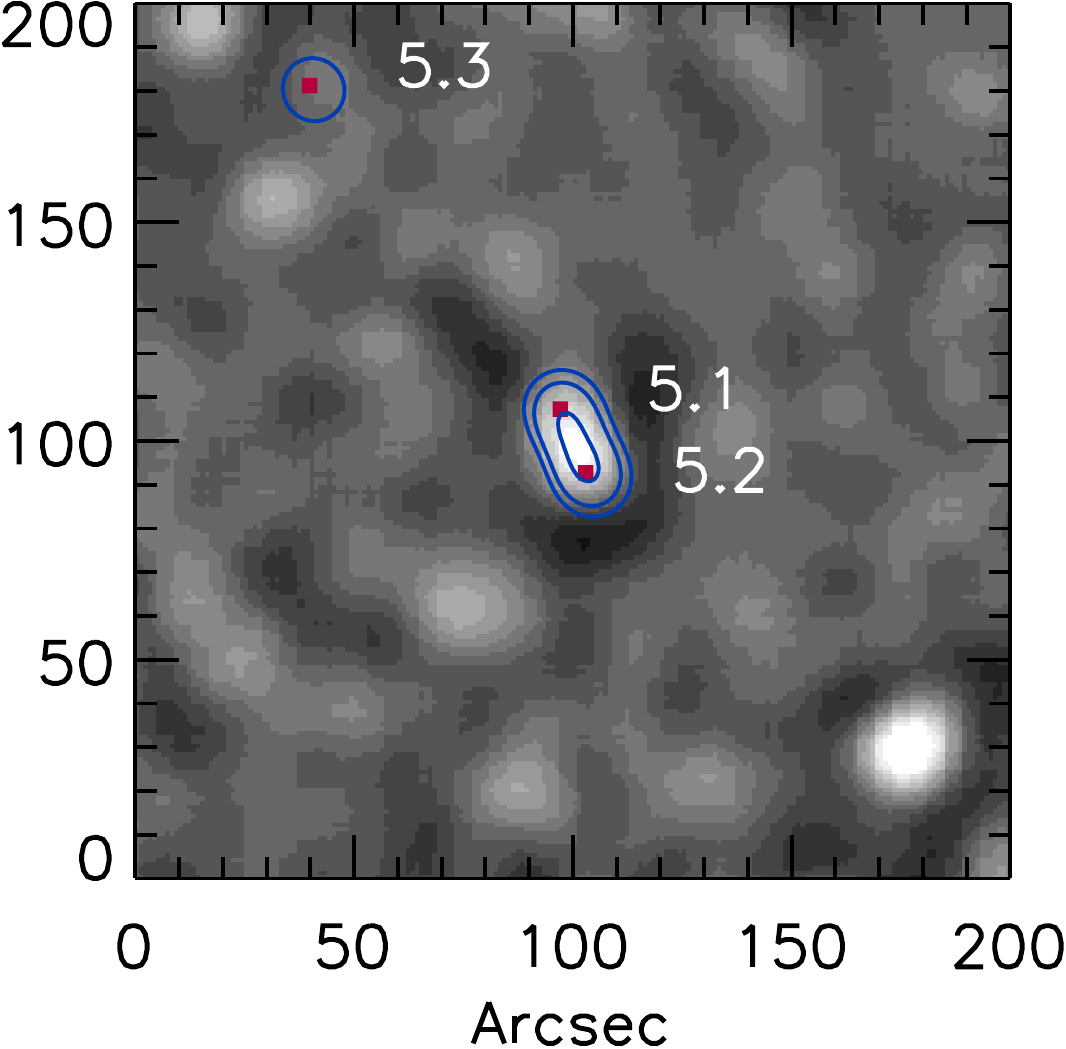}}
\centering{\includegraphics[width=3.5in,angle=0]{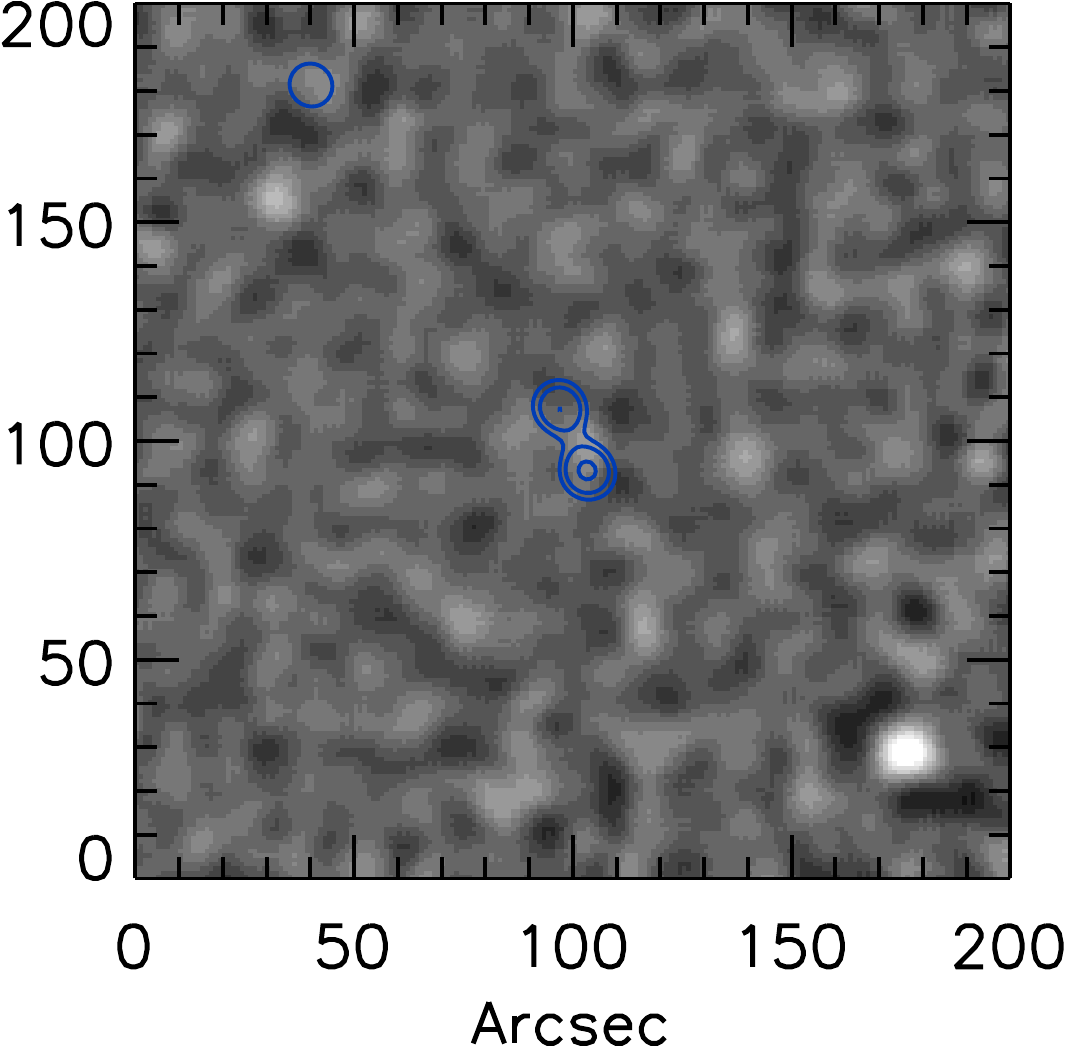}}
\centering{\includegraphics[width=3.5in,angle=0]{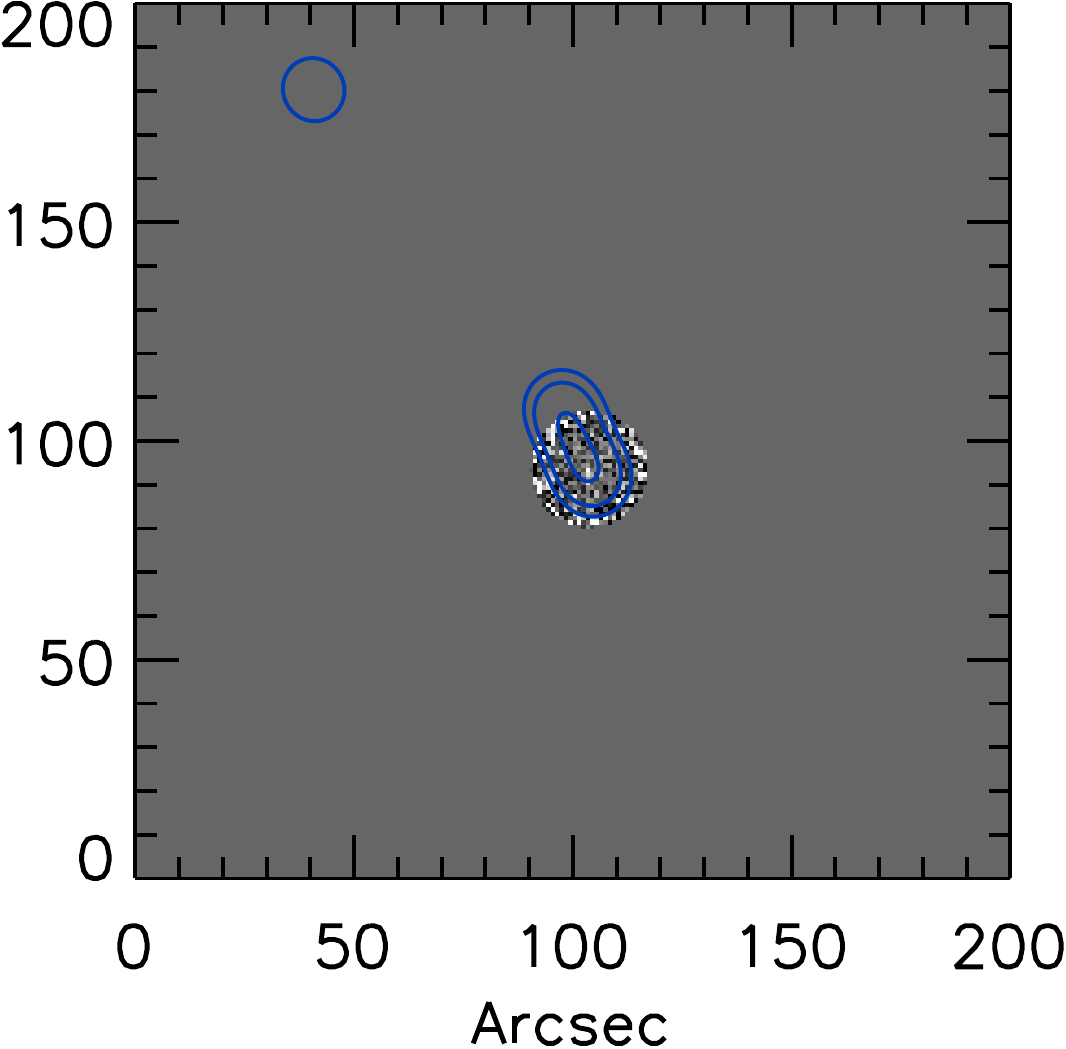}}
\centering{\includegraphics[width=3.5in,angle=0]{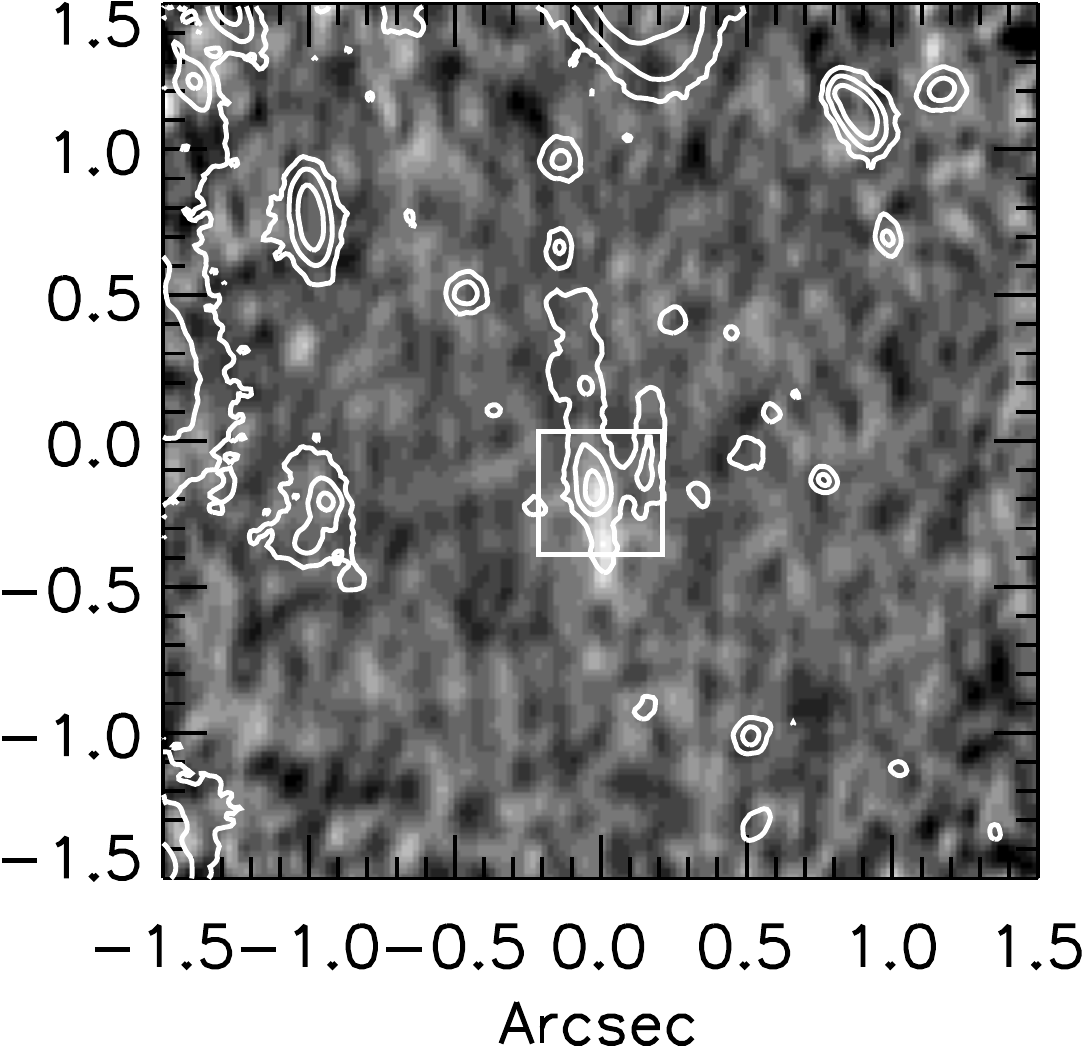}}
\caption{
Lensed source~5 in MACSJ0717 from Zitrin et al.\ (2009).
The overlaid blue contours in three of the four panels
show the predicted shape of the lensed source
based on the positions of images 5.1, 5.2, and 5.3 and their magnifications
of 6.8, 7.5, 3.0, respectively.
{\em (Upper left panel)\/} 
The SCUBA-2 \afluxb image with the predicted SCUBA-2 \afluxb shape overlaid.
In order to match the observed \afluxb flux,, 
we need a demagnified flux of 0.42~mJy for the source.
The positions of the lensed components are shown with red squares.  
The 5.1 and 5.2 images are marginally blended at the SCUBA-2
\afluxb FWHM. 
{\em (Upper right panel)\/} 
The \afluxa image with the predicted SCUBA-2 \afluxa shape
overlaid. The 5.1 and 5.2 images are resolved, but neither is detected.
{\em (Lower left panel)\/}
The ALMA 870~$\mu$m field containing the 5.2 image
with the predicted SCUBA-2 \afluxb shape overlaid.
The 5.1 image is off the top of the ALMA field. 
{\em (Lower right panel)\/} A blow-up of the
central region of the ALMA field, with the overlaid white
contours showing the HFF F160W image. The large
white square shows the position of the 5.2 image.  The
ALMA image shows two peaks, one associated with
the brightest position of 5.2 and a second associated
with a fainter region to the south.
\label{alma_lens1}
}
\end{figure*}
%---------------------------------------------------------------------

%%%%%%%%
\vskip 0.5cm
{\bf MACS\,J0717.5+37450:}
%%%%%%%%
The SCUBA-2 data show that the central region of this cluster
contains a surprisingly low number of sources (three)
with \afluxb fluxes greater
than 2~mJy within a $2'$ radius from the cluster center. 
Two of these sources are a close pair and
correspond to lensed source~5 from Zitrin et al.\ (2009).

Only half of the ALMA exposure time for our band~7 (870~$\mu$m) observations in this cluster 
were completed, 
and hence our sensitivity is considerably poorer than for the other two clusters,
with a typical central rms of 0.4~mJy. Only the southern source
of the close pair (corresponding to component 5.2 of the lensed source) 
is detected when our data are
combined with deeper archival band~7 data from ALMA program \#2017.1.00091.S
(PI:~A.~Pope). Pope et al.\ (2017) also detected
the source at 1.1~mm with the AzTEC camera on the Large Millimeter Telescope.
We summarize the source's properties in
Table~\ref{macsj0717_band7}, and we show the system in more detail in Figure~\ref{alma_lens1}.

This field was only partially observed in the ALMA band~6 program of 
Gonz\'alez-L\'opez et al.\ (2017). 
These observations are less sensitive than the other mosaics on the HFFs
and yielded no detections. Targeted band~6 observations based
on an AzTEC image were made as part of ALMA program \#2016.1.00293 (PI:~A.~Pope), and
we summarize these four bright detections, which lie outside our $2'$ selection radius,
in Table~\ref{macsj0717_band6}.

In the upper left panel of Figure~\ref{alma_images3},
we show both the band~6 (square fields) and band~7 (circular fields) targeted ALMA 
observations. 
In the upper right panel, we show the Gonz\'alez-L\'opez et al.\ (2017) band~6 mosaic. 
In the lower panels, we show the SCUBA-2 \afluxar, HLS 100~$\mu$m, 
and BUFFALO F160W images, respectively.
In all panels, we mark with white circles the positions of either a band~6 
or a band~7 detection from the targeted observations.

%---------------------------------------------------------------------
% FIGURE 9: Lensed images in RXJ1347
% rxj1347_create_lens and rxj1347_create_lens_450 and alma_52 and opt_52
%---------------------------------------------------------------------
\begin{figure*}
\includegraphics[width=3.5in,angle=0]{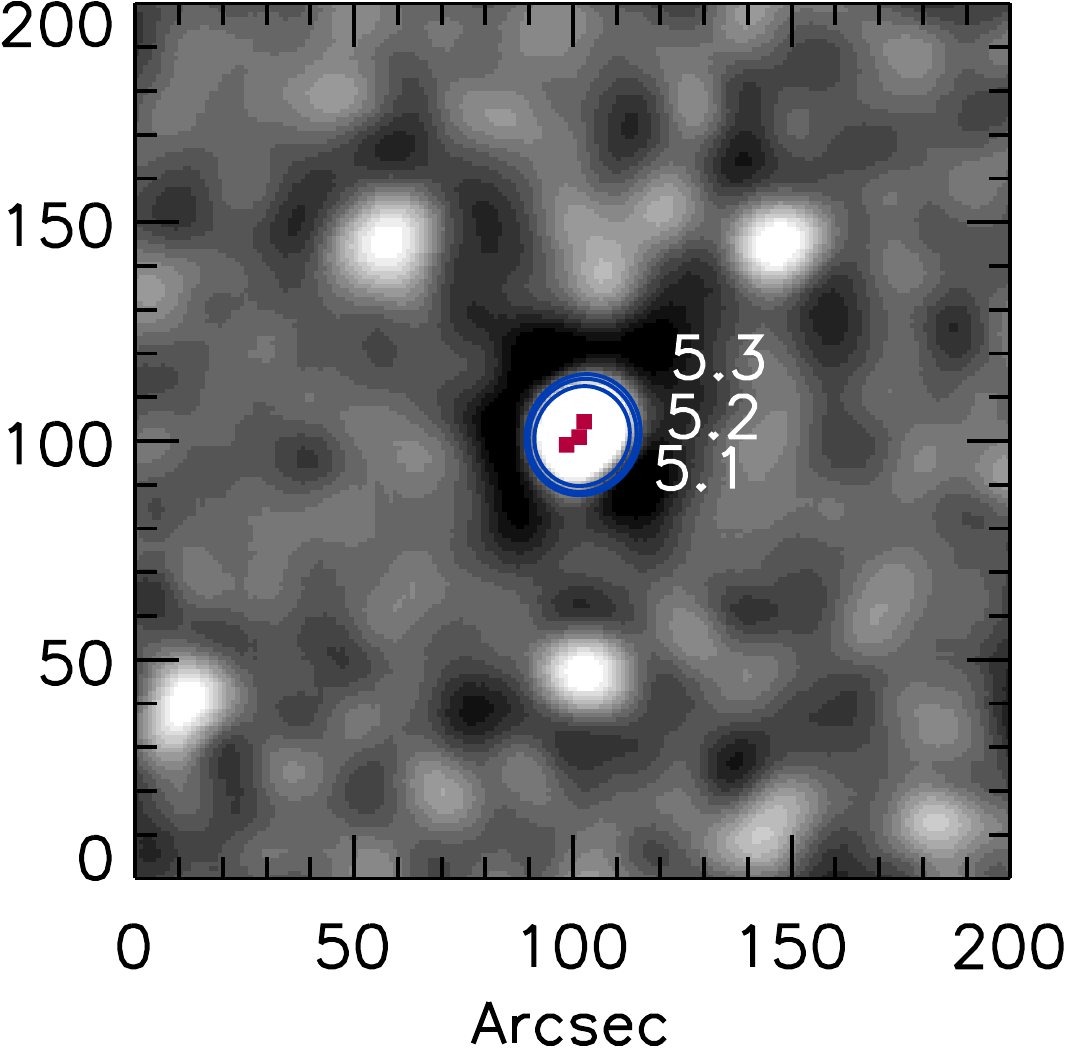}
\includegraphics[width=3.5in,angle=0]{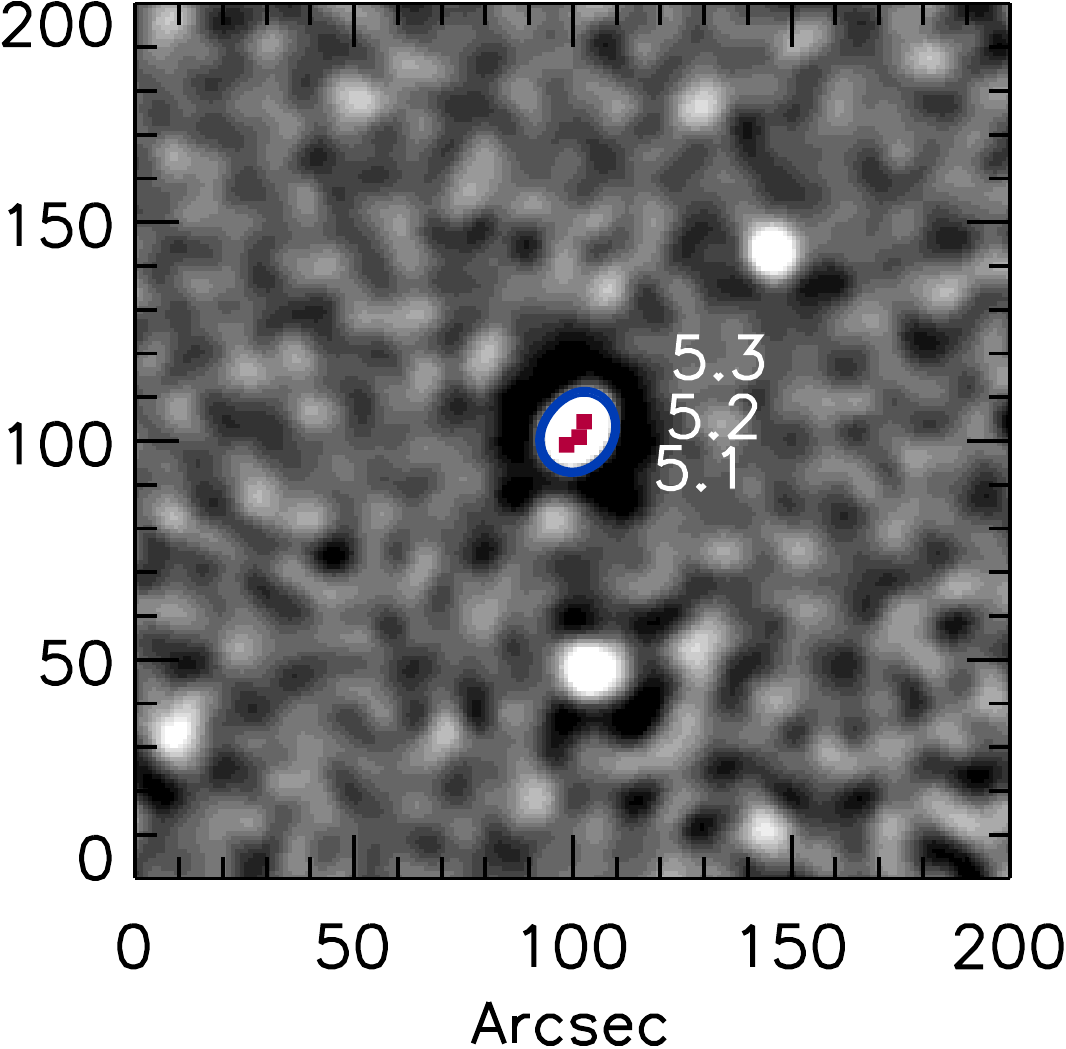}
\includegraphics[width=3.5in,angle=0]{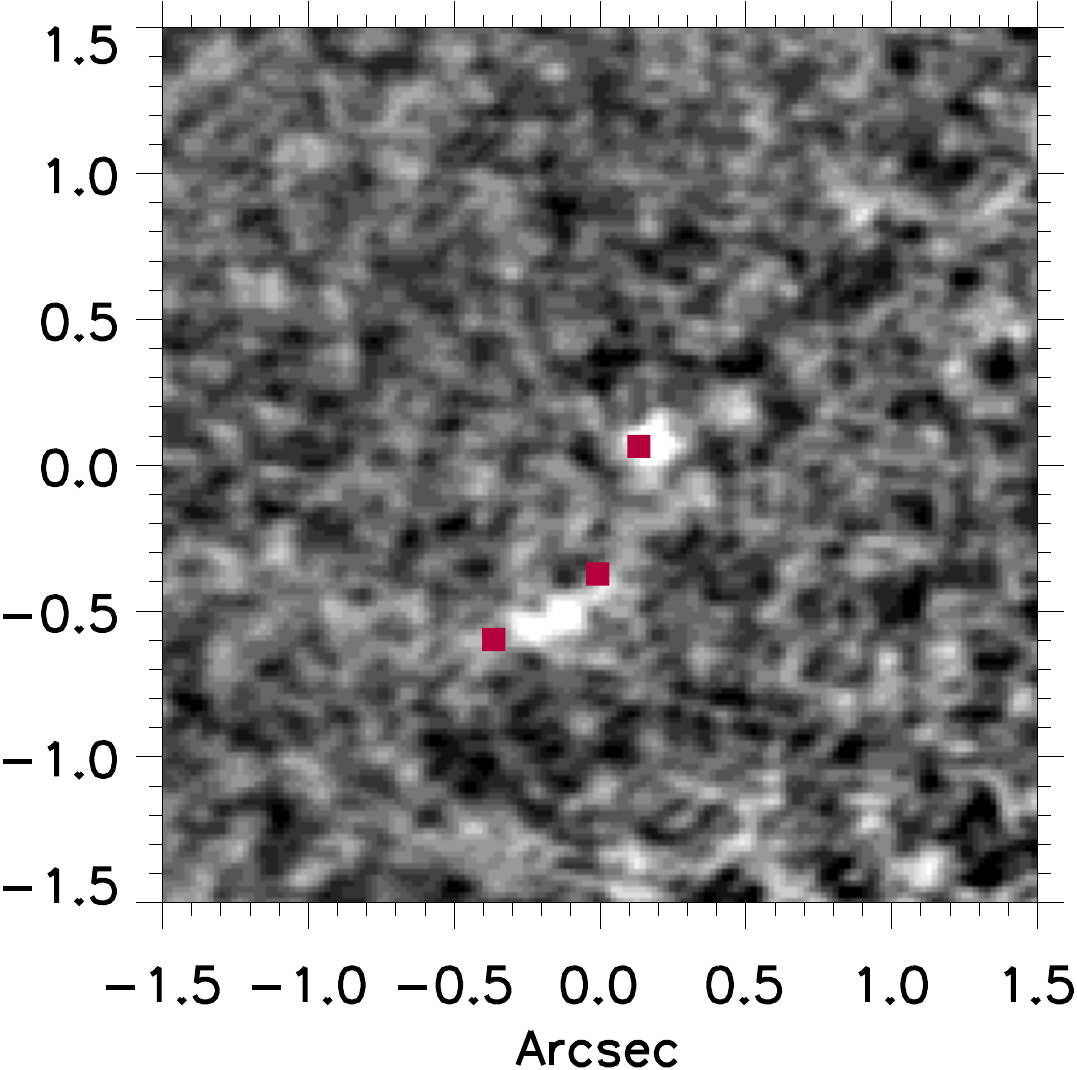}
\includegraphics[width=3.5in,angle=0]{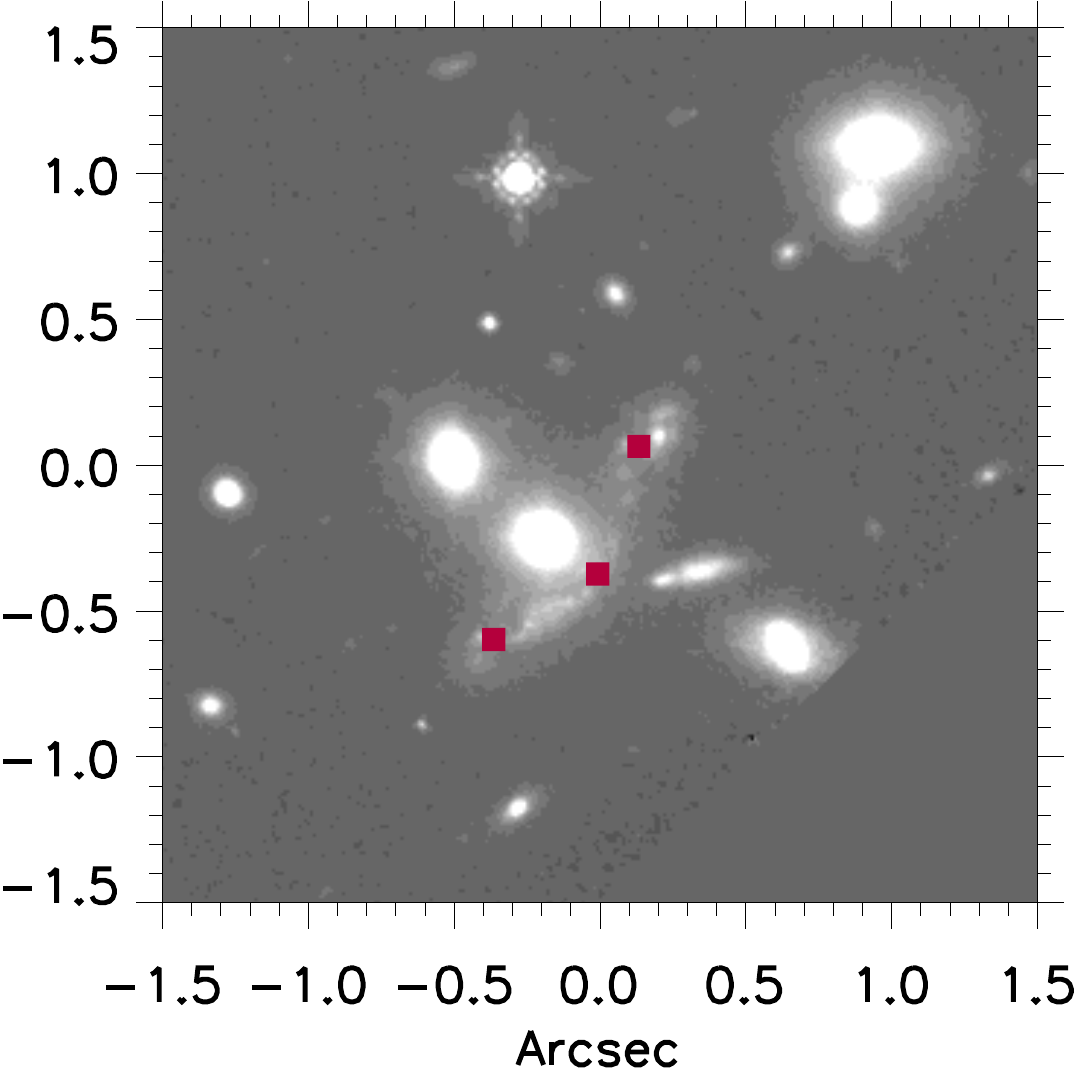}
\caption{
Lensed source~5 in RXJ1347 from Zitrin et al.\ (2015).
The overlaid blue contours in the top two panels
show the predicted shape of the lensed source
based on the positions of the lensed components 5.1, 5.2, and 5.3 and their magnifications
of 17.7, 2.2, and 18.2, respectively.
The positions of the lensed components are shown with red squares in all four panels.
Note that we applied a $0\farcs4$ astrometric correction
in R.A. to the Zitrin et al.\ (2015) coordinates to match them
to the HST image. This correction was used in all four panels.
{\em (Upper left panel)\/} 
The SCUBA-2 \afluxb image with the predicted SCUBA-2 \afluxb shape overlaid.
In order to match the observed \afluxb flux, we need a demagnified value
of 0.4~mJy for the source.
All three lensed components are blended at the SCUBA-2 \afluxb FWHM. 
{\em (Upper right panel)} 
The SCUBA-2 \afluxa image with the predicted SCUBA-2 \afluxa shape
overlaid. In order to match the observed \afluxa flux, we need a demagnified value
of 0.66~mJy for the source. 
All three lensed components are still blended at the SCUBA-2 \afluxa FWHM.
Note that the dark region around the source in the SCUBA-2 images is a result of the
Mexican hat-shape filter matching.
{\em (Lower left panel)\/}
The ALMA 1.2~mm image of the source from ALMA program 
2018.1.00035.L (PI:~K.~Kohno).
{\em (Lower right panel)\/} 
The CLASH F160W image of the source.
\label{alma_lens2}
}
\end{figure*}
%---------------------------------------------------------------------

%---------------------------------------------------------------------
\subsection{Remaining Clusters}
%---------------------------------------------------------------------
We searched the ALMA archives for detected sources
lying within a radius of $5'$ from the cluster centers.
These observations are primarily at millimeter wavelengths, but 
having accurate positions for SMGs is useful for our analysis in 
Section~\ref{interpretALMA}. We summarize the ALMA
positions of the detected sources in Table~\ref{archivetable}.
In some  cases, the ALMA source lies below the
SCUBA-2 detection threshold, but we give the measured 
SCUBA-2 \afluxb and \afluxa fluxes at the ALMA
position for all the ALMA sources.
For the SCUBA-2 errors, we list the white
noise, since we are pre-selecting from ALMA. 

\vskip 0.5cm
Breakdown by cluster:
\vskip 0.5cm

\begin{itemize}

\item[$\bullet$] A2390 and MACSJ1423: Only the BCGs have ALMA archival observations.

\item[$\bullet$] A1689:  There are 4 detected sources in the ALMA archive.
One is the complex, heavily studied $z=7.13$ source A1689-zd1 
(Watson et al.\ 2015; Wong et al.\ 2022). Two others appear to be cluster members
based on the photzs. Only the final source is bright enough
to be detected in the SCUBA-2 sample.

\item[$\bullet$] A2744: There are 9 detected sources in the ALMA archive,
seven of which come from Gonz\'alez-L\'opez et al.\ (2017)
and the remaining 2 from ALMA program \#2017.1.01219.S (PI:~F.~Bauer).
Seven of the nine sources are detected in the SCUBA-2 images.

\item[$\bullet$] MACSJ0416: There are 5 detected sources in the ALMA archive,
four of which come from Gonz\'alez-L\'opez et al.\ (2017);
two of these four have SCUBA-2 detections.
One source is near coincident with the
strong lens component 12.3 in Hoag et al.\ (2016), but we cannot
obtain a consistent solution for the submillimeter fluxes at all three positions.
Thus, we do not believe this SMG is associated with the lensed system.

\hskip 0.35cm
The fifth source is MACS0416\_Y1, which lies at a redshift
of $z=8.311$ based on ALMA fine structure line measurements
(Tamura et al.\ 2019; Bakx et al.\ 2020).
Consistent with the ALMA measured 870~$\mu$m flux of 0.14~mJy 
(Bakx et al.\ 2020), this source is not detected in the SCUBA-2 imaging.

\item[$\bullet$] RXJ1347: 
Apart from the BCG, the only other ALMA detections 
from ALMA program \#2018.1.00035.L (PI:~K.~Kohno)
are two components of a lensed system (5.1, 5.2, 5.3 of Zitrin et al.\ 2015).
We give their positions in Table~\ref{archivetable}.
Zitrin et al.\  measured a photz of 1.28 for the lensed system, which we give
in the table, but note that Brada\v{c} et al.\ (2008)
placed it at a redshift of $z=4$ based on their gravitational lens modeling.
We show the system in more detail in Figure~\ref{alma_lens2}.

\end{itemize}

%---------------------------------------------------------------------
\section{Interpreting the ALMA sample}
\label{interpretALMA}
%---------------------------------------------------------------------
In this section, we analyze the redshift distributions, demagnified flux distributions, 
and flux ratios of the ALMA cluster sample (we exclude the BCGs)
with either speczs or photzs. This includes nearly all our band~7 A370, MACSJ1149, and
MACSJ0717 sources 
(Tables~\ref{a370_band7}, \ref{macsj1149_band7}, and \ref{macsj0717_band7})
and nearly all the (mostly band~6) remaining cluster sources in our
Table~\ref{archivetable}.

%---------------------------------------------------------------------
% FIGURE 10 ; plot_m160_z
%---------------------------------------------------------------------
\begin{figure}
\vskip -0.1cm
\includegraphics[width=3.4in,angle=0]{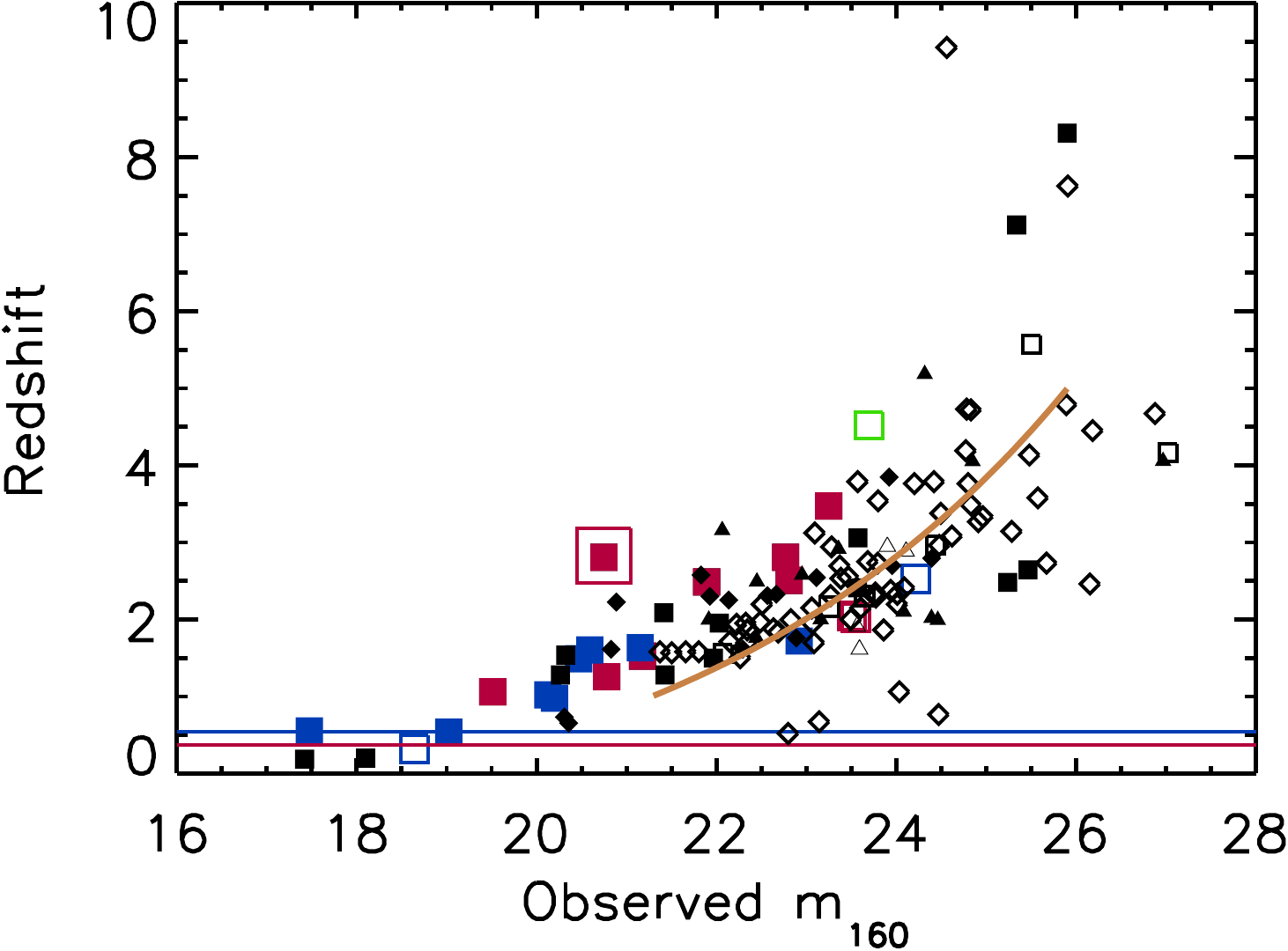}
\caption{
Redshift vs. observed F160W magnitude. The red squares
show A370, blue MACSJ1149, green MACSJ0717,
and black the remaining cluster fields. The AGN in A370 is marked
with a larger open square. The black diamonds and triangles
show the CDF-S and CDF-N fields, respectively. In all fields,
solid symbols show speczs and open symbols show photzs.
The horizontal lines show the redshifts of two of the clusters
(A370 in red and MACSJ1149 in blue).
\label{m160_plot_z}
}
\end{figure}
%---------------------------------------------------------------------

%---------------------------------------------------------------------
\subsection{Redshifts}
%---------------------------------------------------------------------
Measuring the redshifts of SMGs is key 
to determining their physical sizes, luminosities, and stellar masses, along
with accurate magnifications for the lensed sources.
Where possible, we use the optical/NIR spectroscopic redshifts (hereafter, speczs) 
that we either obtained with the DEIMOS, LRIS, and MOSFIRE instruments
on the Keck telescopes or pulled from the literature.
However, a number of the SMGs are faint at optical/NIR wavelengths and 
hence do not have speczs. For these objects, we use photzs.
We give all the redshifts and their sources in the tables.

%---------------------------------------------------------------------
% FIGURE 11: Redshift distribution
% a2744_redshift_hist and macsj0416_redshift_hist
% a370_redshift_hist and macsj0717_redshift_hist and macsj1149_redshift_hist
%---------------------------------------------------------------------
\begin{figure}[th]
\vskip -0.1cm
\includegraphics[width=2.6in,height=1.5in,angle=0]{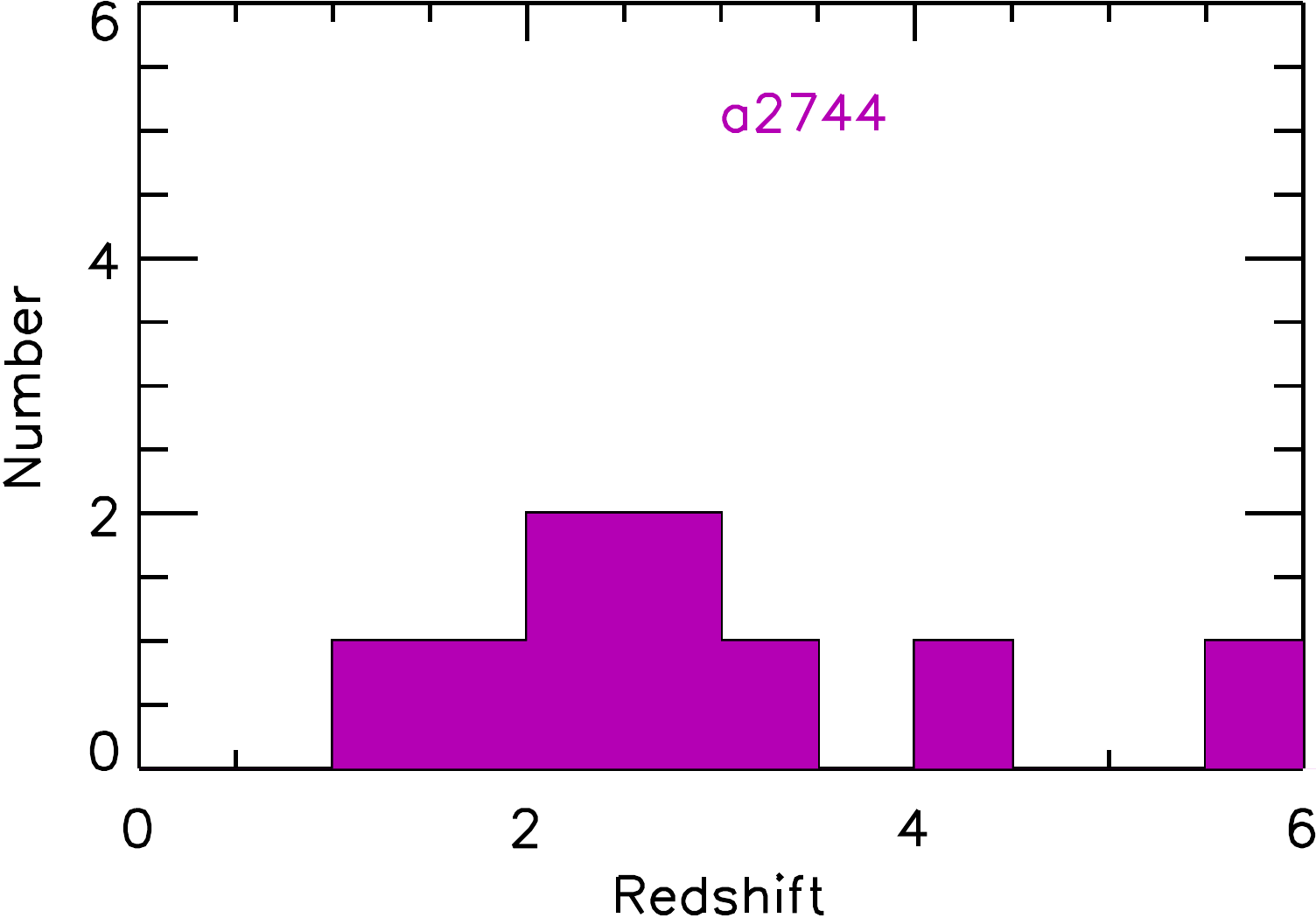}
\includegraphics[width=2.6in,height=1.5in,angle=0]{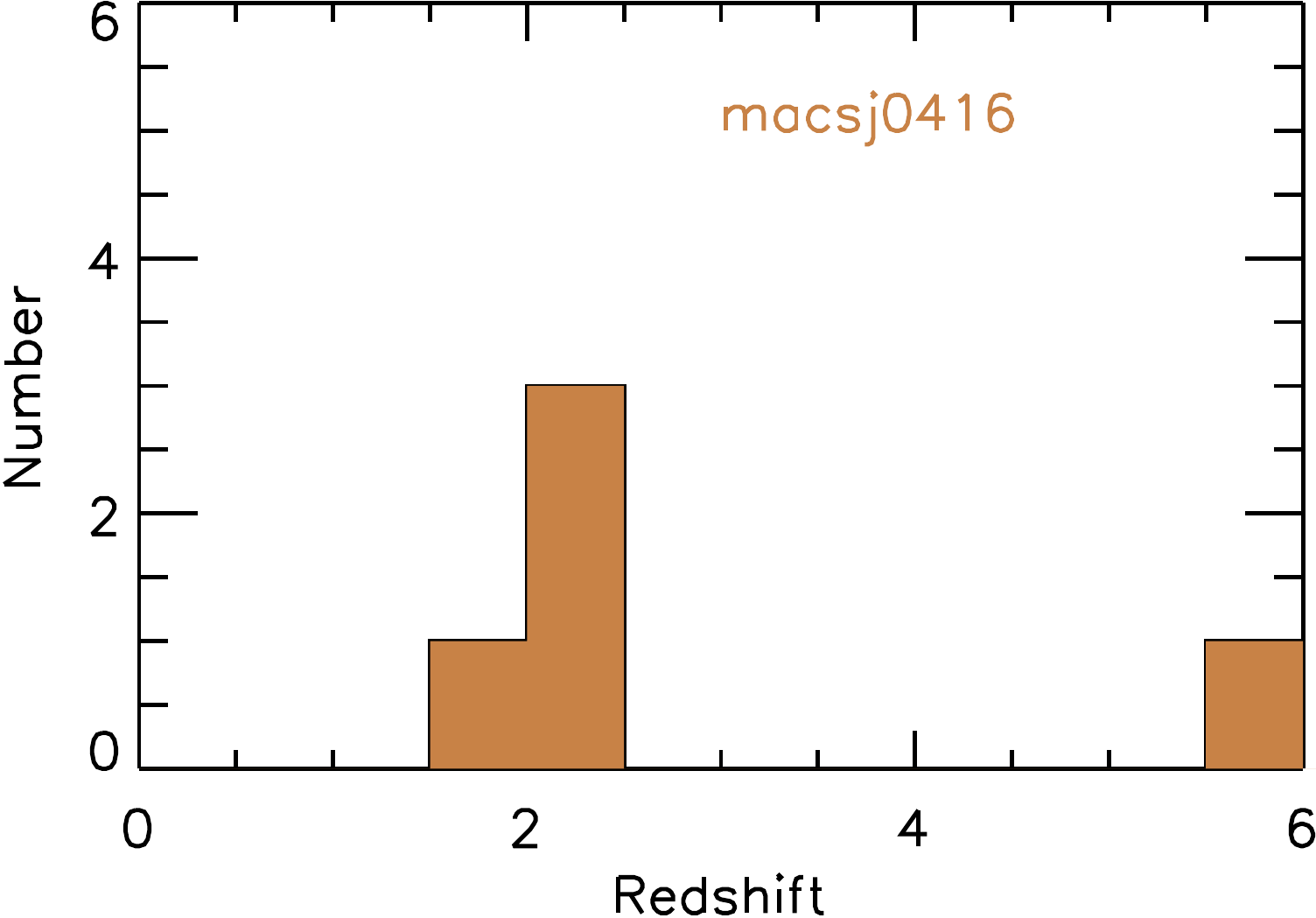}
\includegraphics[width=2.6in,height=1.5in,angle=0]{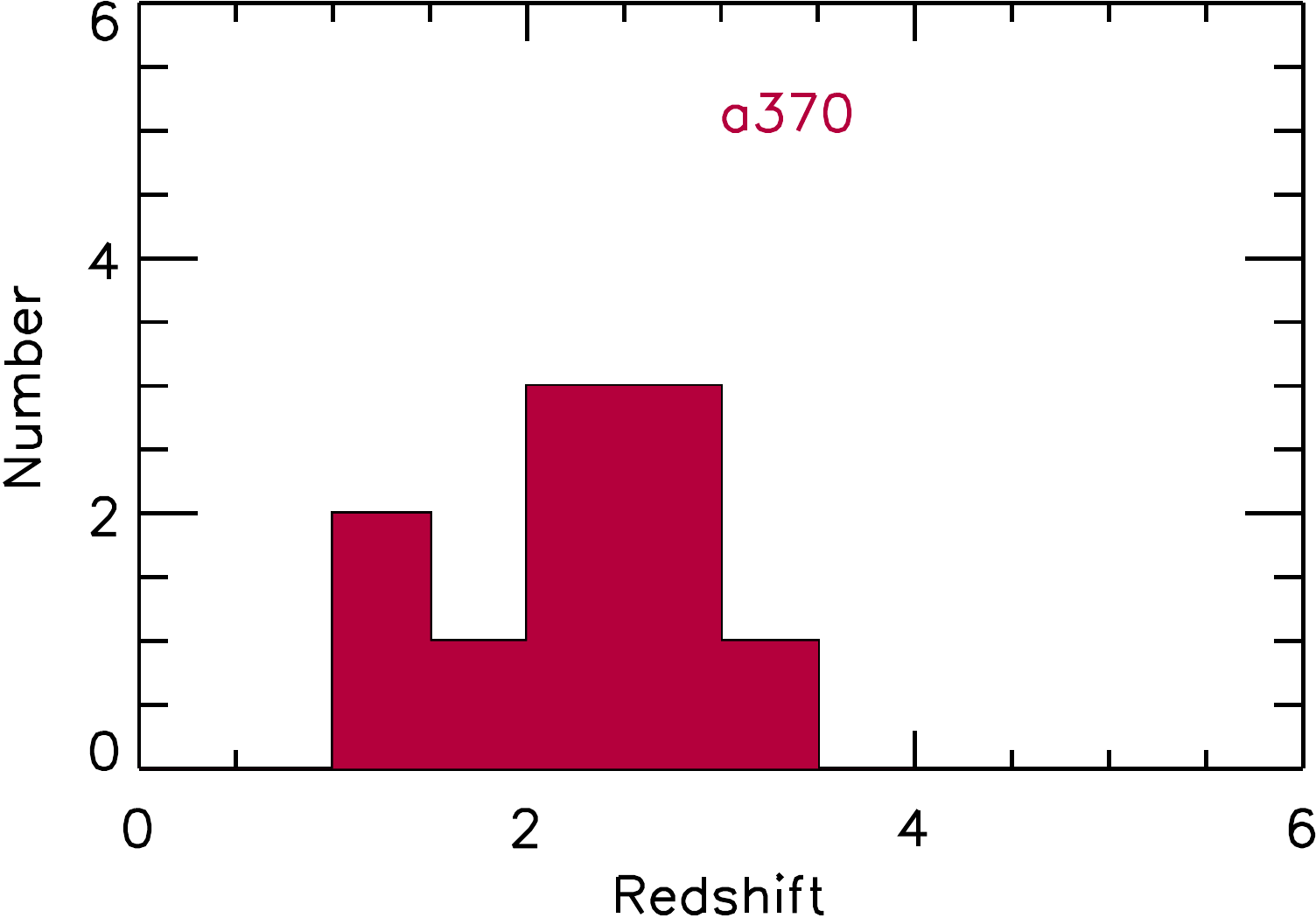}
\includegraphics[width=2.6in,height=1.5in,angle=0]{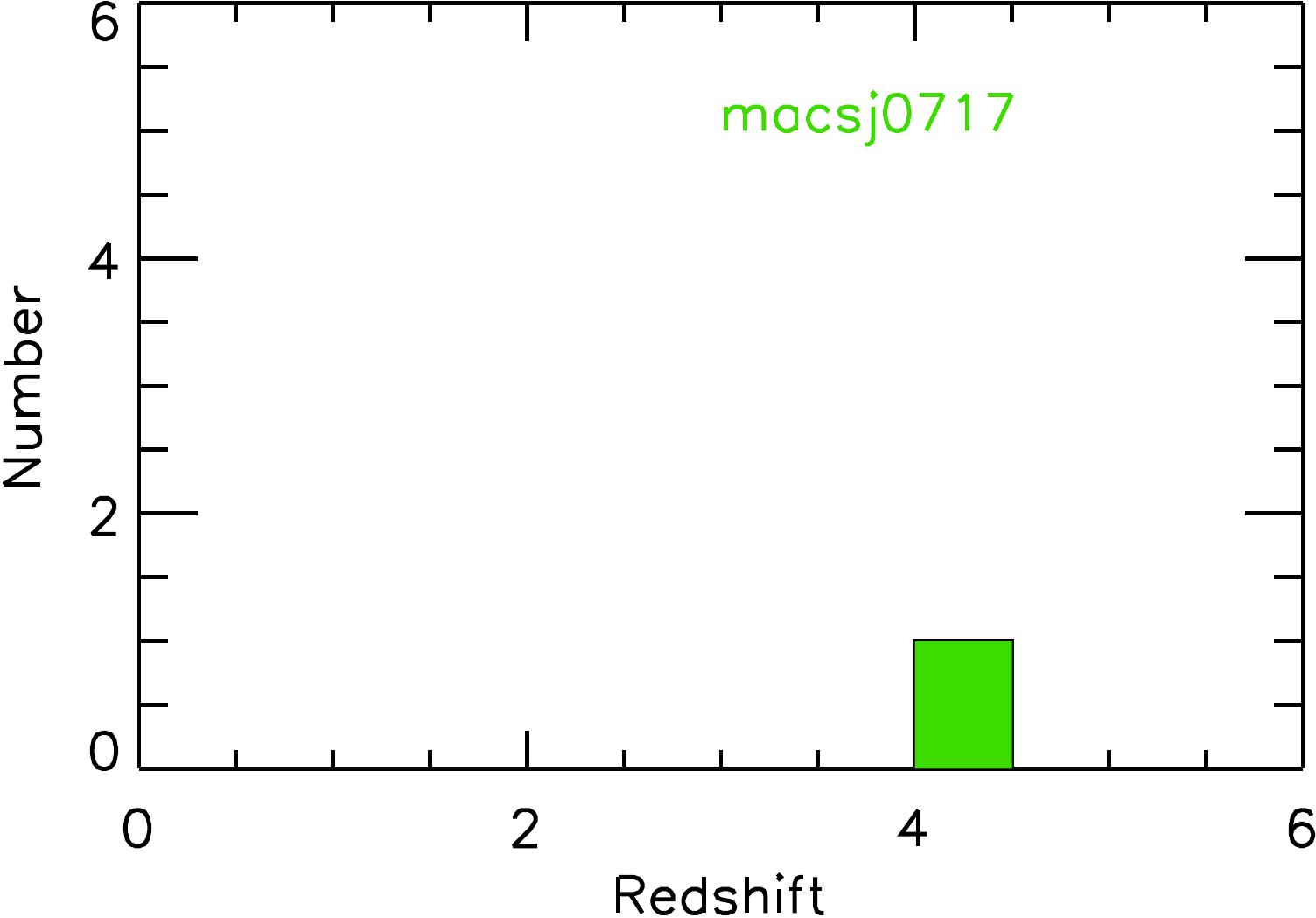}
\includegraphics[width=2.6in,height=1.5in,angle=0]{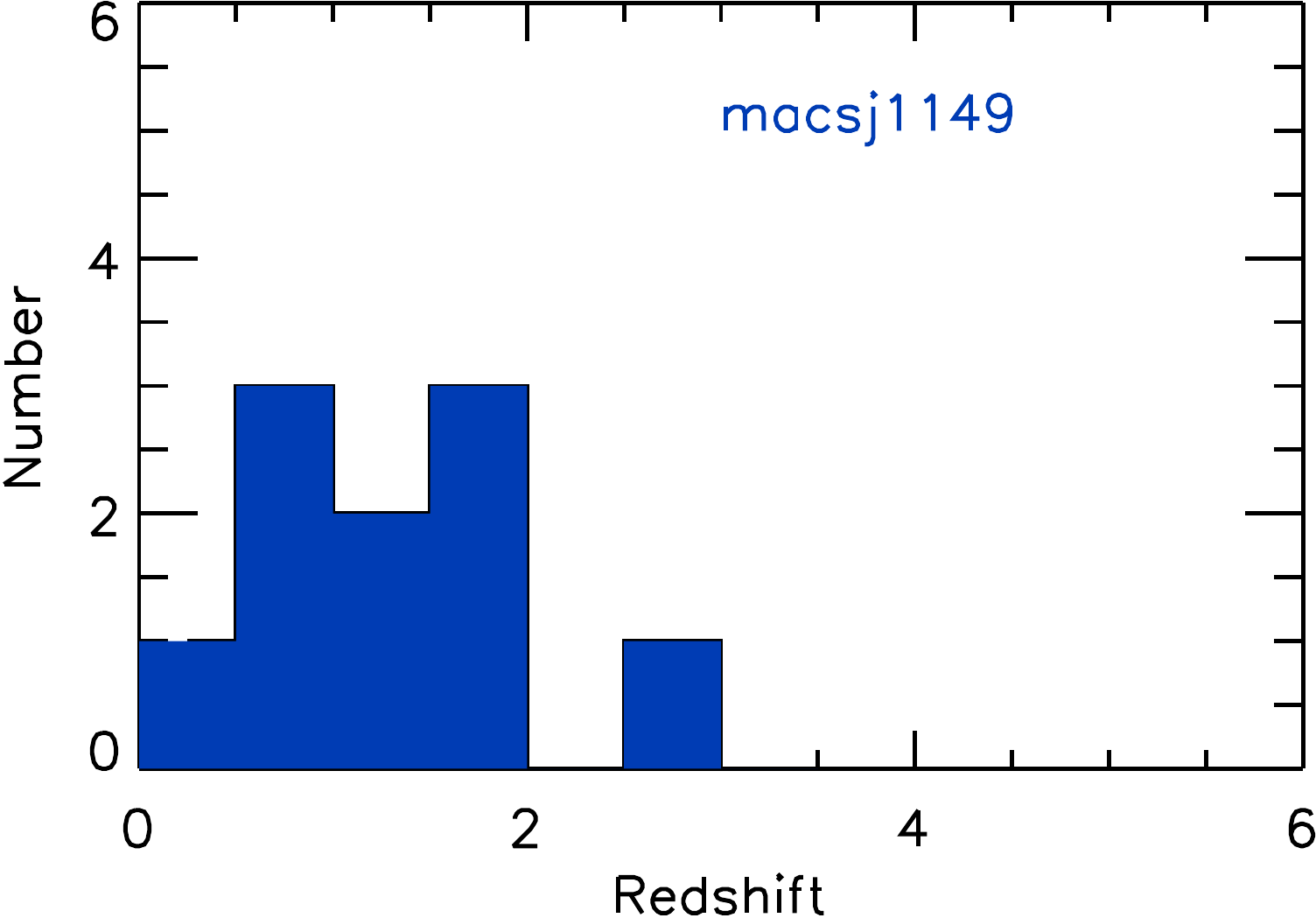}
\caption{
Redshift distributions for the five HFFs contained in the ALMA
cluster sample with either speczs or photzs. 
Sources with $z>6$ are shown in the $z=5.5$--6 bin.
\label{histogram}
}
\end{figure}
%---------------------------------------------------------------------

We plot redshift versus observed HST F160W magnitude 
for the sample in Figure~\ref{m160_plot_z}. 
We also include unlensed sources from the CDF-N and
CDF-S fields (Cowie et al.\ 2017, 2018) for comparison.
We note that there is a strong correlation of the redshift
with the observed F160W magnitude, with fainter sources being
systematically at higher redshift. This means, in turn, that
the spectroscopically identified sources are strongly biased
to lower redshifts, and that the highest redshift sources
may be hard to obtain even photzs for.

Different lensing cluster fields appear to have different redshift distributions.
In Figure~\ref{histogram}, we show the redshift distributions for
the 5 HFFs contained in the sample. Some fields,
such as MACSJ1149, are dominated by low-redshift ($z<2$) sources,
while other fields, such as A2744, have a significant number
of high-redshift ($z>4$) sources. This variance, which is
a consequence of the small field sizes,  emphasizes the need
for multiple fields to obtain properly averaged redshift distributions.

\begin{figure}
\vskip -0.1cm
\includegraphics[width=3.2in,angle=0]{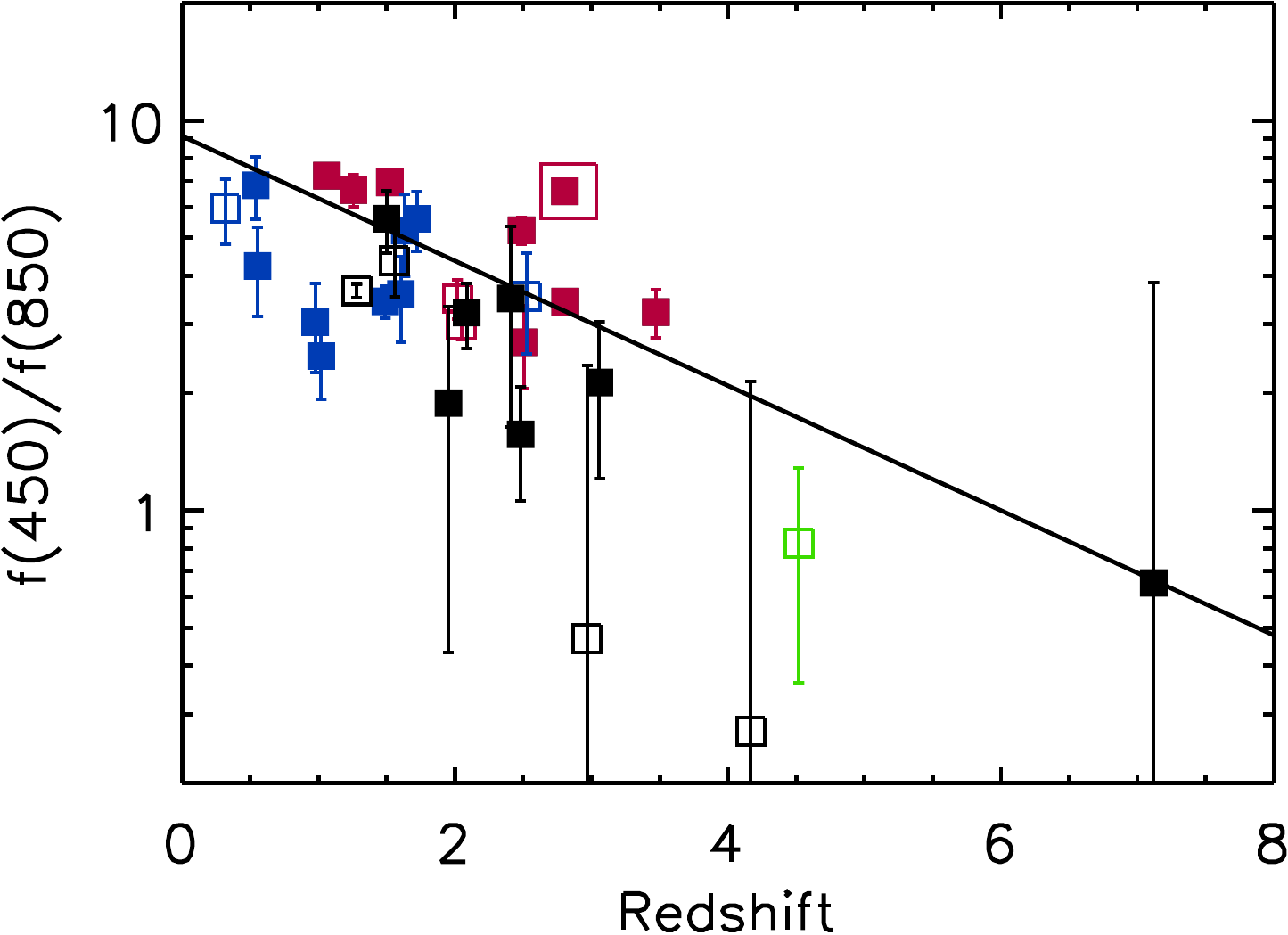}
\caption{
SCUBA-2 \afluxa to \afluxb flux ratio vs.
redshift for A370 (red), MACSJ1149 (blue), MACSJ0717 (green),
and the remaining cluster fields (black). Solid symbols show speczs,
and open symbols show photzs. The AGN in A370 is marked
with a large open square. Only sources detected above
the $2\sigma$ level in the \afluxb images are shown.
The power law fit (solid line) is taken from Barger et al.\ (2022) 
and is based on the CDF-N and CDF-S.
\label{plot_f4f8_z}
}
\end{figure}

The \afluxa to \afluxb flux ratio may be used to make
a rough estimate of the redshift, with lower
ratios corresponding to higher redshift sources (e.g.,
Barger et al.\ 2022). 
We show the \afluxa to \afluxb flux ratio versus redshift
in Figure~\ref{plot_f4f8_z} 
compared with the power law fit from Barger et al.\
for the CDF-N and CDF-S.
High redshifts ($z>4$) generally correspond to sources where 
the flux ratio is less than 2.

%---------------------------------------------------------------------
% FIGURE 12: Redshift vs. 450/850 ratio
% plot_f4f8_z
%---------------------------------------------------------------------
%---------------------------------------------------------------------

%-------------------------------:--------------------------------------
% FIGURE 13: Flux distribution
% newhist
%---------------------------------------------------------------------
\begin{figure}[tbh]
\vskip -0.1cm
\includegraphics[width=3.2in,angle=0]{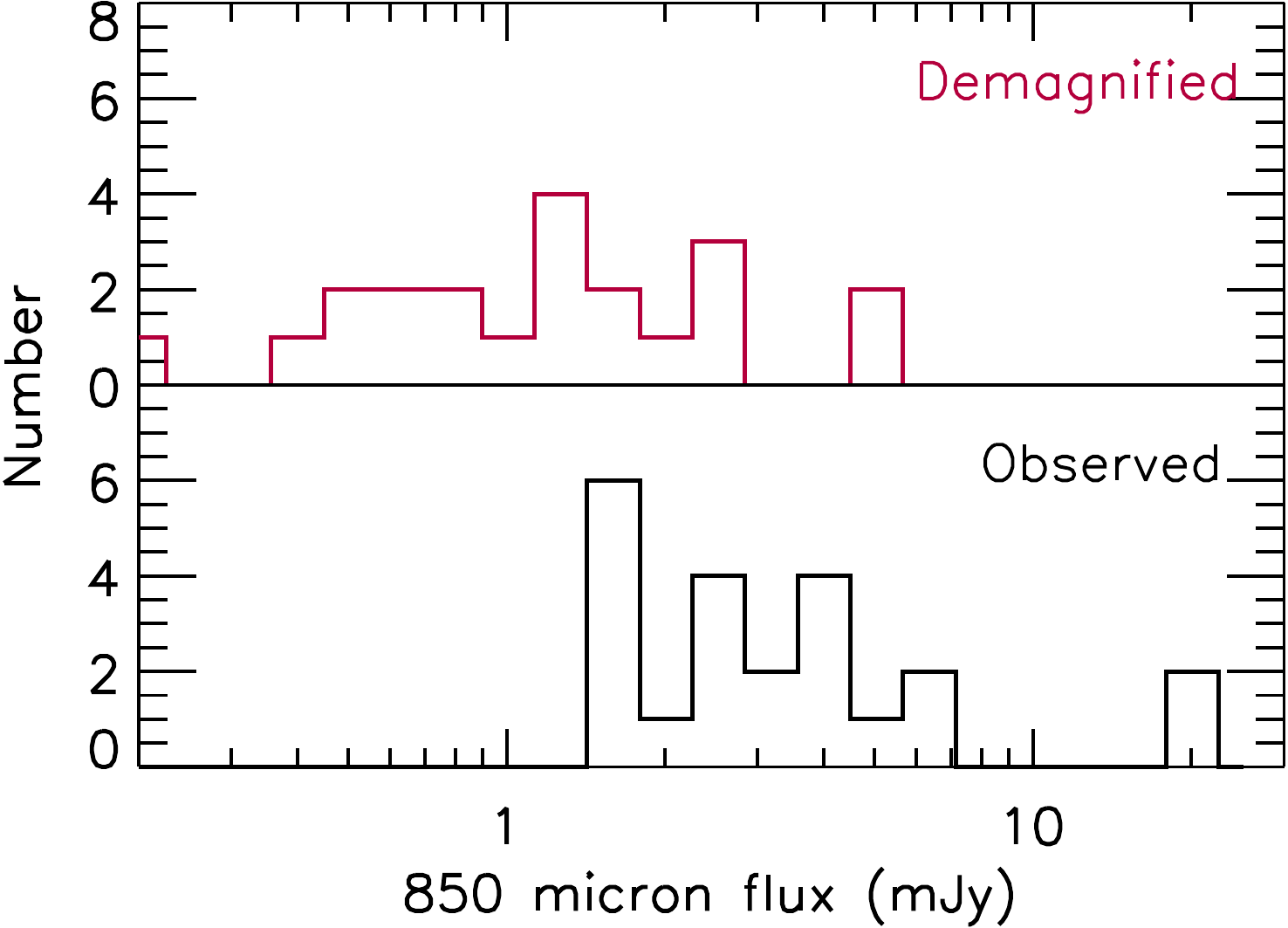}
\caption{
{\em (Lower histogram)\/} Distribution of
observed \afluxb flux for the ALMA
sources with SCUBA-2 \afluxb flux $>1.6$~mJy (black).
{\em (Upper histogram)\/} Distribution of demagnified 
\afluxb flux (red).
\label{newhist}
}
\end{figure}
%---------------------------------------------------------------------

%---------------------------------------------------------------------
\subsection{Magnifications}
%---------------------------------------------------------------------
In order to compute the magnifications and intrinsic flux densities of our faint
SMGs, lens models of the clusters are required. Lens models are available 
for the HFFs from 10 teams. We use the online tool provided by Dan Coe
(\url{https://archive.stsci.edu/prepds/frontier/lensmodels/webtool/magnif.html})
to obtain median magnifications and standard deviations for the ALMA sources
with redshifts in A370, MACSJ1149, and MACSJ0717 
(Tables~\ref{a370_band7}--\ref{macsj0717_band7})
to illustrate the range of magnifications.
Most of the sources in the cluster samples have modest magnifications
in the 1.5--4 range and errors of $\sim10$--20\%.
However, for the very small number of high amplification sources lying
close to the critical lines at $z =$1--6, the errors can be larger.
Meneghetti et al.\ (2017) quote an uncertainty of 10\% at magnifications of
3, with a degradation to 30\% at magnifications of 10. Meanwhile, based on A2744,
Priewe et al.\ (2017) quote a larger uncertainty of 30\% at magnifications of 2,
with a degradation to 70\% at rare high magnifications of 40.

For uniformity, we choose to use the Zitrin et al.\ (2009, 2013, 2014, 2015)
and Zitrin (2021) models for the HFFs and the
CLASH clusters MACSJ1423, MACS2129, 
and RXJ1347 (though note that we only have the one BCG in MACSJ1423, so we do
not consider that cluster further). Thus, due to the lack of Zitrin models,
we do not consider A2390 (for which we only have the one BCG, anyway) or A1689 further in 
our ALMA analysis.

We compute the source magnifications from the appropriate Zitrin models using the median
in a $1''$ box surrounding the source positions; 
however, we note that these values are very similar to the values computed at the
central positions of the sources. For the few sources that lie outside the Zitrin fields in the HFFs
(no sources lie outside the Zitrin fields in the CLASH clusters), we adopt 
the median magnifications from the other models in the HFFs.

In Figure~\ref{newhist}, we show the effects of demagnification.
In the lower histogram, we show the distribution of observed
\afluxb flux for the ALMA sources with SCUBA-2 \afluxb flux
$>1.6$~mJy, while in the upper histogram, we show
the distribution of demagnified \afluxb flux. 
The median observed flux is 3.6~mJy, and the median demagnified
flux is 1.5~mJy. Only three sources have magnifications $>4$.
The highest magnification of 88 (in MACSJ0416;
coordinate R.A. $=4^h 16^m 10.80^s$, decl. $=-24^\circ 4' 47.6''$)
produces the one very faint source in the demagnified histogram.

%---------------------------------------------------------------------
% FIGURE 14: plot_zitrin_magnif and plot_demag_z
%---------------------------------------------------------------------
\begin{figure*}[th]
\includegraphics[width=3.4in,angle=0]{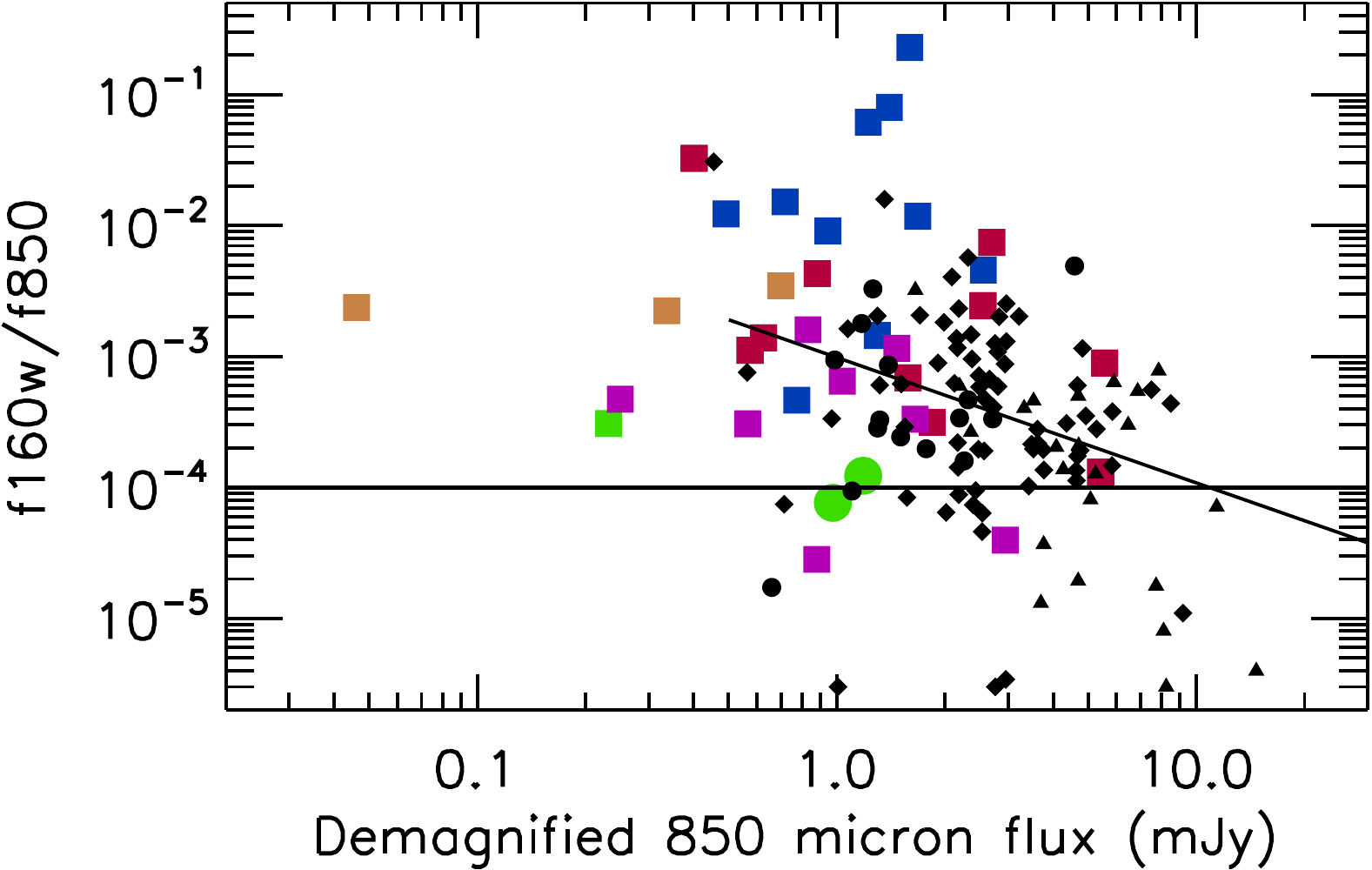}
\includegraphics[width=3.4in,angle=0]{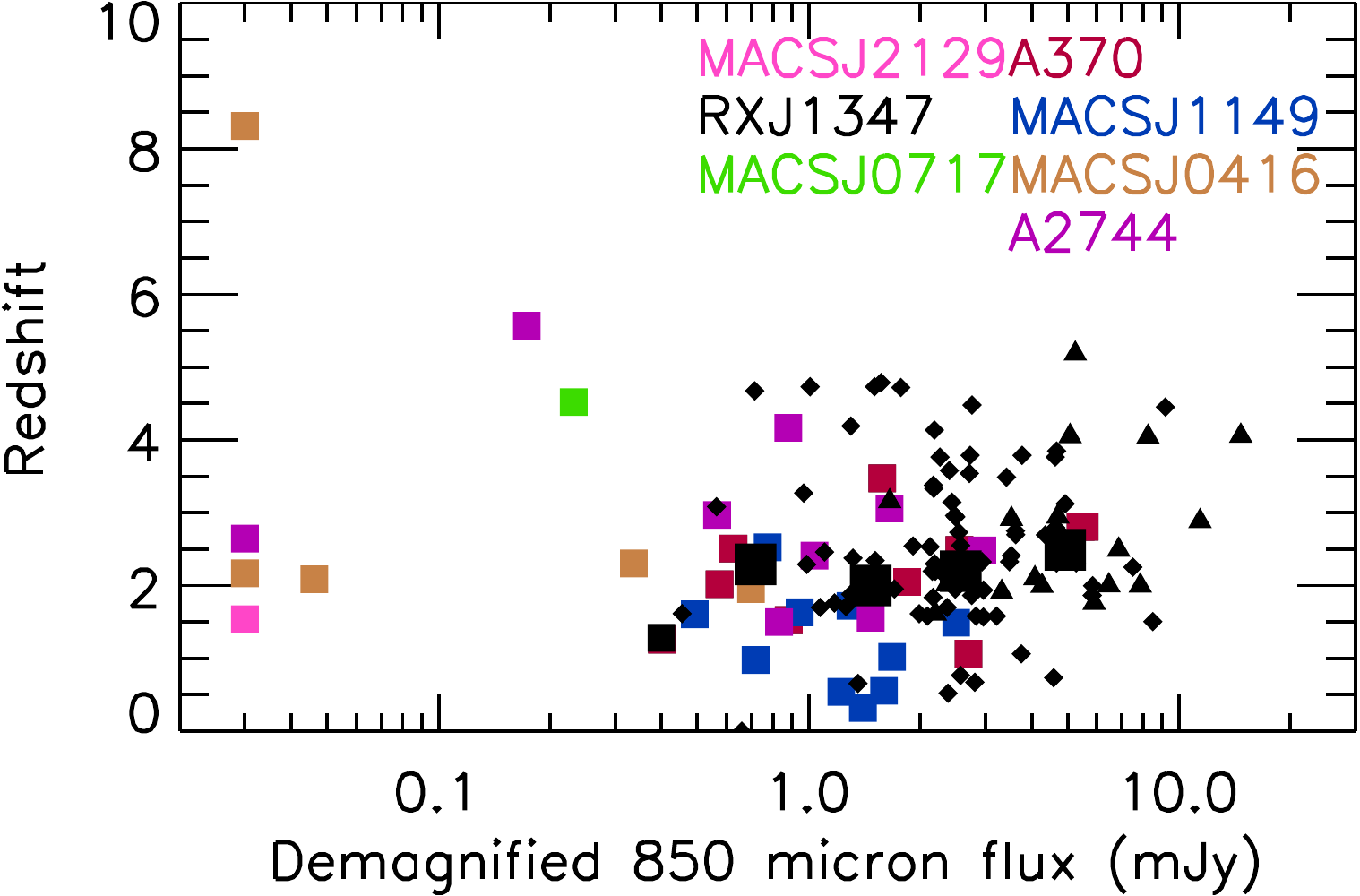}
\caption{
Ratio of F160W flux to SCUBA-2 \afluxb flux (left) and
redshift (right) vs. demagnified SCUBA-2 \afluxb flux
for the ALMA clusters listed in the legend with either speczs or photzs.
We also show SMA sources in the CDF-N
from Cowie et al.\ (2017) (black triangles) and ALMA sources
in the CDF-S from Cowie et al.\ (2018) (black diamonds).
In the left panel, we include two SMA-selected sources 
in MACSJ0717 from Hsu et al.\ (2017; green circles). We also show our
definition of optical/NIR dark SMGs (horizontal line).
Finally, we show a power law fit to the points above 0.5~mJy,
$\log$(flux ratio)=$(-3.00\pm0.11)-(0.96\pm0.23)\log$(demagnified \afluxb flux),
where the errors on the parameters are $1\sigma$.
In the right panel, we show the ALMA cluster sources that are not detected
in the SCUBA-2 850~$\mu$m image at a nominal flux of 0.03~mJy.
\label{plot_zitrin_magnif}
}
\end{figure*}
%---------------------------------------------------------------------

%---------------------------------------------------------------------
\subsection{Optical/NIR Counterparts to the Faint SMGs}
%---------------------------------------------------------------------
The detailed matching of individual faint SMGs with their UV/optical counterparts 
is still poorly understood and will only be resolved with large, well-studied 
samples, such as those presented here. 

Studies of low-redshift starburst galaxies 
(e.g., Chary \& Elbaz 2001; Le Floc'h et al.\ 2005; Reddy et al.\ 2010) have shown that 
fainter sources are generally less dusty. However, some recent work has suggested that 
fainter SMGs are, on average, at lower redshifts 
(e.g., Mobasher et al.\ 2009; Magliocchetti et al.\ 2011; Hsu et al.\ 2016; 
Aravena et al.\ 2016, 2020; Cowie et al.\ 2018). 
In this section, we will argue that the primary evolution is in the extinction rather than
in the redshift distribution.

In Figure~\ref{plot_zitrin_magnif}, we plot for the ALMA
cluster sample with redshifts the F160W 
to \afluxb flux ratio versus demagnified SCUBA-2 \afluxb flux (left panel), 
and redshift versus demagnified SCUBA-2 \afluxb flux (right panel).
We supplement this with blank-field observations of the CDF-N and CDF-S.

There are 23 CDF-N SMGs in the footprint of the HST F160W image
with $>5\sigma$ SMA detections from Cowie et al.\ (2017). 
All of these are also detected in the 
SCUBA-2 \afluxb image, with the lowest S/N being 7.5.

Targeted ALMA imaging in the CDF-S by Cowie et al.\ (2018) yields 74 ALMA 
detected sources in the footprint of the HST F160W image.
Contiguous millimeter mosaics have also been used to generate ALMA samples
in the  CDF-S, but the most recent analysis of these data (G\'omez-Guijarro et al.\ 2022) 
only adds 10 directly detected  sources to those given in Cowie et al.\ (2018),
reflecting the inefficiency of this procedure. 
83 of the 84 sources in this combined ALMA CDF-S sample 
are detected above a 
$2\sigma$ threshold (79 above $3\sigma$, and 76 above $4\sigma$) 
in the SCUBA-2 \afluxb image.
Seven of the 10 additional ALMA sources are detected above the $4\sigma$ level
in the SCUBA-2 image. 

We add these two samples (84 CDF-S and 23 CDF-N SMGs)
to Figure~\ref{plot_zitrin_magnif} using the measured SCUBA-2 fluxes.
We measured the $1.6~\mu$m fluxes using corrected $1''$ diameter apertures
on the Hubble Legacy Fields\footnote{\url{http://archive.stsci.edu/hlsps/hlf}}
combined F160W images (G.~Illingworth et al., in preparation).

In Figure~\ref{plot_zitrin_magnif}(a), we see 
a trend of increasing F160W to \afluxb flux ratio from brighter to fainter SMGs.
However, there is a wide spread in this ratio. Hereafter, we will refer to the 
sources that are extremely faint in this ratio (i.e., $<10^{-4}$) as 
optical/NIR dark SMGs (horizontal line).
For a source with a 2~mJy \afluxb flux, this would correspond
to an F160W magnitude of 25.6.  Based on the figure, such sources are 
more common at brighter \afluxb fluxes but continue to exist
down to \afluxb fluxes fainter than 1~mJy.

In Figure~\ref{plot_zitrin_magnif}(b), some of the lowest flux sources at
are very high redshifts due to ALMA targeting of known high-redshift
sources (the $z=8.311$ source in MACSJ0416 from Tamura et al.\ 2019
and Bakx et al.\ 2020) or
to possibly uncertain photzs (the $z=5.56$ source in A2744 and the
$z=4.52$ source in MACSJ0717).
For sources around 1~mJy,
redshifts range from cluster redshifts to redshifts just above $z=4$. 

The median redshifts for the combined sample are $z=2.29$, 2.00, and 2.20, and 2.49 in the
0.5--1, 1--2, 2--4, and 4--8~mJy flux ranges. These median redshifts are slightly
lower than the strongly-lensed galaxy sample \afluxb curve over this flux range given in 
B{\'e}thermin et al.\ (2015; Figure~3), which is based on their phenomenological model of 
galaxy evolution. Their curve shows a decline from about $z=2.9$ to $z=2.6$, with the
lower fluxes being at lower redshifts. Meanwhile, their full galaxy sample \afluxb curve
declines from about $z=2.7$ to $z=1.9$. 

For the present sample above 0.5~mJy, the field galaxies
have a median redshift of $z=2.27$, while the lensed galaxies
have a median redshift of $z=2.20$. For the sources that are not in our primary
ALMA band~7 sample, there may be biases due to selection effects. However, in our
highly complete band~7 sample in A370, MACSJ1149, and MACSJ1707 (only two of
these sources do not have redshifts, one of which lies off the F160W field), 
above 0.5~mJy we find a median redshift of $z=2.00$, which is quite 
similar to our overall lensed galaxy median redshift of $z=2.20$, as quoted above.

Thus, we see no strong evidence for evolution in the
redshift distribution versus demagnified \afluxb flux over this flux range.
This implies that the trend of increasing F160W to \afluxb flux ratio from brighter to fainter 
SMGs that we observe results from decreasing extinction as we move to fainter SMGs.

\begin{figure}
\vskip -0.1cm
\includegraphics[width=3.3in,angle=0]{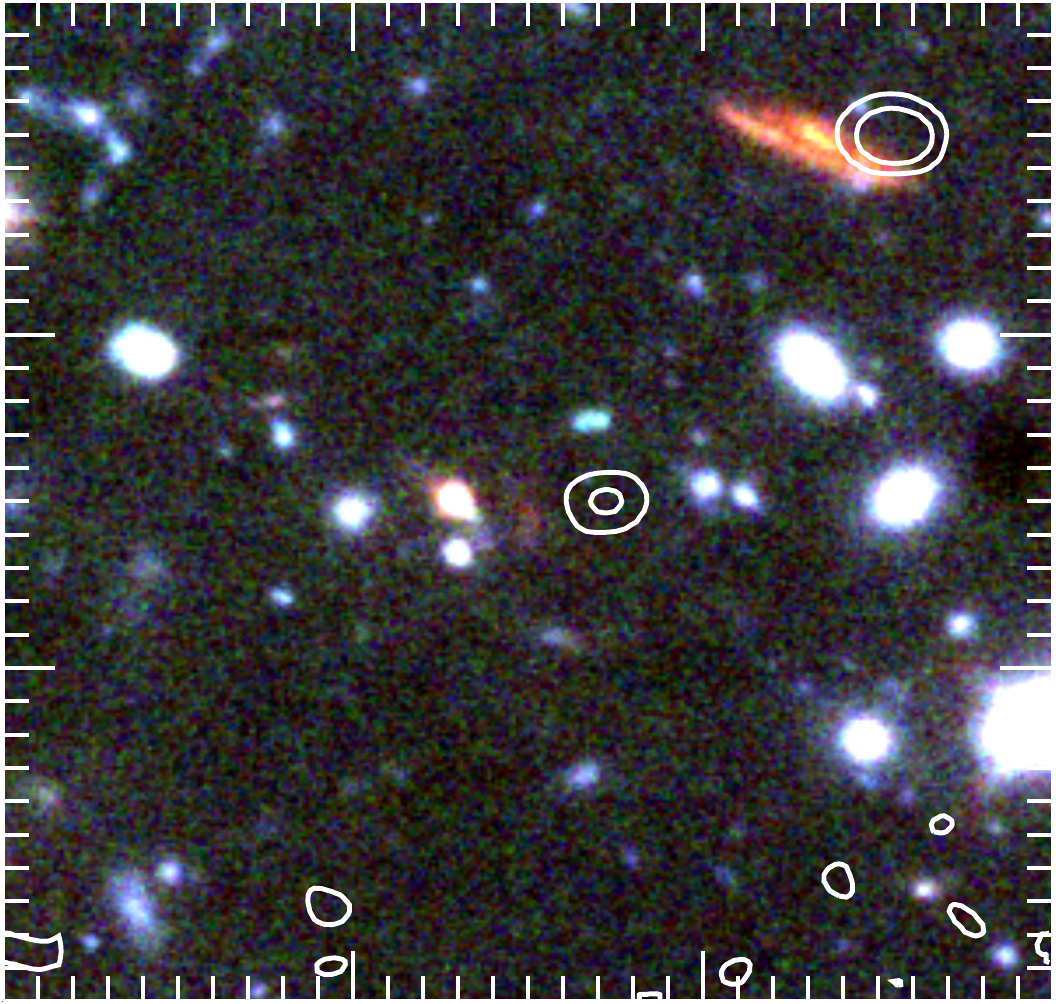}
\caption{Centered in the image is
an example of an optical/NIR dark, faint SMG
at R.A. $=0^h 14^m 19.50^s$, decl. $=-30^\circ 22' 48.7''$
in A2744 with $z_{phot}=4.16$ (see Table~\ref{archivetable}).
It has an F160W magnitude of 27.0 and a demagnified
SCUBA-2 \afluxb flux of 0.9~mJy.
The color image ($18''$ on a side) shows the HFF
data for F105W (blue), F125W (green), and F160W (red).
The white contours show the ALMA 1.2~mm image 
(ALMA program \#2017.1.01219.S; PI: F. Bauer)
offset by $1\farcs2$
in the horizontal direction so its optical/NIR counterpart may be more
clearly seen (i.e., the ALMA source
is coincident with the faint red galaxy at the center
of the image). 
Note that the SMG in the upper right 
(R.A. $=0^h 14^m 19.12^s$, decl. $=-30^\circ 22' 42.2''$)
has an F160W magnitude of 23.5
and a similar demagnified SCUBA-2 flux to the example source.
\label{sample}
}
\end{figure}

However, we emphasize that
there is a wide range of measured F160W to \afluxb flux ratios in the faint SMG
samples, including sources that are extremely
dark in the optical/NIR, as was first noted by Chen et al.\ (2014) and Hsu et al.\ (2017). 
In Figure~\ref{sample}, we show an example of an optical/NIR dark, faint SMG  
in the A2744 field with $z_{phot}=4.16$ (see Table~\ref{archivetable}). 
The source, which has a delensed \afluxb flux of 0.9~mJy,
is extremely faint in the NIR with a F160W magnitude of 27.0.
This suggests that
there is a population of faint SMGs that are also extremely optical/NIR faint,
either because they are very dusty and/or because they are at very high redshifts
(see also, e.g., Wang et al.\ 2019).

%---------------------------------------------------------------------
% FIGURE 15; sample
%---------------------------------------------------------------------
%---------------------------------------------------------------------

%---------------------------------------------------------------------
\subsection{Estimating Redshifts from the F160W Flux}
\begin{figure}[th]
\vskip -0.1cm
\includegraphics[width=3.3in,angle=0]{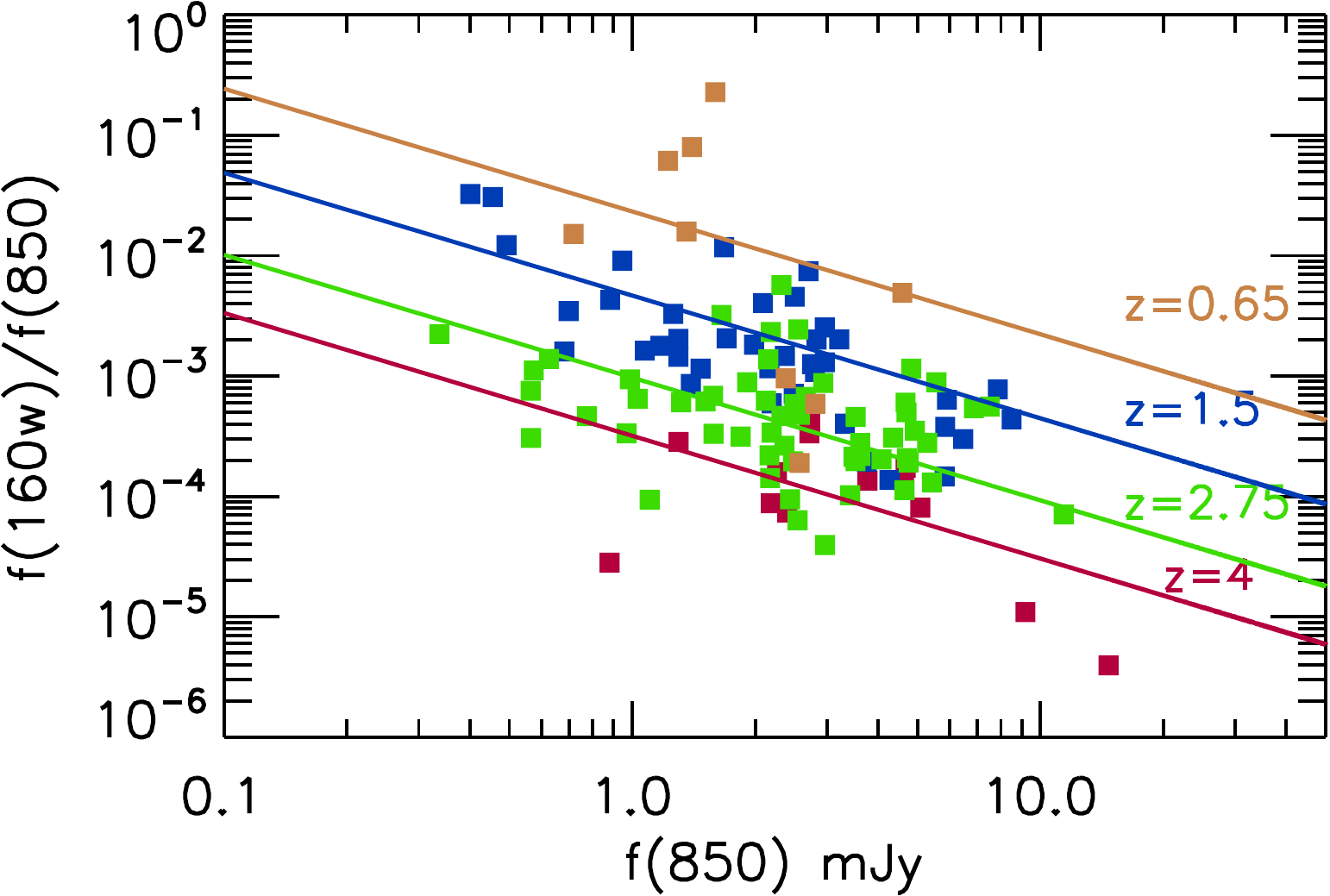}
\caption{
Ratio of $1.6~\mu$m flux to SCUBA-2 \afluxb flux vs. demagnified
SCUBA-2 \afluxb flux, color-coded by redshift
(gold---$z<1$; blue---$z=1$--2; green---$z=2$--3.5;
red---$z=3.5$--4.5). The color-coded lines show the fitted relationship
given in Equation~\ref{eqn_final} computed
at the labeled redshifts.
\label{plot_color_smm_z}
}
\end{figure}

%---------------------------------------------------------------------
Given the observed distribution for the combined sample in Figure~\ref{m160_plot_z},
where we plot redshift versus observed F160W magnitude,
sources with F160W magnitudes $\gtrsim25$ tend to be at $z>4$.
We can further refine the use of F160W magnitude for estimating redshifts
by plotting the F160W to \afluxb flux ratio versus demagnified flux with
the sources color-coded by redshift (see Figure~\ref{plot_color_smm_z}).
We see that, as an ensemble, the sources in the various redshift ranges are 
reasonably well separated and 
can be fit by a power law of the form (gold curve in Figure~\ref{m160_plot_z})
\begin{equation}
1+z = 0.624 (f_{160})^{-0.258} \,,
 \label{eqn_final}
\end{equation}
where $f_{160}$ is in mJy. There does not appear to be a strong dependence on the \afluxb flux. 

\begin{figure}[th]
\vskip -0.1cm
\includegraphics[width=3.1in,angle=0]{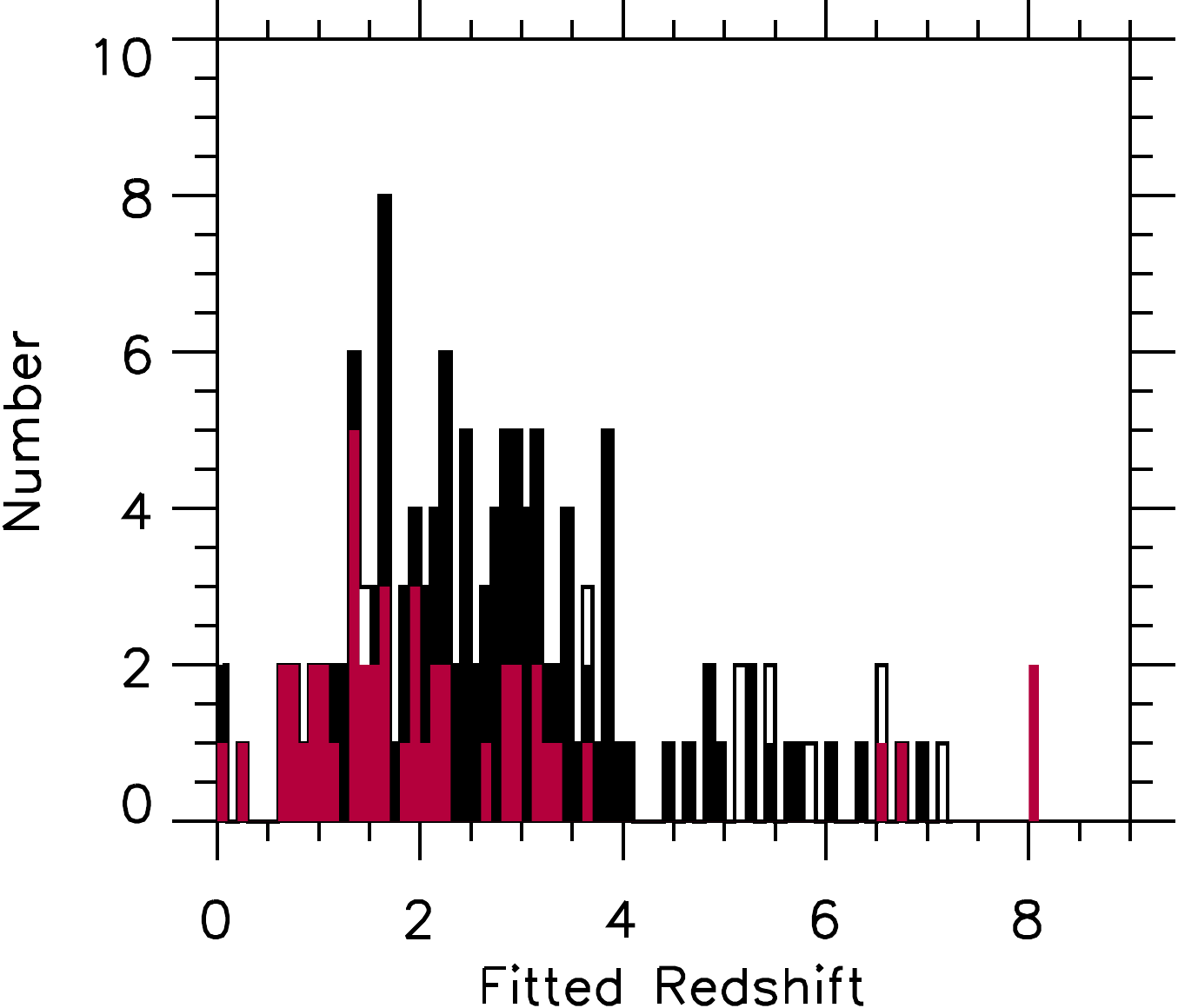}
\caption{
Distribution of fitted redshifts (calculated using Equation~\ref{eqn_final}).
We use shading to denote which sources already have speczs (red),
photzs (black), and neither (open). 
\label{fitted_hist}
}
\end{figure}

In Figure~\ref{fitted_hist}, we show the distribution of redshifts determined from
Equation~\ref{eqn_final}, though we caution that redshifts for 
individual sources may have substantial uncertainty.
We find that for sources above 0.5~mJy, $20\pm4$\%
are at $z>4$. For sources in the range 0.5 to 2~mJy, 16 (10, 24)\%
are at $z>4$, where the numbers in parentheses are the 68\% confidence range.

%---------------------------------------------------------------------
% FIGURE 16; plot_color_smm_z
%---------------------------------------------------------------------
%---------------------------------------------------------------------

%---------------------------------------------------------------------
% FIGURE 17; fitted_redshifts_hist
%---------------------------------------------------------------------
%---------------------------------------------------------------------

%---------------------------------------------------------------------
\section{Interpreting the SCUBA-2 Sample}
%---------------------------------------------------------------------
Now that we are informed by the ALMA data from the previous section, 
we turn to analyzing the SCUBA-2 sample. After developing the SCUBA-2
sample, one of our main goals is to determine candidate high-redshift SMGs. 
This is important, because the space density 
at $z > 4$ remains relatively poorly constrained.

%---------------------------------------------------------------------
\subsection{Positional Uncertainty}
%---------------------------------------------------------------------
We first measured the accuracy of the \afluxa
source positions using the SCUBA-2 cluster sample
together with the CDF-N and CDF-S samples 
(Cowie et al.\ 2017, 2018).
We measured the offset between the $>4\sigma$ \afluxa source position
and the nearest ALMA/SMA counterpart, if one exists within $5''$.
In the well-studied A370 and MACS1149 fields, there are 18 ALMA
sources detected above $4\sigma$ in the \afluxa band.
Of these, there are two pairs that correspond to a single
\afluxa source, one in each field. The remainder are single
matches. This corresponds to a multiplicity of $11^{26}_{4}\%$,
where the upper and lower values give the $68\%$ confidence range.

\begin{figure}
\vskip 1.2cm
\includegraphics[width=3.2in,angle=0]{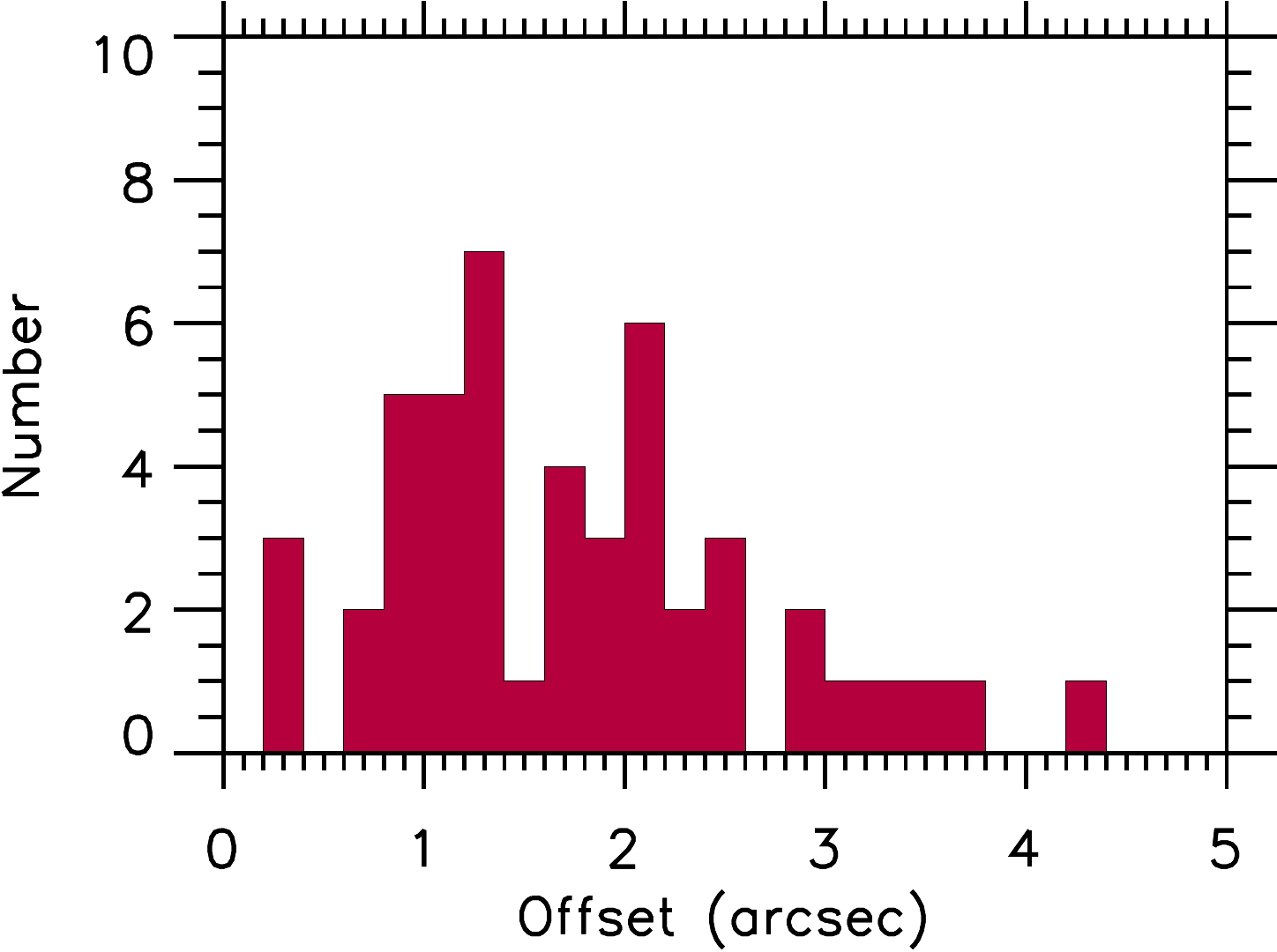}
\caption{
Distribution of offsets between \afluxa source position 
and nearest ALMA/SMA counterpart, if there is one within $5''$,
for sources with $>4\sigma$ \afluxa detections in either the
SCUBA-2 cluster sample or the CDF-N and CDF-S samples.
Only sources with \afluxa rms errors $<3$~mJy are included.
\label{450_offsets}
}
\end{figure}

In Figure~\ref{450_offsets}, we show the distribution of offsets
between the \afluxa SCUBA-2 and ALMA positions.
For 80\% of the sources with a counterpart, the offsets are $<2\farcs5$.
The mean and median offsets 
are both 1\farcs7. This is very similar to the offsets measured in Barger
et al.\ (2022) between the $>4\sigma$ \afluxa source position and 
the nearest 20~cm counterpart.

Since many of the SCUBA-2 cluster sample do not have $>4\sigma$ 
\afluxa detections, we also need the uncertainties for
the \afluxb positions. These scale roughly with
the PSF, and Cowie et al.\ (2017) found that 96\% of the $>4\sigma$
\afluxb source positions are within $4''$ of the nearest 20~cm
counterpart in the CDF-N.  We take this as the uncertainty
in the \afluxb positions.

%---------------------------------------------------------------------
% FIGURE 18: new_450_off
%---------------------------------------------------------------------
%---------------------------------------------------------------------

%---------------------------------------------------------------------
\subsection{Central Sample}
%---------------------------------------------------------------------
Only the central members of the SCUBA-2 cluster sample lie in
regions where lensing magnifications are significant. 
In studying the fainter SMGs, we therefore restrict to sources within a
$1\farcm75$ radius from the cluster center, where amplifications may be $>2$.
Excluding \afluxb sources corresponding to BCGs and pairs (see Table~2), 
this provides a sample of 111 SMGs.

\begin{figure}[hpb!]
\vskip 1.2cm
\includegraphics[width=3.2in,angle=0]{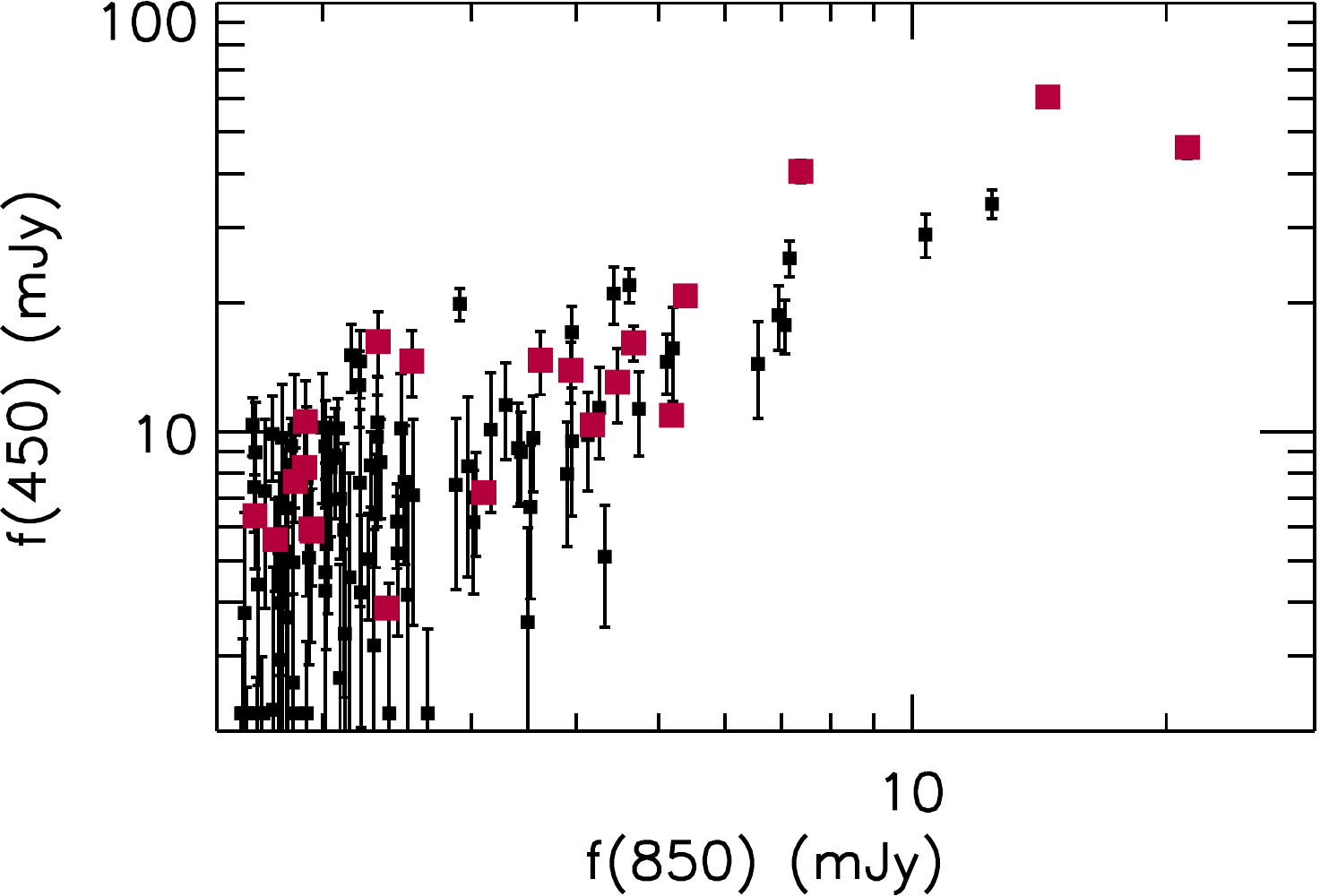}
\centering{\caption{Observed \afluxa flux
vs. observed \afluxb flux for SMGs in the central regions
($<1\farcm75$ radius) of the SCUBA-2 \afluxb cluster sample,
after excluding BCGs and pairs.
The measured \afluxa fluxes are shown with $1\sigma$ errors. 
If the \afluxa flux is negative, then the source is shown at a nominal
\afluxa flux of 2.2~mJy.
SMGs with ALMA counterparts are shown with red large squares.
}
\label{new_matching}}
\end{figure}

In Figure~\ref{new_matching}, we plot the observed \afluxa flux
versus the observed \afluxb flux for
this sample. 16 (14\%) have measured  ALMA counterparts
(red squares), and 39 have $>4\sigma$ detections at \afluxar. 
The ALMA detected sources appear to be well representative of this sample,
so we can assume that information derived in Section~\ref{interpretALMA}
can be applied here.

%---------------------------------------------------------------------
% FIGURE 19: new_matching
%---------------------------------------------------------------------
%---------------------------------------------------------------------

%---------------------------------------------------------------------
% FIGURE 20: compare_magnif
%---------------------------------------------------------------------
\begin{figure}
\vskip 1.2cm
\includegraphics[width=3.2in,angle=0]{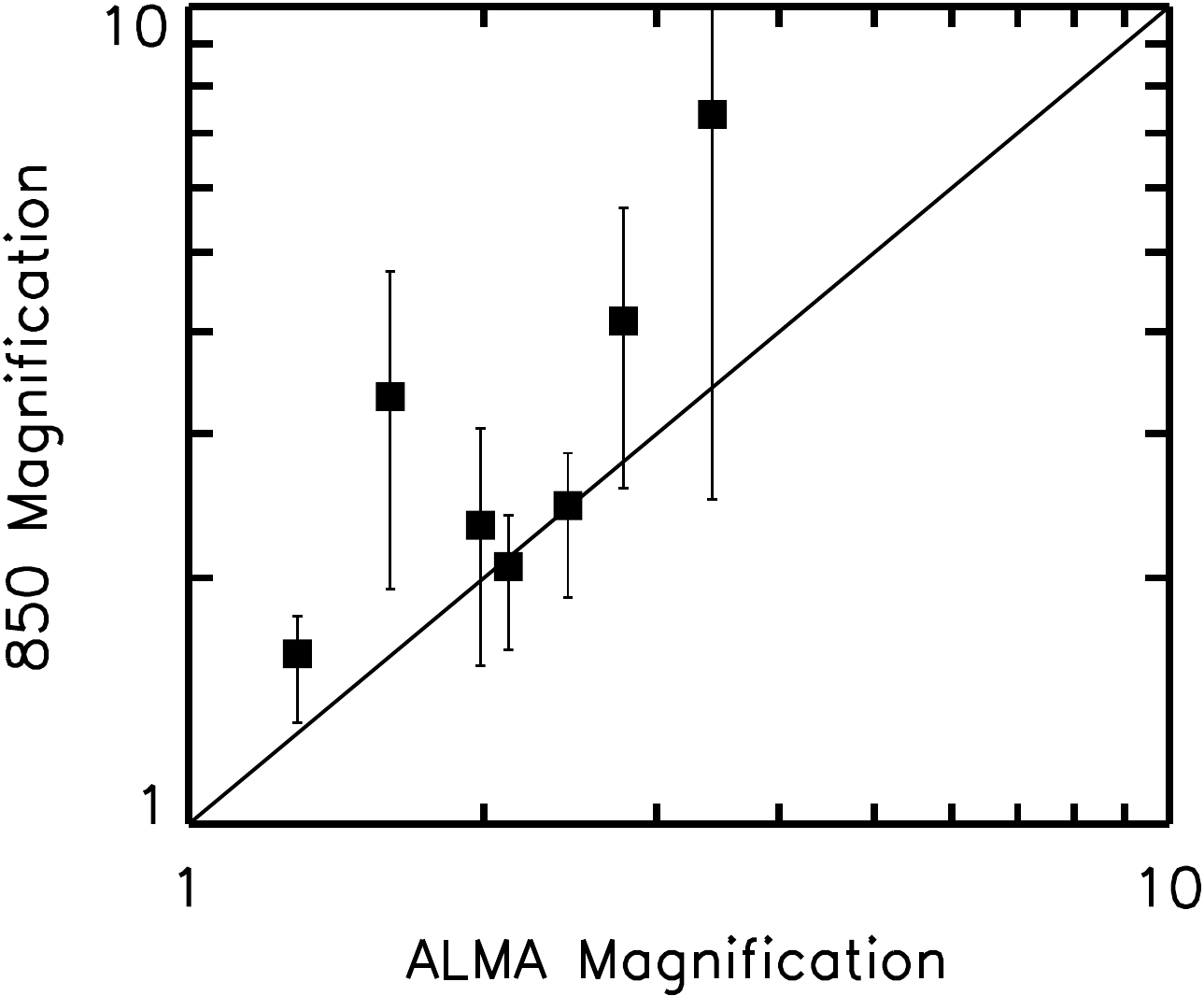}
\caption{Magnification computed at the SCUBA-2 \afluxb position for
$z=2$ (squares) and for a range from $z=1$ to $z=4$ (error bars) 
vs. magnification computed at the
ALMA position and accurate redshift for
sources from A370 and MACSJ1149 with high enough 
\afluxb fluxes to appear in Table~\ref{tab2}.
\label{compare_magnif}
}
\end{figure}
%---------------------------------------------------------------------

%---------------------------------------------------------------------
% FIGURE 21: show_central_f4f8_high, show_central_f4f8, show_outer_f4f8
%---------------------------------------------------------------------
\begin{figure}[th]
\includegraphics[width=3.2in,angle=0]{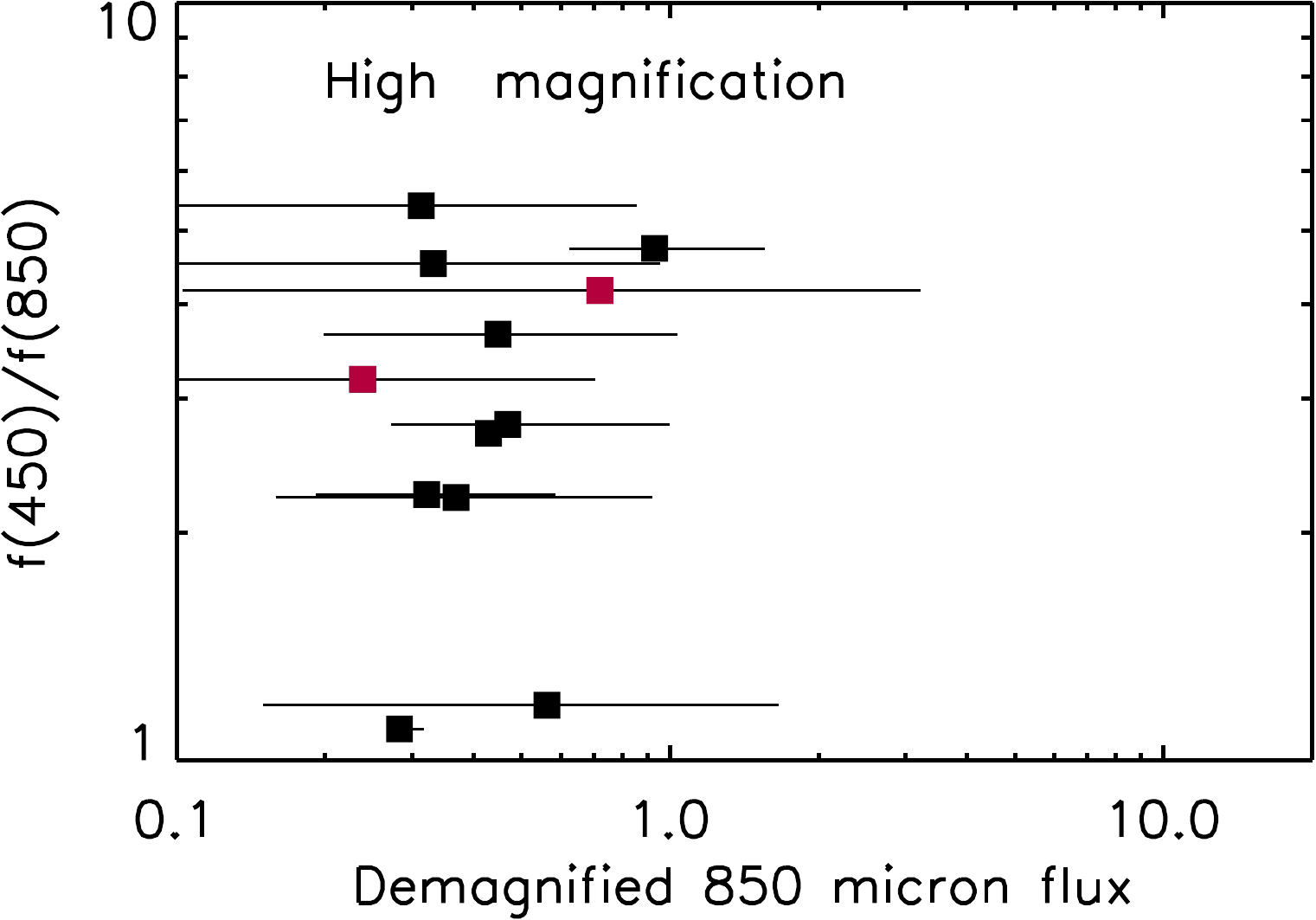}
\includegraphics[width=3.2in,angle=0]{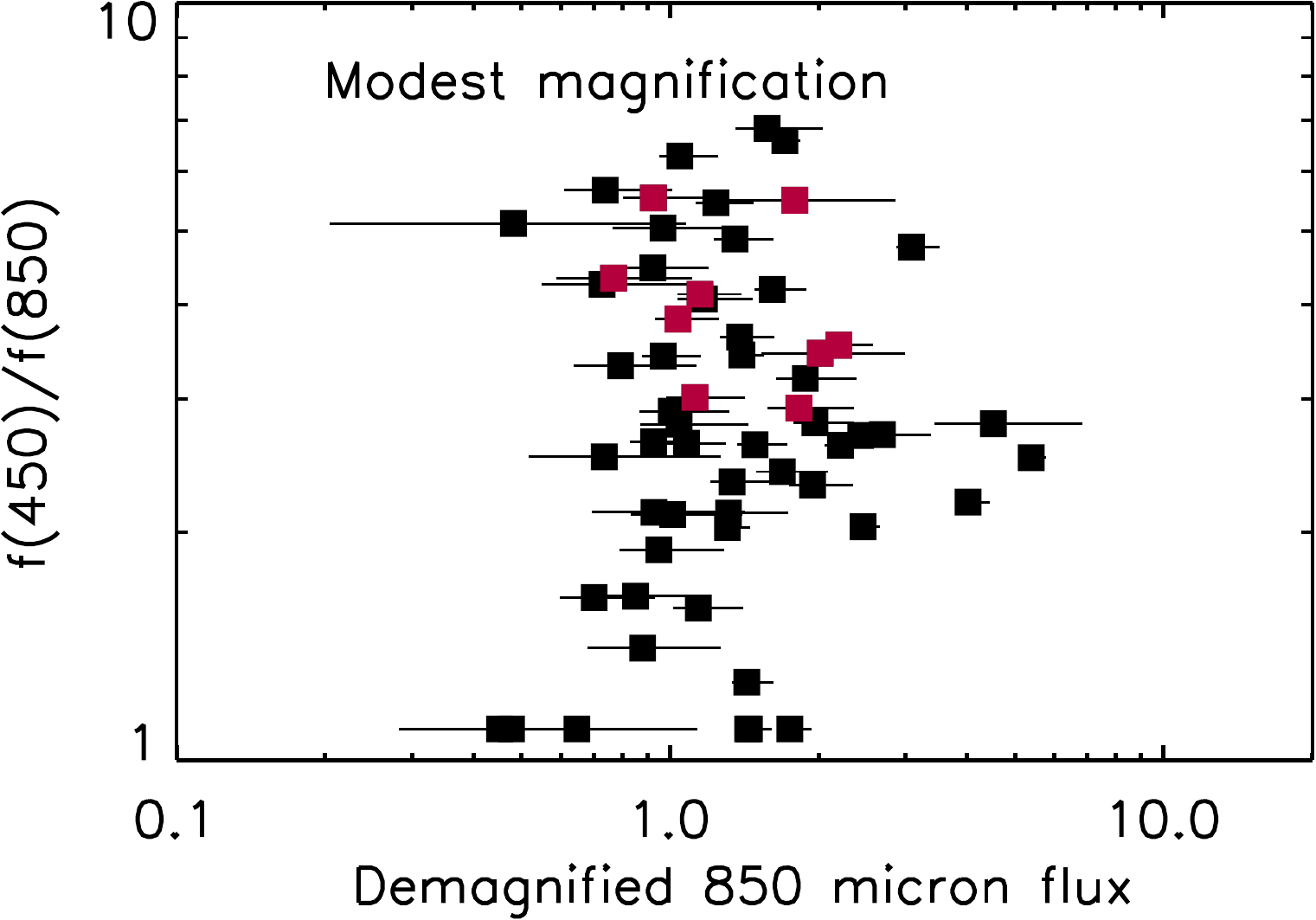}
\includegraphics[width=3.2in,angle=0]{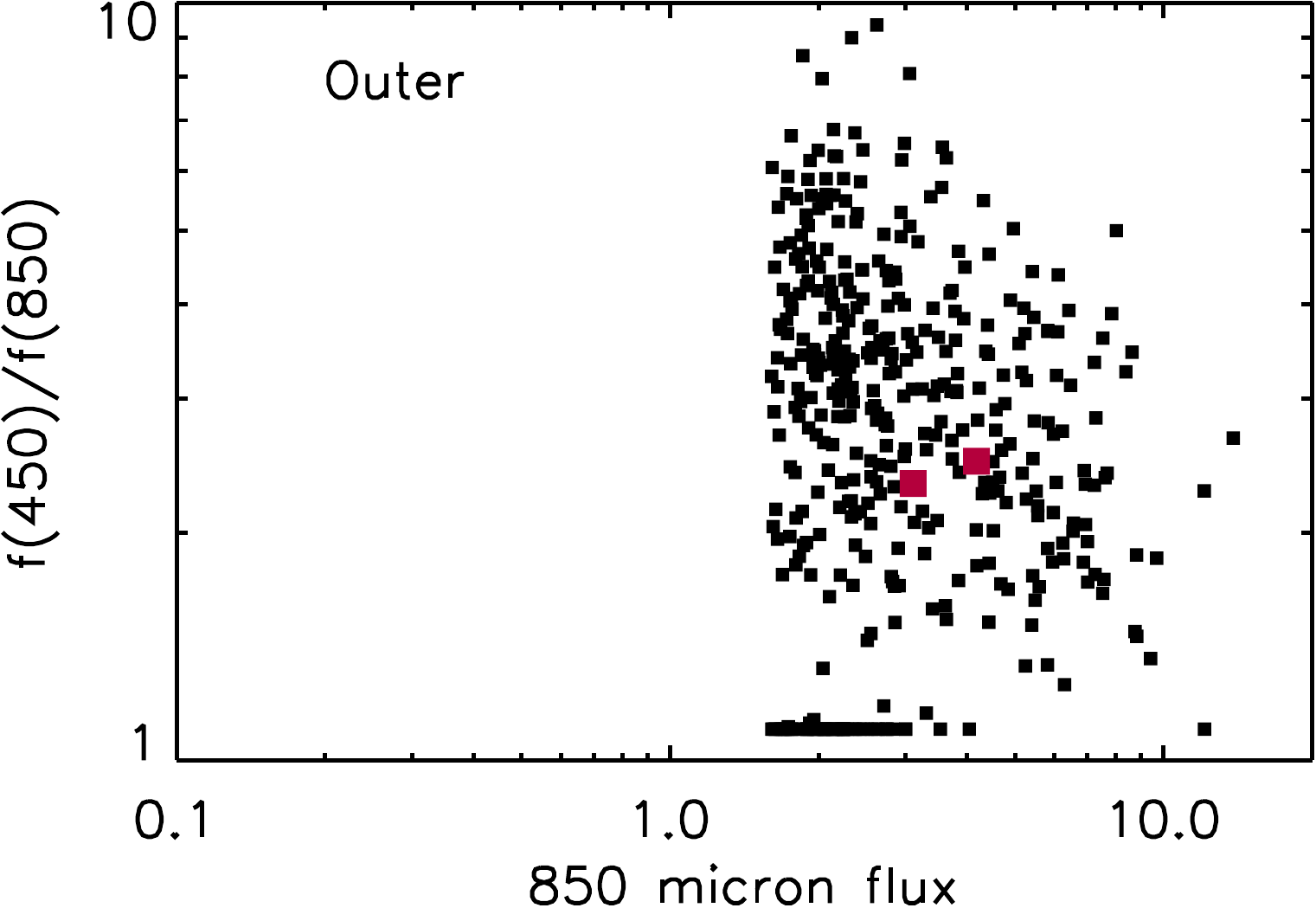}
\centering{\caption{
\afluxa to \afluxb flux ratio vs. \afluxb flux. 
{\em (Top two panels)\/} 
Central sample divided by magnification: $>4.2$ for $z=2$
(top) and $<4.2$ for $z=2$ (middle).
The uncertainties show a range from $z=1$--4. 
{\em (Bottom panel)\/} Full sample outside a $2\farcm5$ radius
from the cluster center, where we do not apply a magnification correction.
In all panels, sources with ALMA identifications are shown in red.}
\label{f4_flux}}
\end{figure}
%---------------------------------------------------------------------

We now again exclude A1689 and A2390 and refer to the
remaining sample as {\em our central \afluxb sample.\/}
Sources with modest magnifications (we take this to mean
$<4.2$) are relatively insensitive to $4''$ 
positional changes, but sources with higher magnifications
can be much more uncertain. In both cases,
the redshift is the primary source of uncertainty, but
this uncertainty is again less for the sources with modest magnifications. 
For the sources in these fields,
we computed the magnification at the SCUBA-2 \afluxb position
for $z=2$ and estimated the uncertainty by considering a 
range from $z=1$--4. To illustrate how well this works,
we show in Figure~\ref{compare_magnif} for A370 and MACSJ1149
these magnifications versus magnifications computed at the ALMA positions
and accurate redshifts. Within the magnification uncertainties
corresponding to the adopted $z=1-4$ range, there is broad agreement.

%---------------------------------------------------------------------
\subsection{450 to 850 Micron Ranges} 
%---------------------------------------------------------------------
In the top two panels of Figure~\ref{f4_flux}, we show the 
\afluxa to \afluxb flux ratio for the central sample versus the 
demagnified \afluxb flux, with the magnification uncertainties 
corresponding to the adopted $z = 1$--4 range.
We use two panels to split the SMGs into those with high
magnifications (here $>4.2$ for $z=2$),
where the uncertainties are large, and those with modest magnifications,
where the uncertainties are smaller. 

In the bottom panel, we
show the sample outside a $2\farcm5$ radius from the cluster center
(again excluding BCGs and close pairs, but using all of the clusters),
where we expect the magnification to be near one. Here
we use the observed \afluxb flux.

%---------------------------------------------------------------------
% FIGURE 22: f4f8_hist
%---------------------------------------------------------------------
\begin{figure}
\vskip 1.2cm
\includegraphics[width=3.2in,angle=0]{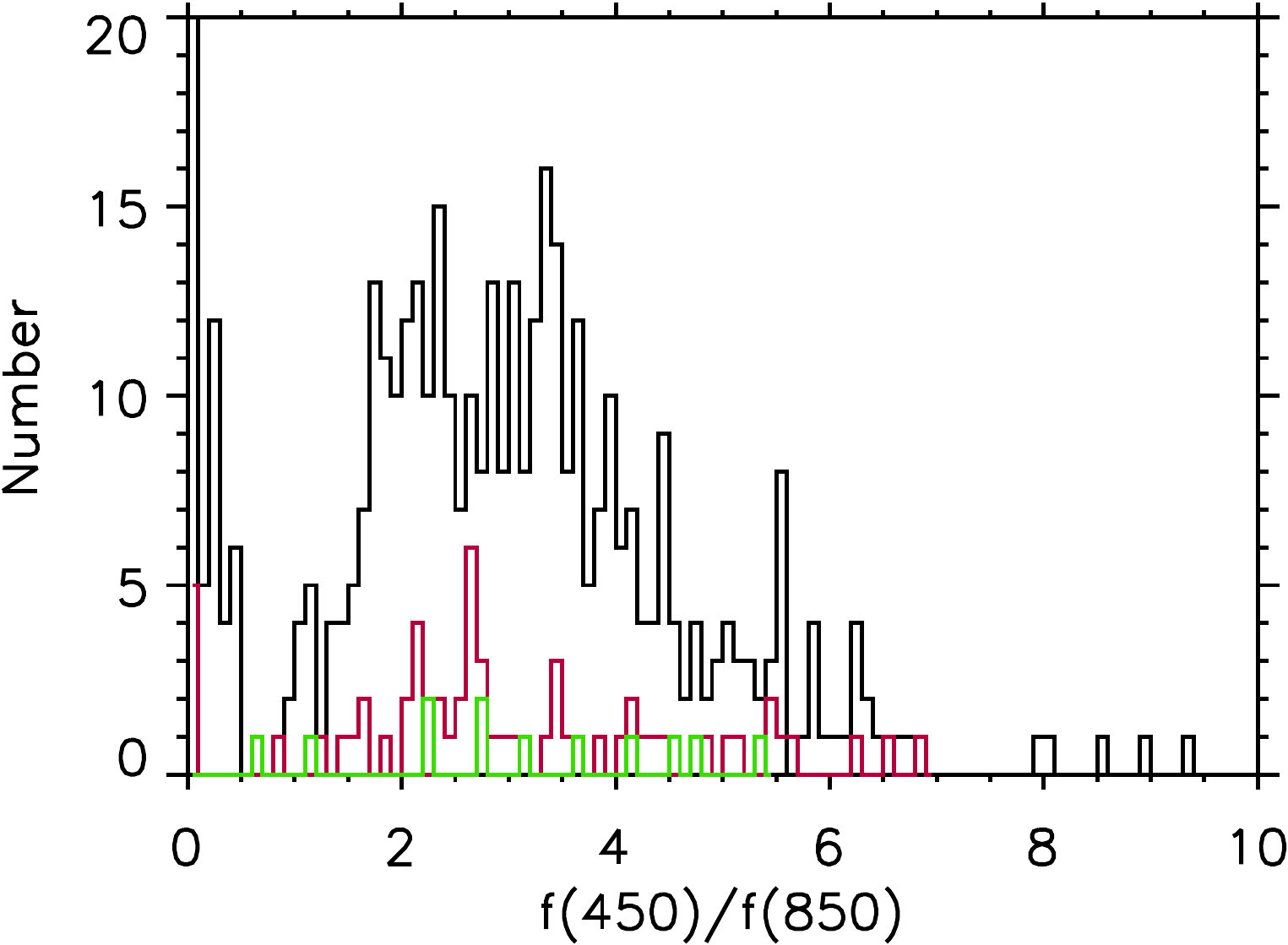}
\centering{\caption{
Histogram of the \afluxa to \afluxb
flux ratio for the outer sample (black), the central
sample with moderate magnifications (red), and the
central sample with high magnifications (green).
}
\label{f4f8_hist}}
\end{figure}

%---------------------------------------------------------------------

These three samples have very similar distributions
of \afluxa to \afluxb flux ratios, as we show in Figure~\ref{f4f8_hist}.
A Mann-Whitney test does not show a significant difference between the three samples.
If, as we have argued, this flux ratio is a good redshift indicator, then the redshift
distribution of this large \afluxb sample is invariant
as a function of the \afluxb flux over the 0.5--10~mJy flux range. 
This is consistent with the smaller ALMA sample with direct redshift 
measurements (see Figure~\ref{plot_zitrin_magnif}).

%---------------------------------------------------------------------
\subsection{Candidate High-Redshift SMGs}
%---------------------------------------------------------------------
We restrict to the $>5\sigma$ \afluxb full cluster sample, after including confusion noise, 
to improve the fidelity of our candidate high-redshift SMG selection. We determine the sources
in this sample
that have \afluxa to \afluxb flux ratios $<2$, which Barger et al.\ (2022) give as a criterion 
to select candidate high-redshift SMGs with $z>4$. We list these candidates, which make up 
21\% of the $>5\sigma$ \afluxb full cluster sample, in Table~\ref{faintsample}.
At present, only one
has a measured ALMA counterpart (source~2 in MACSJ1149; see Table~\ref{macsj1149_band7}).
This source does not have a specz or a photz.

The 55 SMGs in Table~\ref{faintsample} are quite bright, with \afluxb fluxes from 
2.3~mJy to 12.1~mJy and a median flux of 4.3~mJy. They should be straightforward
targets for future interferometric observations. Seven lie within the
central high-magnification regions ($<1\farcm75$ radius) of 
clusters observable with ALMA, and
these represent the most interesting targets.

%---------------------------------------------------------------------
\section{Summary}
%---------------------------------------------------------------------
In this paper, we presented deep SCUBA-2 \afluxa and \afluxb imaging of ten strong lensing clusters.
We constructed a catalog of 404 sources with \afluxb fluxes detected above $4\sigma$ that lie within
a radius of $4\farcm$5 from the cluster centers. 
We also presented catalogs of $>4.5\sigma$ ALMA band~6 (1.2~mm) 
and band~7 (870~$\mu$m) detections from our observations and from the archive, for which
we gave, where available, spectroscopic or photometric redshifts from new Keck observations or 
from the literature. We supplemented our cluster lensed ALMA sample with a CDF-S ALMA sample 
and a CDF-N SMA sample from previous work to aid with interpretation. Our main results based on 
these samples are as follows:

\vskip 0.25cm

$\bullet$ We noted a correlation between redshift and observed F160W magnitude,
with fainter sources being at higher redshifts. Higher redshift sources may therefore be hard 
to get even photometric redshifts for.

$\bullet$ We found that different cluster fields have
different redshift distributions, emphasizing the need for observations of
multiple fields to obtain properly averaged redshift distributions.

$\bullet$ We confirmed the use of the \afluxa to \afluxb flux ratio for estimating redshifts,
with $z>4$ generally corresponding to a flux ratio $<2$. 

$\bullet$ We used publicly available lens models for the clusters to determine magnifications,
most of which are in the 1.5--4 range with modest errors of 10--20\%, though there are a very
small number of high amplification sources where the errors can be large.

$\bullet$ Utilizing both our cluster lensed sample and the CDF samples, we found 
a trend of increasing F160W to \afluxb flux ratio from brighter to fainter SMGs.

$\bullet$ We found no evidence that the fainter SMGs, which we probe primarily through our cluster lensed 
sample, have a different redshift distribution than the brighter SMGs, which we probe primarily through the 
CDF samples.

$\bullet$ Since we did not find a change in the redshift distribution as a function of the 
demagnified \afluxb flux, 
we concluded that the observed trend in the F160W to \afluxb flux ratio as a function of the
demagnified \afluxb flux results from decreasing extinction in the fainter SMGs.

$\bullet$ 
We caution, however, that there is wide spread in the F160W to \afluxb flux ratio relation.
For example, although optical/NIR dark SMGs, which we defined 
as being extremely faint in the F160W to \afluxb flux ratio (i.e., $<10^{-4}$),
are more common at brighter
\afluxb fluxes, we found that they continue to exist down to \afluxb fluxes fainter than 1~mJy.

$\bullet$ We argued that roughly 20\% of the SMGs are at $z>4$, independent of submillimeter flux.

\vskip 0.25cm

Finally, informed by the ALMA results, we separately analyzed the SCUBA-2 cluster sample
and identified 55 $z>4$ candidates selected on the basis of the 
\afluxa to \afluxb flux ratio. These are bright at \afluxb and hence good
targets for future interferometric observations, including seven that lie in
the central high magnification regions of clusters that are observable with ALMA.

%\clearpage
%======================================================================
%   Acknowledgements
%======================================================================
%\begin{acknowledgments}
\vskip 0.75cm
We thank the anonymous referee for constructive comments that helped us to improve 
the manuscript.
We gratefully acknowledge support for this research from 
NASA grants NNX17AF45G and 80NSSC22K0483 (L.~L.~C.), 
the William F. Vilas Estate (A.~J.~B.),
a Kellett Mid-Career Award and a WARF Named Professorship from the 
University of Wisconsin-Madison Office of the 
Vice Chancellor for Research and Graduate Education with funding from the 
Wisconsin Alumni Research Foundation (A.~J.~B.),
the Millennium Science Initiative Program -- ICN12\_009 (F.~E.~B), 
CATA-Basal -- FB210003 (F.~E.~B), and FONDECYT Regular -- 1190818 (F.~E.~B) 
and 1200495 (F.~E.~B, C.~O.),
the Ministry of Science and Technology of Taiwan (MOST 1092112-M-001-016-MY3)
(C.-C.~C.),
WSGC Graduate and Professional Research Fellowships (L.~H.~J., A.~J.~T.),
and a Sigma Xi Grant in Aid of Research (A.~J.~T.).

This work utilizes gravitational lensing models produced by PIs Brada\v{c}, 
Natarajan \& Kneib (CATS), Merten \& Zitrin, Sharon, Williams, Keeton, Bernstein 
and Diego, and the GLAFIC group. This lens modeling was partially funded by the 
HST Frontier Fields program conducted by STScI. STScI is operated by the 
Association of Universities for Research in Astronomy, Inc. under NASA contract 
NAS 5-26555. The lens models were obtained from the Mikulski Archive for Space 
Telescopes (MAST).

The National Radio Astronomy Observatory is a facility of the National Science
Foundation operated under cooperative agreement by Associated Universities, Inc.
This paper makes use of the following ALMA data: 
ADS/JAO.ALMA\,\#2013.1.00999.S, \\ %(FB)
ADS/JAO.ALMA\,\#2015.1.01425.S, \\ %(FB)
ADS/JAO.ALMA\,\#2016.1.00293.S, \\ %(AP)
ADS/JAO.ALMA\,\#2017.1.00091.S, \\ %(AP)
ADS/JAO.ALMA\,\#2017.1.00341.S, \\ %(us)
ADS/JAO.ALMA\,\#2017.1.01219.S, \\ %(FB)
ADS/JAO.ALMA\,\#2018.1.00003.S, \\ %(us)
and ADS/JAO.ALMA\,\#2018.1.00035.L. %(KK)
ALMA is a partnership of ESO (representing its member states), NSF (USA), and NINS (Japan), 
together with NRC (Canada), MOST and ASIAA (Taiwan), and KASI (Republic of Korea),
 in cooperation with the Republic of Chile. The Joint ALMA Observatory is operated by 
 ESO, AUI/NRAO, and NAOJ.
 
 The James Clerk Maxwell Telescope is operated by the East Asian Observatory on 
behalf of The National Astronomical Observatory of Japan, Academia Sinica Institute 
of Astronomy and Astrophysics, the Korea Astronomy and Space Science Institute, 
the National Astronomical Observatories of China and the Chinese Academy of 
Sciences (grant No. XDB09000000), with additional funding support from the Science 
and Technology Facilities Council of the United Kingdom and participating universities 
in the United Kingdom and Canada. 
 
The W.~M.~Keck Observatory is operated as 
a scientific partnership among the California Institute of Technology, the 
University of California, and NASA, and was made possible by the generous financial 
support of the W.~M.~Keck Foundation.

We wish to recognize and acknowledge 
the very significant cultural role and reverence that the summit of Maunakea has always 
had within the indigenous Hawaiian community. We are most fortunate to have the 
opportunity to conduct observations from this mountain.
%\end{acknowledgments}

\facilities{ALMA, JCMT, KeckI, KeckII}

%======================================================================
%   References
%======================================================================

%-----------------------------------------------------------------------------

%---------------------------------------------------------------------
% TABLE 2
%---------------------------------------------------------------------
\begin{deluxetable*}{lccrcrrcrlcc}
\setcounter{table}{1}
\renewcommand\baselinestretch{1.0}
\tablewidth{0pt}
\tablecaption{SCUBA-2 \afluxb Sample ($>4\sigma$) \label{tab2}}
\scriptsize
\tablehead{No. and Name & R.A. & Decl.&  $f_{850}$  & Error & S/N  & $f_{450}$ & Error & S/N & $f_{450}/f_{850}$ & Error & Offset \\ 
& J2000.0 & J2000.0 & \multicolumn{2}{c}{(mJy)} &  & \multicolumn{2}{c}{(mJy)} &  & & & (arcmin)  \\ 
(1) & (2) & (3) & (4) & (5) & (6) & (7) & (8) & (9) & (10) & (11) & (12)}
\startdata
   1   SMM131126-11906 &       13       11 26.3 &       -1 19 06.3 &   12.0 &  0.5 &   24.0 &   33.9 &   2.6 &      12.0 &    2.73 &  0.23  &   1.2\cr
   2   SMM131121-11951 &       13       11 21.8 &       -1       19 51.3 &   11.0 &  0.5 &   21.0 &   14.7 &   2.7 &      5.3 &    1.31 &  0.25  &   1.7\cr
   3   SMM131131-11804 &       13       11 31.8 &       -1 18 04.3 &   9.5 &  0.6 &   17.0 &   38.3 &   2.8 &      13.0 &    3.99 &  0.37  &   2.2\cr
   4   SMM131117-12239 &       13       11 17.5 &       -1       22 39.3 &   7.6 &  0.6 &   12.0 &   18.3 &   3.7 &      4.8 &    2.39 &  0.52  &   3.7\cr
   5   SMM131121-12108 &       13       11 21.8 &       -1 21 08.3 &   6.2 &  0.5 &   11.0 &   18.2 &   2.8 &      6.4 &    2.92 &  0.52  &   1.9\cr
   6   SMM131118-12213 &       13       11 18.6 &       -1       22 13.3 &   5.4 &  0.6 &   8.9 &   15.3 &   3.6 &      4.2 &    2.80 &  0.73  &   3.2\cr
   7   SMM131119-12151 &       13       11 19.2 &       -1       21 51.3 &   5.4 &  0.6 &   9.1 &   9.5 &   3.4 &      2.7 &    1.75 &  0.66  &   2.9\cr
   8   SMM131129-12048 &       13       11 29.2 &       -1       20 48.4 &   5.1 &  0.5 &   10.0 &   14.5 &   2.3 &      6.2 &    2.84 &  0.52  &  0.6\cr
   9   SMM131138-11636 &       13       11 38.0 &       -1       16 36.3 &   4.8 &  0.7 &   7.2 &   8.1 &   3.3 &      2.4 &    1.68 &  0.73  &   4.3\cr
  10   SMM131128-12224 &       13       11 28.4 &       -1       22 24.3 &   4.7 &  0.5 &   8.8 &   21.5 &   2.9 &      7.3 &    4.50 &  0.79  &   2.2\cr
  11   SMM131113-11858 &       13       11 13.2 &       -1       18 58.3 &   4.7 &  0.6 &   7.9 &   14.0 &   3.6 &      3.8 &    2.95 &  0.85  &   4.0\cr
  12   SMM131134-12017 &       13       11 34.8 &       -1       20 17.3 &   4.7 &  0.5 &   9.0 &   11.2 &   2.5 &      4.4 &    2.38 &  0.59  &   1.5\cr
  13   SMM131114-12229 &       13       11 14.7 &       -1       22 29.4 &   4.4 &  0.6 &   7.4 &   15.1 &   3.6 &      4.1 &    3.43 &  0.95  &   4.1\cr
  14   SMM131126-12317 &       13       11 26.0 &       -1       23 17.4 &   4.3 &  0.6 &   7.7 &   23.6 &   3.1 &      7.4 &    5.48 &   1.02  &   3.1\cr
  15   SMM131127-11847 &       13       11 27.1 &       -1       18 47.3 &   4.2 &  0.5 &   8.2 &   11.3 &   2.7 &      4.1 &    2.67 &  0.71  &   1.4\cr
  16   SMM131119-12257 &       13       11 19.8 &       -1       22 57.3 &   4.2 &  0.6 &   7.0 &   13.1 &   3.6 &      3.5 &    3.10 &  0.97  &   3.5\cr
  17   SMM131123-12046 &       13       11 23.9 &       -1       20 46.3 &   4.1 &  0.5 &   8.1 &   9.8 &   2.5 &      3.8 &    2.37 &  0.67  &   1.3\cr
  18   SMM131133-11712 &       13       11 33.3 &       -1       17 12.3 &   3.8 &  0.6 &   6.3 &   18.0 &   3.0 &      5.9 &    4.70 &   1.08  &   3.1\cr
  19   SMM131116-12307 &       13       11 16.2 &       -1 23 07.3 &   3.8 &  0.6 &   6.3 &   11.6 &   3.7 &      3.1 &    3.06 &   1.09  &   4.2\cr
  20   SMM131141-11824 &       13       11 41.1 &       -1       18 24.3 &   3.7 &  0.7 &   5.3 &   9.3 &   3.8 &      2.4 &    2.50 &   1.12  &   3.6\cr
  21   SMM131130-11911 &       13       11 30.8 &       -1       19 11.3 &   3.6 &  0.5 &   7.1 &   14.6 &   2.4 &      5.9 &    4.04 &  0.88  &   1.1\cr
  22   SMM131128-11816 &       13       11 28.1 &       -1       18 16.4 &   3.5 &  0.5 &   6.6 &   13.6 &   2.9 &      4.6 &    3.79 &  0.99  &   1.9\cr
  23   SMM131133-12023 &       13       11 33.8 &       -1       20 23.3 &   3.5 &  0.5 &   6.9 &   9.6 &   2.4 &      3.9 &    2.71 &  0.78  &   1.3\cr
  24   SMM131132-11952 &       13       11 32.2 &       -1       19 52.3 &   3.5 &  0.5 &   6.9 &   3.6 &   2.3 &      1.5 &    1.02 &  0.69  &   1.0\cr
  25   SMM131121-11946 &       13       11 21.0 &       -1       19 46.3 &   3.3 &  0.5 &   6.3 &   15.0 &   2.7 &      5.3 &    4.50 PAIR &   1.09  &   1.9\cr
  26   SMM131113-12115 &       13       11 13.2 &       -1       21 15.3 &   3.3 &  0.8 &   5.7 &   6.8 &   3.4 &      1.9 &    2.02 &   1.09  &   4.0\cr
  27   SMM131124-12212 &       13       11 24.1 &       -1       22 12.4 &   3.2 &  0.6 &   5.7 &   8.0 &   3.0 &      2.6 &    2.50 &   1.04  &   2.3\cr
  28   SMM131131-12409 &       13       11 31.9 &       -1 24 09.4 &   3.1 &  0.6 &   5.3 &   15.3 &   3.4 &      4.4 &    4.83 &   1.41  &   4.0\cr
  29   SMM131118-12326 &       13       11 18.8 &       -1       23 26.3 &   3.0 &  0.6 &   5.1 &   24.6 &   3.6 &      6.6 &    8.06 &   1.98  &   4.0\cr
  30   SMM131132-11821 &       13       11 32.7 &       -1       18 21.4 &   2.8 &  0.6 &   5.0 &   5.4 &   2.7 &      1.9 &    1.88 &   1.03  &   2.0\cr
  31   SMM131139-12034 &       13       11 39.0 &       -1       20 34.3 &   2.8 &  0.6 &   4.9 &   9.5 &   2.8 &      3.3 &    3.38 &   1.23  &   2.5\cr
  32   SMM131133-12424 &       13       11 33.5 &       -1       24 24.3 &   2.7 &  0.6 &   4.4 &   11.9 &   3.7 &      3.2 &    4.29 &   1.64  &   4.3\cr
  33   SMM131134-11804 &       13       11 34.4 &       -1 18 04.3 &   2.7 &  0.6 &   4.5 &   10.0 &   2.9 &      3.3 &    3.60 &   1.34  &   2.6\cr
  34   SMM131134-11904 &       13       11 34.0 &       -1 19 04.3 &   2.7 &  0.5 &   5.0 &   11.3 &   2.6 &      4.2 &    4.10 &   1.27  &   1.7\cr
  35   SMM131121-11634 &       13       11 21.8 &       -1       16 34.4 &   2.7 &  0.7 &   4.0 &  0.4 &   4.2 &    0.1 &   0.13 &   1.57  &   4.0\cr
  36   SMM131128-11907 &       13       11 28.0 &       -1 19 07.4 &   2.6 &  0.5 &   5.3 &  0.9 &   2.5 &     0.4 &   0.34 &  0.95  &   1.1\cr
  37   SMM131129-11818 &       13       11 29.3 &       -1       18 18.3 &   2.6 &  0.5 &   4.8 &   5.2 &   2.7 &      1.8 &    2.00 &   1.14  &   1.8\cr
  38   SMM131134-11721 &       13       11 34.5 &       -1       17 21.4 &   2.5 &  0.6 &   4.1 &   8.7 &   3.0 &      2.8 &    3.39 &   1.45  &   3.1\cr
  39   SMM131122-11746 &       13       11 22.8 &       -1       17 46.4 &   2.5 &  0.6 &   4.0 &   5.2 &   3.7 &      1.3 &    2.05 &   1.55  &   2.7\cr
  40   SMM131117-12234 &       13       11 17.8 &       -1       22 34.4 &   2.4 &  0.6 &   4.0 &   17.3 &   3.7 &      4.6 &    6.97 PAIR &   2.28  &   3.6\cr
  $\cdots$ & $\cdots$ & $\cdots$ & $\cdots$ & $\cdots$ & $\cdots$ & $\cdots$ & $\cdots$ & $\cdots$ & $\cdots$ & $\cdots$ &$\cdots$ \cr
\enddata
\tablecomments{
The columns are (1) number and name, (2) and (3) SCUBA-2 \afluxb R.A. and decl.,
(4), (5), and (6) SCUBA-2 \afluxb flux, error
(we have included a confusion noise of 0.33~mJy added in quadrature to
the white noise here),
and S/N, (7), (8), and (9) SCUBA-2 \afluxa flux, error, and S/N, 
(10) and (11) SCUBA-2 \afluxa to \afluxb flux ratio and error,
and (12) offset from the cluster center in arcmin.
In Column~(10), we also indicate whether the source is part of a close pair 
(labeled ``PAIR'') or corresponds to the brightest cluster galaxy (labeled ``BCG"). 
(This table is available in its entirety in machine-readable form.)
}
\end{deluxetable*}
%---------------------------------------------------------------------

%---------------------------------------------------------------------
% TABLE 3
%---------------------------------------------------------------------
\begin{deluxetable*}{lccrcrrcclccc}
\setcounter{table}{2}
\renewcommand\baselinestretch{1.0}
\tablewidth{0pt}
\tablecaption{A370 Band~7 (870~$\mu$m) ALMA $>4.5\sigma$ \label{a370_band7}}
\scriptsize
\tablehead{No. and Name & R.A. & Decl. & Pk Flux  & Error & S/N & Tot Flux & SCUBA-2 & m160 & $z$ & Median & $\sigma$ & Zitrin \\
& & & & & & & $f_{850}:f_{450}$ & & & Magnif. & & Magnif. \\
& J2000.0 & J2000.0 & \multicolumn{2}{c}{(mJy)} & & (mJy)  & (mJy) &  & & & \\ 
(1) & (2) & (3) & (4) & (5) & (6) & (7) & (8) & (9) & (10) & (11) & (12) & (13)}
\startdata
   1  ALMA023951--13559 &   2  39 51.94 &  -1  35 59.0 &  11.00 & 0.35 &  30.8  &   14.3 & 19.0 :   37.0  &    22.7 &   2.808 & 3.93 & 0.52 & \nodata \cr
   2  ALMA023951--13558 &   2  39 51.80 &  -1  35 58.5 &  6.16 & 0.37 &  16.4  &   8.0 & 18.0 :   40.0  &    20.7 &   2.806$^a$ & 3.61 & 0.46 & \nodata \cr
   3  ALMA023957--13453 &   2  39 57.58 &  -1  34 53.7 &  4.05 & 0.24 &  16.6  &   4.2 & 4.1 :   10.0  &    23.5 &   2.04 & 2.10 & 0.41 & 2.44 \cr
   4  ALMA023956--13426 &   2  39 56.57 &  -1  34 26.3 &  5.25 & 0.25 &  20.4  &   6.2 & 6.8 :   37.0  &    19.5 &   1.062$^a$ & 2.87 & 1.10 & 2.78 \cr
   5  ALMA023949--13551 &   2  39 49.12 &  -1  35 51.8 &  1.55 & 0.24 &  6.3  &   1.9 & 1.8 :   4.6  &    22.8 &   2.507 & 1.87 & 0.46 & 3.09 \cr
   6  ALMA023950--13542 &   2  39 50.19 &  -1  35 42.1 &  2.49 & 0.24 &  10.0  &   2.6 & 1.2 :   9.4  &    23.5 &   2.01 & 2.48 & 2.96 & 2.12 \cr
   7  ALMA023947--13517 &   2  39 47.11 &  -1  35 17.7 &  2.20 & 0.39 &  5.5  &   2.1 & 2.4 :   7.0  &    23.2 & 3.472 & 1.56 & 0.38 & 1.65 \cr
   8  ALMA023958--13424 &   2  39 58.16 &  -1  34 24.7 &  1.58 & 0.25 &  6.1  & 2.1 &  0.6 :   11.0  &    20.7 & 1.256 & 1.33 & 0.27 & \nodata \cr 
   9  ALMA023946--13332 &   2  39 46.88 &  -1  33 32.8 &  2.43 & 0.27 &  8.7  &  2.6 &  2.2 :   13.0  &    21.8 & 2.487 & 1.55 & 0.29 & 1.00 \cr
  10  ALMA023954--13320 &   2  39 54.24 &  -1  33 20.7 &  2.04 & 0.24 &  8.3  &   3.5 & 2.4 :   13.0  &    21.1 & 1.523 & 3.17 & 0.97 & \nodata \cr 
\enddata
\tablecomments{
The columns are (1) number and name, (2) and (3) ALMA R.A. and decl., 
(4), (5), and (6) ALMA peak flux, error, and S/N,
(7) ALMA $2''$ aperture flux (except for blended source numbers 1 and 2, which are peak flux times 1.3),
(8) SCUBA-2 \afluxb and \afluxa fluxes,
(9) m160 magnitude, 
(10) redshift (speczs have three
digits after the decimal point, and photzs have two),
 (11) and (12) median magnification and standard deviation for all the models submitted by the
HFF teams, as obtained from the online tool provided by Dan Coe
(\url{https://archive.stsci.edu/prepds/frontier/lensmodels/webtool/magnif.html}),
and (13) adopted Zitrin magnification.
Note that all speczs are from our Keck observations. 
Sources~1 and 2 also had previous speczs from Ivison et al.\ (1998), 
and source~4 had a previous specz from Barger et al.\ (1999) and Soucail et al.\ (1999).
The m160 magnitudes and photzs are from Brada\v{c} et al.\ (2019).
$^a$Type~2 AGN.
}
\end{deluxetable*}
%---------------------------------------------------------------------

%---------------------------------------------------------------------
% TABLE 4
%---------------------------------------------------------------------
\begin{deluxetable*}{lccccrcclccc}
\setcounter{table}{3}
\renewcommand\baselinestretch{1.0}
\tablewidth{0pt}
\tablecaption{A370 Band~6 (1.2~mm) ALMA $>4.5\sigma$ \label{a370_band6}}
\scriptsize
\tablehead{No. and Name & R.A. & Decl. &  Pk Flux  & Error & S/N & SCUBA-2 & m160 & $z$ & Median & $\sigma$ & Zitrin \\ 
& & & & & & $f_{850}:f_{450}$ & & & Magnif. & & Magnif. \\
& J2000.0 & J2000.0 & \multicolumn{2}{c}{(mJy)} &  & (mJy) &  & & & & \\ 
(1) & (2) & (3) & (4) & (5) & (6) & (7) & (8) & (9) & (10) & (11) & (12)}
\startdata
1 ALMA023956--13426$^b$ & 2 39 56.57 & -1 34 26.3 &  1.50 & 0.06 &  23.4  &   6.8 :   37.0  &    19.5 &   1.062$^a$ & 2.87 & 1.10 & 2.78 \cr
2 ALMA023950--13542$^b$ & 2 39 50.19 & -1 35 42.1 &  0.92 &  0.18 &   5.2  &   1.2 :   9.4  &    23.5 &   2.01 & 2.48 & 2.96 & 2.12 \cr
3 ALMA023958--13424$^b$ & 2 39 58.14 & -1 34 24.6 &  0.55 &  0.12 &   4.5  &  0.6 :   11.0  &    20.7 & 1.256 & 1.33 & 0.27 & \nodata \cr 
\enddata
\tablecomments{
The columns are (1) number and name, (2) and (3) ALMA R.A. and decl., 
(4), (5), and (6) ALMA peak flux, error, and S/N,
(7) SCUBA-2 \afluxb and \afluxa fluxes,
(8) m160 magnitude, 
(9) redshift (speczs have three 
digits after the decimal point, and photzs have two),
 (10) and (11) median magnification and standard deviation for all the models submitted by the
HFF teams, as obtained from the online tool provided by Dan Coe
(\url{https://archive.stsci.edu/prepds/frontier/lensmodels/webtool/magnif.html}),
and (12) adopted Zitrin magnification.
Note that both speczs are from our Keck observations.
The m160 magnitudes and photz are from Brada\v{c} et al.\ (2019).
$^a$Type~2 AGN.
$^b$This source is also in the band~7 sample (see Table~\ref{a370_band7}).
}
\end{deluxetable*}
%---------------------------------------------------------------------

%---------------------------------------------------------------------
% TABLE 5
%---------------------------------------------------------------------
\begin{deluxetable*}{lccccrccclccc} [ht]
\setcounter{table}{4}
\renewcommand\baselinestretch{1.0}
\tablewidth{0pt}
\tablecaption{MACSJ1149.5+2223 Band~7 (870~$\mu$m) ALMA $>4.5\sigma$ \label{macsj1149_band7}}
\scriptsize
\tablehead{No. and Name & R.A. & Decl. & Pk Flux  & Error & S/N & Tot Flux & SCUBA-2 & m160 & $z$ & Median & $\sigma$ & Zitrin \\
& & & & & & & $f_{850}:f_{450}$ & & & Magnif. & & Magnif. \\
& J2000.0 & J2000.0 & \multicolumn{2}{c}{(mJy)} & & (mJy)  & (mJy) & & & & \\ 
(1) & (2) & (3) & (4) & (5) & (6) & (7) & (8) & (9) & (10) & (11) & (12) & (13)} 
\startdata
1 ALMA014941-222316 &  11 49 41.48 &   22  23 16.1 & 0.94 & 0.14 &  6.8  &  1.3  &  1.4 :  5.0  &    24.2 &    2.52 & 1.97 & 0.48 & 2.04 \cr
2 ALMA014943-222400 &  11 49 43.64 &  22  24 0.39 &  1.08 & 0.19 &  5.7  &   1.0 &  2.2 :  1.9  &  \nodata & \nodata & \nodata & \nodata & \nodata \cr
3 ALMA014933-222226 &  11 49 33.87 &   22  22 26.9 & 0.85 & 0.15 &  5.5  &   1.0 & 1.4 :  6.1  &    17.4 &   0.554 & 1.00 & 0.02 & \nodata \cr 
4 ALMA014942-222339 &  11 49 42.37 &  22  23 39.5 & 0.60 & 0.13 &  4.6  &   0.6  & 1.3 :  6.4  &    21.1 &    1.632 & 1.43 & 0.21 & \nodata \cr 
5 ALMA014941-222436 &  11 49 41.46 &   22  24 36.1 &  1.49 & 0.17 &  9.0  &  2.0 &  1.5 :  8.5  &    22.9 &    1.720 & 1.20 & 0.14 & 1.29 \cr
6 ALMA014930-222427 &  11 49 30.67 &   22  24 27.7 &  2.57 & 0.15 &  17.0  &  4.3 &  4.6 :  15.0  &    20.4 &    1.489 & 1.68 & 0.29 & 2.03 \cr
7 ALMA014937-222430 &  11 49 37.28 &  22  24 30.3 & 0.67 & 0.14 &  4.7  &   0.2 &  2.4 :  6.0  &    20.1 &    1.020 & 1.53 & 0.16 & 1.60 \cr
8 ALMA014936-222424 &  11 49 36.09 &  22  24 24.2 & 0.91 & 0.11 &  8.3  &  1.6 & 1.5 :  5.4  &  20.5 &  1.603 & 3.07 & 0.72 & 3.42 \cr
9 ALMA014934-222445 &  11 49 34.42 &   22  24 45.3 & 0.81 & 0.16 &  5.1  &  1.3 &  1.8 :  5.3  &     20.1 &   0.976 & 3.89 & 2.11 & 2.72 \cr
10 ALMA014935-222231 & 11 49 35.46 &  22  22 31.9 &  1.02 & 0.11 &  9.2  &  1.2 & 1.8 :  5.5  &   22.0 & \nodata & \nodata & \nodata & \nodata \cr
11 ALMA014930-222253 & 11 49 30.82 &  22  22 53.7 &  1.26 & 0.15 &  8.5 &   1.2 & 1.4 : 8.3 & 18.6 & 0.31 & 1.05 & 0.11 & 1.11 \cr
12 ALMA014931-222252 & 11 49 31.30 &  22  22 52.2 & 0.62 & 0.13 &  4.9  &  1.2 & 1.3 : 8.8  &   19.0  &  0.540 & 1.01 & 0.06 & 1.16 \cr 
\enddata
\tablecomments{
The columns are as in Table~\ref{a370_band7}.
Note that all of the speczs are from our Keck observations, except for sources~7 and 12, which
are from the HST grism reductions of Rawle et al.\ (2016).
The m160 magnitudes and photzs are from the 
ASTRODEEP catalog of Di Criscienzo et al.\ (2017),
except for sources~4 and 5, where we measured corrected $2''$ diameter m160 magnitudes 
from the BUFFALO F160W image. Source~2 lies outside the BUFFALO image, but it is 
extremely faint in the Subaru/Suprime-Cam $z'$ image (Umetsu et al.\ 2014).
}
\end{deluxetable*}
%---------------------------------------------------------------------

%---------------------------------------------------------------------
% TABLE 6
%---------------------------------------------------------------------
\begin{deluxetable*}{lccccrcccccc}
\setcounter{table}{5}
\renewcommand\baselinestretch{1.0}
\tablewidth{0pt}
\tablecaption{MACSJ1149.5+2223 Band~6 (1.2~mm) ALMA $>4.5\sigma$ \label{macsj1149_band6}}
\scriptsize
\tablehead{No. and Name & R.A. & Decl.&  Pk Flux  & Error & S/N & SCUBA-2 & m160 & $z$ & Median & $\sigma$ & Zitrin \\ 
& & & & & & $f_{850}:f_{450}$ & & & Magnif. & & Magnif. \\
& J2000.0 & J2000.0 & \multicolumn{2}{c}{(mJy)} &  & (mJy) &  & & & & \\ 
(1) & (2) & (3) & (4) & (5) & (6) & (7) & (8) & (9) & (10) & (11) & (12)}
\startdata
1 ALMA014941-222436$^a$ & 11 49 41.46 &  22  24 36.2 & 0.51 & 0.06 &  7.8  &   1.4 :   7.5  &    22.9 &   1.720 & 1.20 & 0.14 & 1.29 \cr
 2 ALMA014926-222455 & 11 49 26.57 &  22  24 55.7 &  3.05 & 0.07 &  42.6  &   6.8 :   12.0  &  \nodata & \nodata & \nodata & \nodata & \nodata \cr
 3 ALMA014930-222427$^a$ &  11 49 30.69 &  22  24 27.7 & 0.55 & 0.07 &  7.5  &   4.7 :   15.0  &    20.4 &   1.489 & 1.68 & 0.29 & 2.03 \cr
 4 ALMA014933-222552 & 11 49 33.33 &   22  25 52.2 &  1.89 & 0.07 &  26.7  &   3.7 :   8.9  &  \nodata & \nodata & \nodata & \nodata & \nodata \cr
 5 ALMA014946-222542 & 11 49 46.82 &  22  25 42.1 &  1.25 & 0.08 &  16.1  &   6.0 :   20.0  &  \nodata & \nodata & \nodata & \nodata & \nodata \cr
 6 ALMA014945-222232 & 11 49 45.01 &  22  22 32.6 & 0.87 & 0.09 &  9.2  &   2.9 :   8.2  &  \nodata & \nodata & \nodata & \nodata & \nodata \cr
 7 ALMA014927-222402 & 11 49 27.52 &  22  24 2.80 & 0.53 & 0.07 &  7.8  &   2.3 :   7.5  &    21.1 & \nodata & \nodata & \nodata & \nodata \cr
 8 ALMA014928-222300 & 11 49 28.07 &  22 23 0.69 & 0.82 & 0.08 &  10.5  &   3.0 :   8.4  &    25.2 & \nodata & \nodata & \nodata & \nodata \cr
\enddata
\tablecomments{
The columns are as in Table~\ref{a370_band6}.
Note that both speczs are from our Keck observations.
The m160 magnitudes are from the 
ASTRODEEP catalog of Di Criscienzo et al.\ (2017),
except for source~1, where we measured the corrected $2''$ diameter m160 magnitude
from the BUFFALO F160W image.
$^a$This source is also in the band~7 sample (see Table~\ref{macsj1149_band7}).
}
\end{deluxetable*}
%---------------------------------------------------------------------

%---------------------------------------------------------------------
% TABLE 7
%---------------------------------------------------------------------
\begin{deluxetable*}{lccccccccccc}
\setcounter{table}{6}
\renewcommand\baselinestretch{1.0}
\tablewidth{0pt}
\tablecaption{MACSJ0717.5+3745 Band~7 (870~$\mu$m) ALMA $>4.5\sigma$ \label{macsj0717_band7}}
\scriptsize
\tablehead{No. and Name & R.A. & Decl. &  Pk Flux  & Error & S/N & SCUBA-2 & m160 & $z$ & Median & $\sigma$ & Zitrin \\ 
& & & & & & $f_{850}:f_{450}$ & & & Magnif. & & Magnif. \\
& J2000.0 & J2000.0 & \multicolumn{2}{c}{(mJy)} &  & (mJy) &  & & & & \\
(1) & (2) & (3) & (4) & (5) & (6) & (7) & (8) & (9) & (10) & (11) & (12)}
\startdata
1 ALMA071730-374433 & 7 17 30.69 & 37 44 33.0 & 0.52 & 0.09 &  6.0  &   2.8 :   2.7  &  23.7 & 4.52 & 6.94 & 2.02 & 8.71 \cr
\enddata
\tablecomments{
The columns are as in Table~\ref{a370_band6}.
The m160 magnitude and photz are from the 
ASTRODEEP catalog of Di Criscienzo et al.\ (2017).
We do not include the Brada{\v c} models in the median, as they give a very high magnification.
}
\end{deluxetable*}
%---------------------------------------------------------------------

%---------------------------------------------------------------------
% TABLE 8
%---------------------------------------------------------------------
\begin{deluxetable*}{lccccrcccccc}
\setcounter{table}{7}
\renewcommand\baselinestretch{1.0}
\tablewidth{0pt}
\tablecaption{MACSJ0717.5+3745 Band~6 (1.2~mm) ALMA $>4.5\sigma$ \label{macsj0717_band6}}
\scriptsize
\tablehead{No. and Name & R.A. & Decl. &  Pk Flux  & Error & S/N & SCUBA-2 & m160 & $z$ & Median & $\sigma$ & Zitrin \\ 
& & & & & & $f_{850}:f_{450}$ & & & Magnif. & & Magnif. \\
& J2000.0 & J2000.0 & \multicolumn{2}{c}{(mJy)} &  & (mJy) &  & & & & \\
(1) & (2) & (3) & (4) & (5) & (6) & (7) & (8) & (9) & (10) & (11) & (12)}
\startdata
1 ALMA071724-374329 &  7 17 24.55 & 37 43 29.5 &  2.62 & 0.18 &  13.8  &   4.9 :   18.0  &  \nodata & \nodata & \nodata & \nodata & \nodata \cr
2 ALMA071742-374225 &  7 17 42.86 & 37 42 25.7 &  3.02 & 0.20 &  14.6  &   2.8 :   5.0  &  \nodata & \nodata & \nodata & \nodata & \nodata \cr
3 ALMA071742-374231 &  7 17 42.55 & 37 42 31.0 &  2.26 & 0.22 &  10.1  &   2.4 :   6.5  &  \nodata & \nodata & \nodata & \nodata & \nodata \cr
4 ALMA071726-374247 &  7 17 26.01 & 37 42 47.5 &  1.03 & 0.17 &  6.0  &   3.8 :   11.0  &  \nodata & \nodata & \nodata & \nodata & \nodata \cr
\enddata
\tablecomments{
The columns are as in Table~\ref{a370_band6}.}
\end{deluxetable*}
%--------------------------------------------------------------------

%---------------------------------------------------------------------
% TABLE 9
%---------------------------------------------------------------------
\begin{deluxetable*}{lrrrcrcccl}
\setcounter{table}{8}
\tablewidth{0pt}
\tablecaption{ALMA Archive Sample \label{archivetable} }
\scriptsize
\tablehead{Cluster & R.A. & Decl. &  $f_{850}$ & Error & $f_{450}$ & Error & Note & m160& Redshift
\\ & J2000.0 & J2000.0 & \multicolumn{2}{c}{(mJy)} & \multicolumn{2}{c}{(mJy)} & &  &  \\ 
(1) & (2) & (3) & (4) & (5) & (6) & (7) & (8) & (9) & (10)}
\startdata
A1689&      13       11 30.80&      -1       19 11.3&  3.6 & 0.3 &  13.0 & 2.4 & &  24.4& \nodata \cr
A1689&      13       11 31.45&      -1       19 32.4& -1.2 & 0.3 & 0.6 &  2.3 & &  17.4& 0.187 \cr
A1689&      13       11 30.03&      -1       20 28.4&-0.7 & 0.3 &0.1 &  2.2 & &  18.1& 0.200 \cr
A1689&      13       11 29.93&      -1       19 18.7& 0.8 & 0.3 & 0.4 & 2.4 & &  25.3&  7.130 \cr
A2390&      21       53 36.78&      17       41 42.3&  7.2 & 0.3 &  4.9 & 2.4 & BCG &\nodata&\nodata\cr
A2744&       0       14 19.80&     -30       23 07.7 &  1.7 & 0.3 & 0.8 & 3.3 & &  24.4&  2.96\cr
A2744&       0       14 18.35&     -30       24 47.3&  6.7 & 0.3 &  10.0 & 3.4 & &  25.2&  2.482\cr
A2744&       0       14 20.40&     -30       22 54.3&  3.6 & 0.4 &  7.8 & 3.3 & &  23.5&  3.058\cr
A2744&       0       14 17.58&     -30       23 00.5 &  3.3 & 0.3 &  18.0 & 3.4 & &  21.9&  1.498\cr
A2744&       0       14 19.12&     -30       22 42.2&  1.8 & 0.4 &  6.5 & 3.4 & &  23.5&  2.409\cr
A2744&       0       14 17.28&     -30       22 58.5&  4.1 & 0.3 &  18.0 & 3.4 & &  22.0&  1.55\cr
A2744&       0       14 22.10&     -30       22 49.6&-0.4 & 0.4 &  5.2 & 3.4 & &  25.4&  2.644\cr
A2744&       0       14 19.50&     -30       22 48.7&  1.8 & 0.4 & -6.0 & 3.4 & &  27.0&  4.16\cr
A2744&       0       14 19.79&     -30       22 37.8& 0.5 & 0.4 &  5.1 & 3.5 & &  25.5&  5.56\cr
MACSJ0416&       4       16 10.79&     -24 04 47.4&  3.7 & 0.3 &  11.0 & 2.2 & &  21.4&  2.086\cr
MACSJ0416&       4       16 06.96&     -24 04 00.0 &  1.5 & 0.3 &  2.8 & 2.2 & &  22.0&  1.953\cr
MACSJ0416&       4       16 08.81&     -24 05 22.5& 0.5 & 0.3 &  2.0 & 2.4 & &  23.6&  1.50\cr
MACSJ0416&       4       16 11.66&     -24 04 19.4&-0.8 & 0.3 &  5.4 & 2.2 & &  23.2&  2.13\cr
MACSJ0416&       4       16 09.44&     -24 05 35.3&-0.1 & 0.4 & -2.8 & 2.5 & &  25.9& 8.311\cr
MACSJ1423&      14       23 47.87&      24 04 42.2&  1.7 & 0.2 &  6.5 & 1.6 & BCG &\nodata&\nodata\cr
MACSJ2129&      21       29 21.32&      -7       41 15.4&  8.8 & 0.4 &  28.0 & 3.3 & &\nodata&\nodata\cr
MACSJ2129&      21       29 22.35&      -7       41 31.0&-0.6 & 0.4 & -2.8 & 3.2 & &  20.3&  1.537\cr
RXJ1347&      13       47 30.62&     -11       45 09.6&  3.7 & 0.3 &  3.2 & 1.9 & BCG &\nodata&\nodata\cr
RXJ1347&      13       47 27.82&     -11       45 55.9&  12.0 & 0.3 &  47.0 & 2.0 & LENS &  20.2&  1.28\cr
RXJ1347&      13       47 27.64&     -11       45 51.0&  14.0 & 0.3 &  51.0 & 1.9 & LENS &  21.4&  1.28\cr
\enddata
\tablecomments{
The columns are (1) cluster, (2) and (3) ALMA R.A. and decl., 
(4) and (5) SCUBA-2 \afluxb flux and error
(we have included a confusion noise of 0.33~mJy added in quadrature to
the white noise here),
(6) and (7) SCUBA-2 \afluxa flux and error,
(8) note as to whether the source is a brightest cluster galaxy (labeled ``BCG'') or 
a multiply-lensed source (labeled ``LENS"; see Figure~\ref{alma_lens2}),
(9) m160 magnitude, and
(10) redshift (speczs have three or more digits after the decimal point, and photzs have two).
Note that in A1689, the speczs
for the second, third, and fourth sources are from Colless et al.\ (2003),
Lin et al.\ (2018), and Wong et al.\ (2022), respectively.
For A2744, the speczs are from Mu\~{n}oz Arancibia et al.\ (2022),
and the photzs for the remaining sources
are from the ASTRODEEP catalogs of Merlin et al.\ (2016) and Castellano et al.\ (2016).
For MACSJ0416, the first two sources have speczs from the 
Grism Lens-Amplified Survey from Space (GLASS) survey
(Treu et al.\ 2015), as quoted in Laporte et al.\ (2017).
The fifth source has a specz from Bakx et al.\ (2020).
The third and fourth sources have photzs from 
the ASTRODEEP catalogs of Merlin et al.\ (2016) and Castellano et al.\ (2016).
For MACSJ2129, the specz is from Molino et al.\ (2017).
For RXJ1347, the photz for the multiply-lensed source is from Zitrin et al.\ (2015).
However, Brada\v{c} et al.\ (2008)
placed it at $z=4$ based on their gravitational lens modeling.
}
\end{deluxetable*}
%---------------------------------------------------------------------
\clearpage

%---------------------------------------------------------------------
% TABLE 10
%---------------------------------------------------------------------
\startlongtable
\begin{deluxetable*}{lrrrcrrcrrc}
\centerwidetable
\renewcommand\baselinestretch{1.0}
\tablewidth{0pt}
\tablecaption{SCUBA-2 High-Redshift Candidates\label{faintsample}}
\scriptsize
\tablehead{No. and Name & R.A. & Decl.& $f_{850}$ & Error & S/N  & $f_{450}$ & Error & S/N & $f_{450}/f_{850}$ & Radius \\ 
& J2000.0 & J2000.0 & \multicolumn{2}{c}{(mJy)} & &  \multicolumn{2}{c}{(mJy)} &  & &  (arcmin)  \\ 
(1) & (2) & (3) & (4) & (5) & (6) & (7) & (8) & (9) & (10) & (11) }
\startdata
   1 A1689   &       13       11 21.8 &       -1       19 51.3 &   11.0 &  0.5 &   21.0 &   14.7 &   2.7 &      5.3 &    1.31 &   1.7\cr
   2 A1689   &       13       11 19.2 &       -1       21 51.3 &   5.4 &  0.6 &   9.1 &   9.5 &   3.4 &      2.7 &    1.75 &   2.9\cr
   3 A1689   &       13       11 38.0 &       -1       16 36.3 &   4.8 &  0.7 &   7.2 &   8.1 &   3.3 &      2.4 &    1.68 &   4.3\cr
   4 A1689   &       13       11 32.2 &       -1       19 52.3 &   3.5 &  0.5 &   6.9 &   3.6 &   2.3 &      1.5 &    1.02 &   1.0\cr
   5 A1689   &       13       11 32.7 &       -1       18 21.4 &   2.8 &  0.6 &   5.0 &   5.4 &   2.7 &      1.9 &    1.88 &   2.0\cr
   6 A1689   &       13       11 28.0 &       -1       19 07.4 &   2.6 &  0.5 &   5.3 &  0.9 &   2.5 &     0.4 &   0.34 &   1.1\cr
   7 A2390   &       21       53 23.5 &       17       40 07.2 &   5.8 &  0.5 &   10.0 &   11.0 &   3.7 &      2.9 &    1.90 &   3.4\cr
   8 A2390   &       21       53 37.4 &       17       43 39.2 &   3.5 &  0.5 &   7.2 &   6.7 &   2.5 &      2.5 &    1.89 &   1.6\cr
   9 A2390   &       21       53 17.9 &       17       41 27.0 &   3.3 &  0.6 &   5.8 &   3.8 &   3.6 &      1.0 &    1.15 &   4.1\cr
  10 A2390   &       21       53 32.6 &       17       44 32.1 &   3.1 &  0.5 &   6.1 &   6.2 &   2.7 &      2.2 &    1.99 &   2.4\cr
  11 A2390   &       21       53 27.2 &       17       40 40.2 &   3.1 &  0.5 &   6.0 & -1.0 &   3.4 &    -0.3 &  -0.31 &   2.4\cr
  12 A2390   &       21       53 26.1 &       17       44 21.1 &   2.8 &  0.5 &   5.3 &   5.5 &   3.5 &      1.5 &    1.90 &   3.1\cr
  13 A2744  &        0       14 19.4 &      -30       23 07.1 &   2.5 &  0.5 &   5.1 &   4.2 &   3.1 &      1.3 &    1.64 &  0.8\cr
  14 A370   &        2       39 57.0 &       -1       37 18.9 &   6.2 &  0.6 &   10.0 &   7.9 &   3.8 &      2.0 &    1.25 &   3.2\cr
  15 A370   &        2       40 3.03 &       -1       31 16.0 &   5.4 &  0.7 &   8.2 &   8.9 &   3.3 &      2.6 &    1.62 &   4.0\cr
  16 A370   &        2       39 51.6 &       -1       30 33.0 &   5.4 &  0.6 &   9.4 &   8.2 &   3.6 &      2.2 &    1.50 &   3.7\cr
  17 A370   &        2       39 45.3 &       -1       38 07.0 &   4.0 &  0.6 &   6.9 & -0.7 &   3.5 &    -0.2 &  -0.18 &   4.1\cr
  18 A370   &        2       39 35.2 &       -1       34 55.0 &   3.6 &  0.6 &   6.3 &   5.8 &   4.1 &      1.4 &    1.60 &   4.3\cr
  19 A370   &        2       39 47.1 &       -1       32 20.0 &   2.8 &  0.6 &   5.0 &  0.5 &   3.5 &     0.1 &   0.16 &   2.3\cr
  20 MACSJ0416   &        4       16 5.61 &      -24 07 18.7 &   8.8 &  0.6 &   14.0 &   12.8 &   3.5 &      3.5 &    1.45 &   2.9\cr
  21 MACSJ0416   &        4       16 19.1 &      -24 03 59.7 &   4.4 &  0.6 &   7.5 &   8.0 &   2.8 &      2.7 &    1.82 &   2.6\cr
  22 MACSJ0717   &        7       17 54.2 &       37       42 53.9 &   12.0 &  0.5 &   23.0 &   12.6 &   2.6 &      4.8 &    1.04 &   4.1\cr
  23 MACSJ0717   &        7       17 43.0 &       37       40 35.9 &   8.8 &  0.5 &   16.0 &   16.4 &   2.4 &      6.8 &    1.86 &   4.4\cr
  24 MACSJ0717   &        7       17 34.3 &       37       48 01.0 &   6.2 &  0.5 &   12.0 &   12.0 &   2.3 &      5.0 &    1.93 &   3.2\cr
  25 MACSJ0717   &        7       17 51.3 &       37       44 50.9 &   5.9 &  0.5 &   12.0 &   10.8 &   2.3 &      4.5 &    1.82 &   3.2\cr
  26 MACSJ0717$^a$  &        7       17 30.8 &       37       44 37.9 &   4.3 &  0.4 &   9.8 &   5.1 &   1.6 &      3.1 &    1.18 &  0.8\cr
  27 MACSJ0717   &        7       17 40.6 &       37       47 10.0 &   2.8 &  0.5 &   5.9 &   4.8 &   2.2 &      2.1 &    1.69 &   2.6\cr
  28 MACSJ0717   &        7       17 38.7 &       37       40 27.0 &   2.7 &  0.5 &   5.1 & -0.2 &   2.4 &   -0.1 & -0.08 &   4.4\cr
  29 MACSJ0717   &        7       17 53.5 &       37       46 26.9 &   2.5 &  0.5 &   5.0 &   3.8 &   2.5 &      1.4 &    1.47 &   4.0\cr
  30 MACSJ1149   &       11       49 44.8 &       22       21 45.7 &   9.4 &  0.5 &   19.0 &   12.8 &   1.7 &      7.3 &    1.36 &   2.9\cr
  31 MACSJ1149   &       11       49 17.5 &       22       23 19.7 &   7.5 &  0.5 &   14.0 &   13.1 &   2.0 &      6.3 &    1.73 &   4.3\cr
  32 MACSJ1149   &       11       49 49.5 &       22       24 41.7 &   7.2 &  0.5 &   14.0 &   12.7 &   1.9 &      6.6 &    1.76 &   3.1\cr
  33 MACSJ1149   &       11       49 46.3 &       22       26 31.7 &   7.0 &  0.5 &   14.0 &   13.6 &   1.9 &      7.1 &    1.94 &   3.4\cr
  34 MACSJ1149   &       11       49 28.9 &       22       27 10.7 &   6.2 &  0.5 &   12.0 &   11.5 &   1.9 &      6.0 &    1.84 &   3.5\cr
  35 MACSJ1149   &       11       49 27.6 &       22       26 02.8 &   5.5 &  0.5 &   11.0 &   9.5 &   1.8 &      5.1 &    1.69 &   2.8\cr
  36 MACSJ1149   &       11       49 38.8 &       22       27 54.7 &   5.2 &  0.6 &   9.0 &   7.0 &   2.6 &      2.6 &    1.33 &   3.9\cr
  37 MACSJ1149   &       11       49 50.3 &       22       22 02.8 &   4.6 &  0.5 &   9.3 &   8.0 &   2.1 &      3.7 &    1.70 &   3.8\cr
  38 MACSJ1149   &       11       49 49.2 &       22       27 06.8 &   4.1 &  0.5 &   8.1 &   7.6 &   2.1 &      3.4 &    1.80 &   4.3\cr
  39 MACSJ1149   &       11       49 55.3 &       22       23 52.7 &   2.6 &  0.5 &   5.0 & -0.1 &   2.0 &   -0.1 & -0.06 &   4.4\cr
  40 MACSJ1149   &       11       49 43.7 &       22       24 02.8 &   2.3 &  0.5 &   5.0 &   3.9 &   1.6 &      2.3 &    1.62 &   1.7\cr
  41 MACSJ1423   &       14       23 40.6 &       24 07 39.1 &   7.5 &  0.5 &   15.0 &   12.5 &   2.2 &      5.5 &    1.66 &   3.0\cr
  42 MACSJ1423   &       14       23 28.4 &       24 05 38.9 &   5.8 &  0.5 &   11.0 &   7.8 &   2.7 &      2.7 &    1.33 &   4.4\cr
  43 MACSJ1423   &       14       24 3.17 &       24 02 53.1 &   3.8 &  0.5 &   7.4 &   6.6 &   2.4 &      2.6 &    1.72 &   4.1\cr
  44 MACSJ1423   &       14       23 28.4 &       24 04 24.0 &   3.5 &  0.5 &   6.7 &   3.9 &   2.6 &      1.4 &    1.09 &   4.4\cr
  45 MACSJ1423   &       14       23 50.0 &       24 08 36.1 &   3.2 &  0.6 &   5.8 &   6.1 &   3.1 &      1.9 &    1.87 &   3.6\cr
  46 MACSJ1423   &       14       24 3.84 &       24 04 52.9 &   2.8 &  0.6 &   5.1 &   4.3 &   2.6 &      1.6 &    1.52 &   3.6\cr
  47 MACSJ1423   &       14       24 7.40 &       24 04 30.0 &   2.8 &  0.5 &   5.2 &   4.9 &   2.5 &      1.9 &    1.72 &   4.4\cr
  48 MACSJ1423   &       14       23 40.4 &       24 06 08.1 &   2.3 &  0.5 &   5.0 &   4.1 &   2.0 &      1.9 &    1.76 &   2.0\cr
  49 MACSJ2129   &       21       29 26.1 &       -7       45 29.8 &   8.7 &  0.7 &   12.0 &   12.9 &   4.9 &      2.6 &    1.47 &   3.9\cr
  50 RXJ1347   &       13       47 21.4 &      -11       48 59.9 &   9.6 &  0.6 &   15.0 &   17.9 &   3.1 &      5.6 &    1.84 &   4.4\cr
  51 RXJ1347   &       13       47 17.4 &      -11       45 50.0 &   7.0 &  0.6 &   12.0 &   12.0 &   2.7 &      4.4 &    1.71 &   3.0\cr
  52 RXJ1347   &       13       47 41.8 &      -11       47 20.0 &   6.8 &  0.7 &   10.0 &   12.5 &   3.2 &      3.8 &    1.82 &   3.7\cr
  53 RXJ1347   &       13       47 14.0 &      -11       44 34.9 &   4.4 &  0.6 &   7.2 &   6.7 &   2.9 &      2.2 &    1.52 &   3.8\cr
  54 RXJ1347   &       13       47 31.2 &      -11       48 12.9 &   3.6 &  0.6 &   6.3 &   5.6 &   2.8 &      1.9 &    1.53 &   3.1\cr
  55 RXJ1347   &       13       47 23.0 &      -11       48 59.0 &   3.4 &  0.6 &   5.4 &   5.4 &   3.1 &      1.7 &    1.58 &   4.2\cr
\enddata
\tablecomments{
The columns are (1) number and name, (2) and (3) SCUBA-2 \afluxb R.A. and decl., 
(4), (5), and (6) SCUBA-2 \afluxb flux, error
(we have included a confusion noise of 0.33~mJy added in quadrature to
the white noise here),
and S/N, (7), (8), and (9) SCUBA-2 \afluxa flux, error, and S/N, 
(10) SCUBA-2 \afluxa to \afluxb flux ratio,
and (11) offset from the cluster center in arcmin.
$^a$This is a multiply-lensed source that is blended at the SCUBA-2 resolution 
(see Figure~\ref{alma_lens1}).
}
\end{deluxetable*}

\appendix

Here we show the \afluxb (left) and \afluxa (right) SCUBA-2 images for the remaining 9 clusters 
(MACSJ1149 is shown in Figure~\ref{macs1149_sample}), with the detected ($>4\sigma$) \afluxb sources
marked (white circles).

%---------------------------------------------------------------------
% FIGURE A1 Images 
%---------------------------------------------------------------------
\begin{figure*}
\includegraphics[width=7.2in,angle=0]{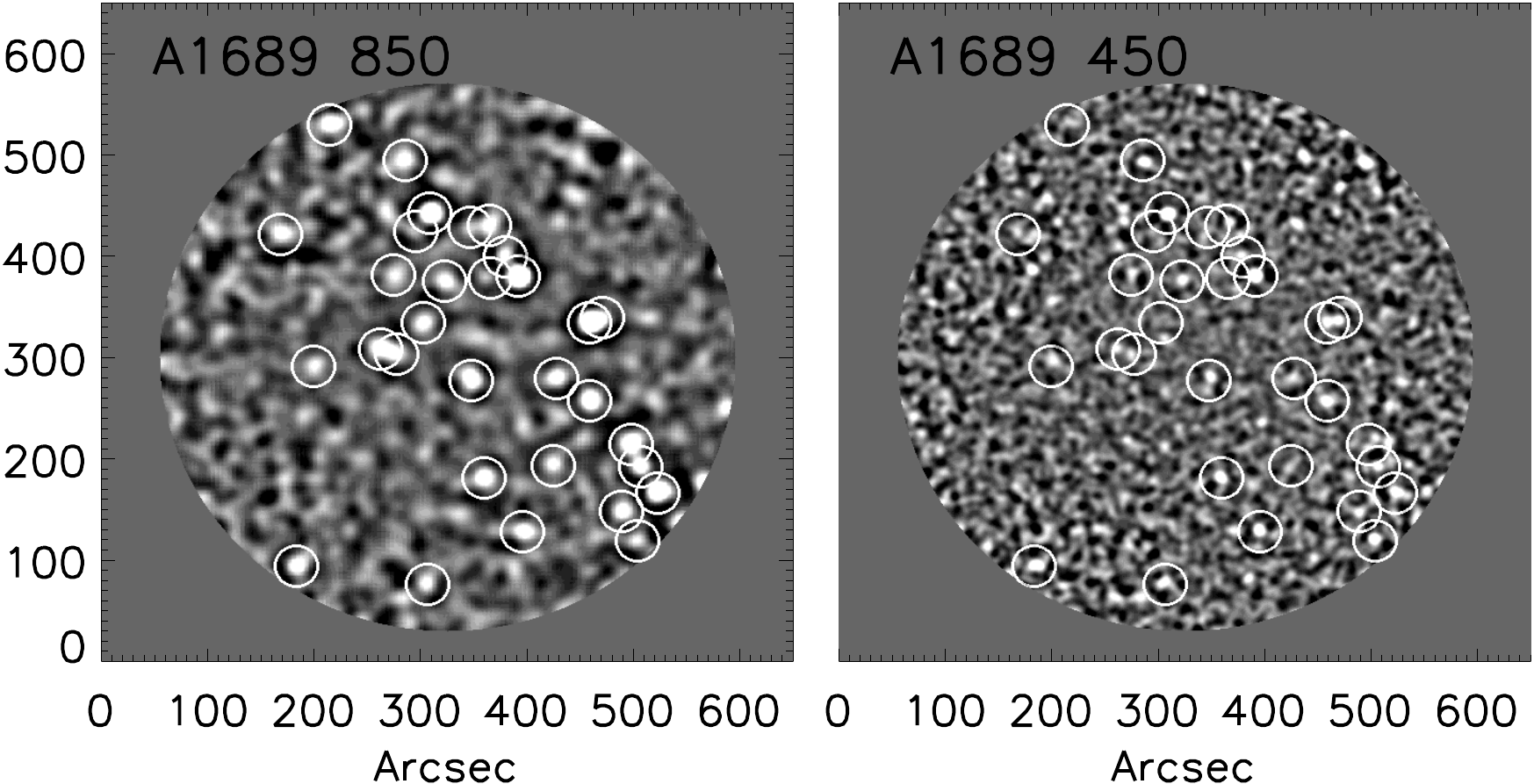}
\vskip 0.4cm
\includegraphics[width=7.2in,angle=0]{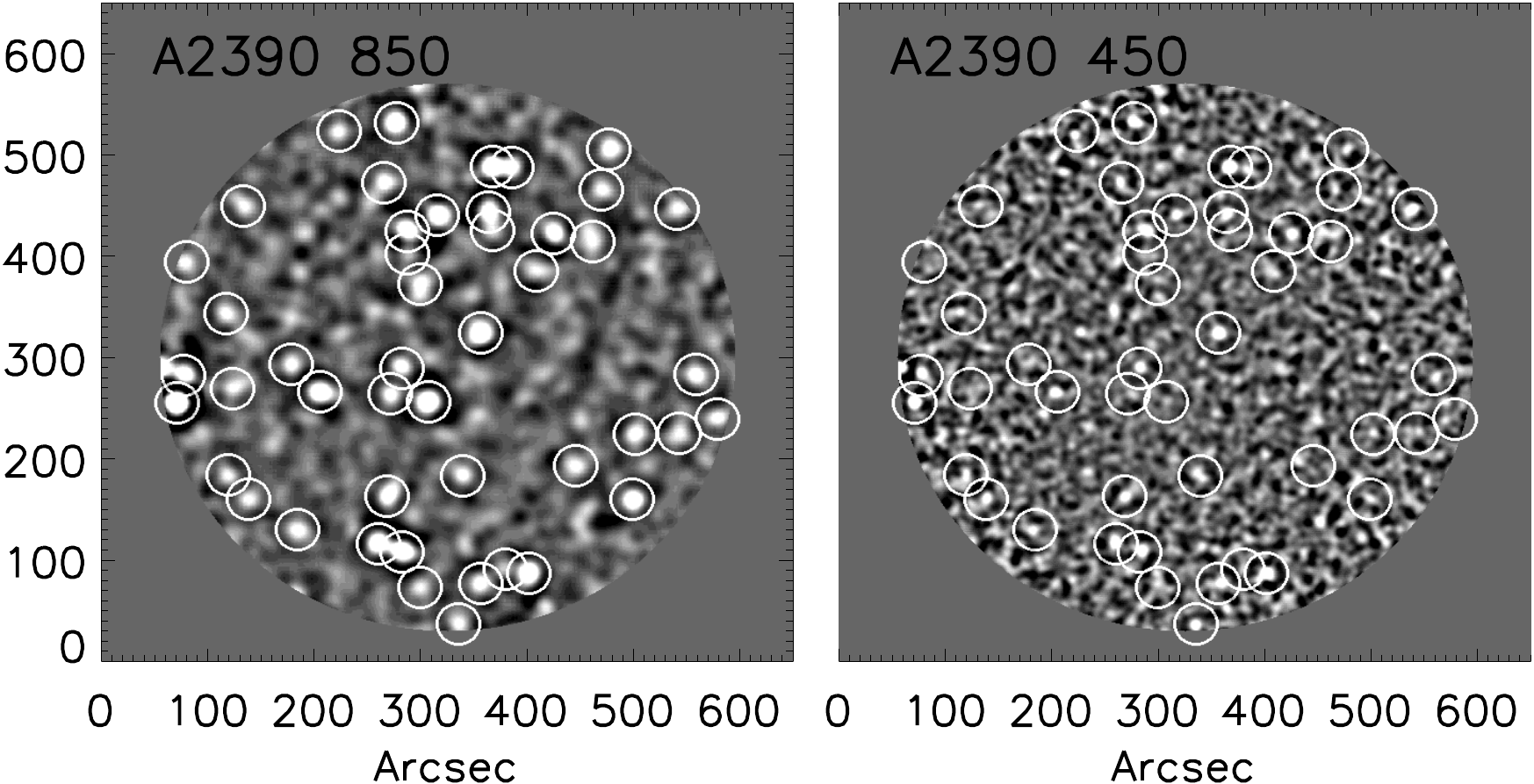}
\centering{\caption{
Images of the {\em (left)} \afluxb and {\em (right)} \afluxa
SCUBA-2 data for nine of the ten clusters (MACSJ1149 is shown
in Figure~\ref{macs1149_sample}). In each
panel, we show the central $4\farcm5$ radius region
of the SCUBA-2 field. In both panels,
the white circles show the $>4\sigma$ \afluxb selected
sample, including a confusion noise of 0.33~mJy.
The sources are summarized in Table~2.}
\label{850_sample}}
\vskip 2cm
\end{figure*}
%---------------------------------------------------------------------

%---------------------------------------------------------------------
% FIGURE A1 Images 
%---------------------------------------------------------------------
\begin{figure*}
\includegraphics[width=7.2in,angle=0]{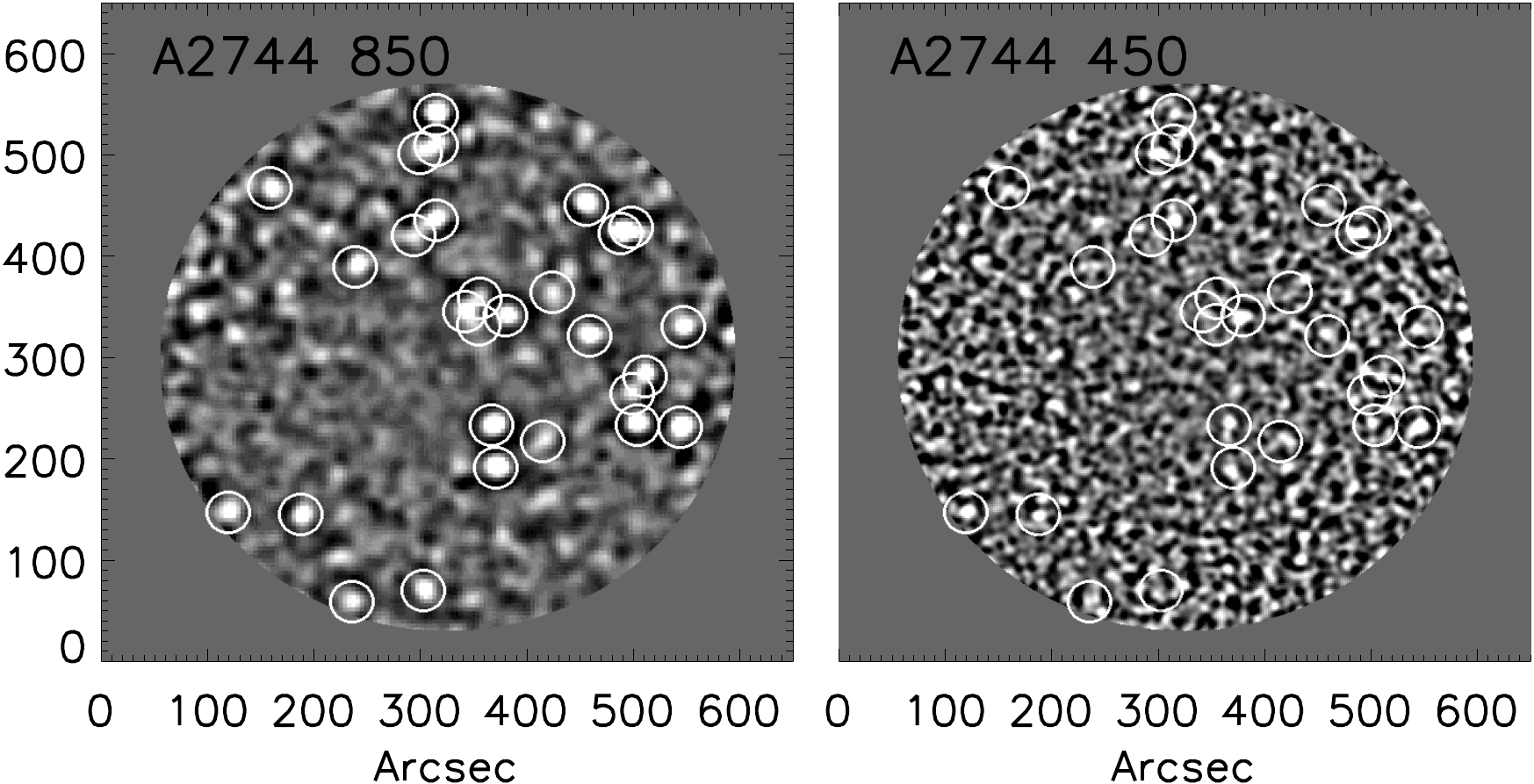}
\vskip 0.4cm
\includegraphics[width=7.2in,angle=0]{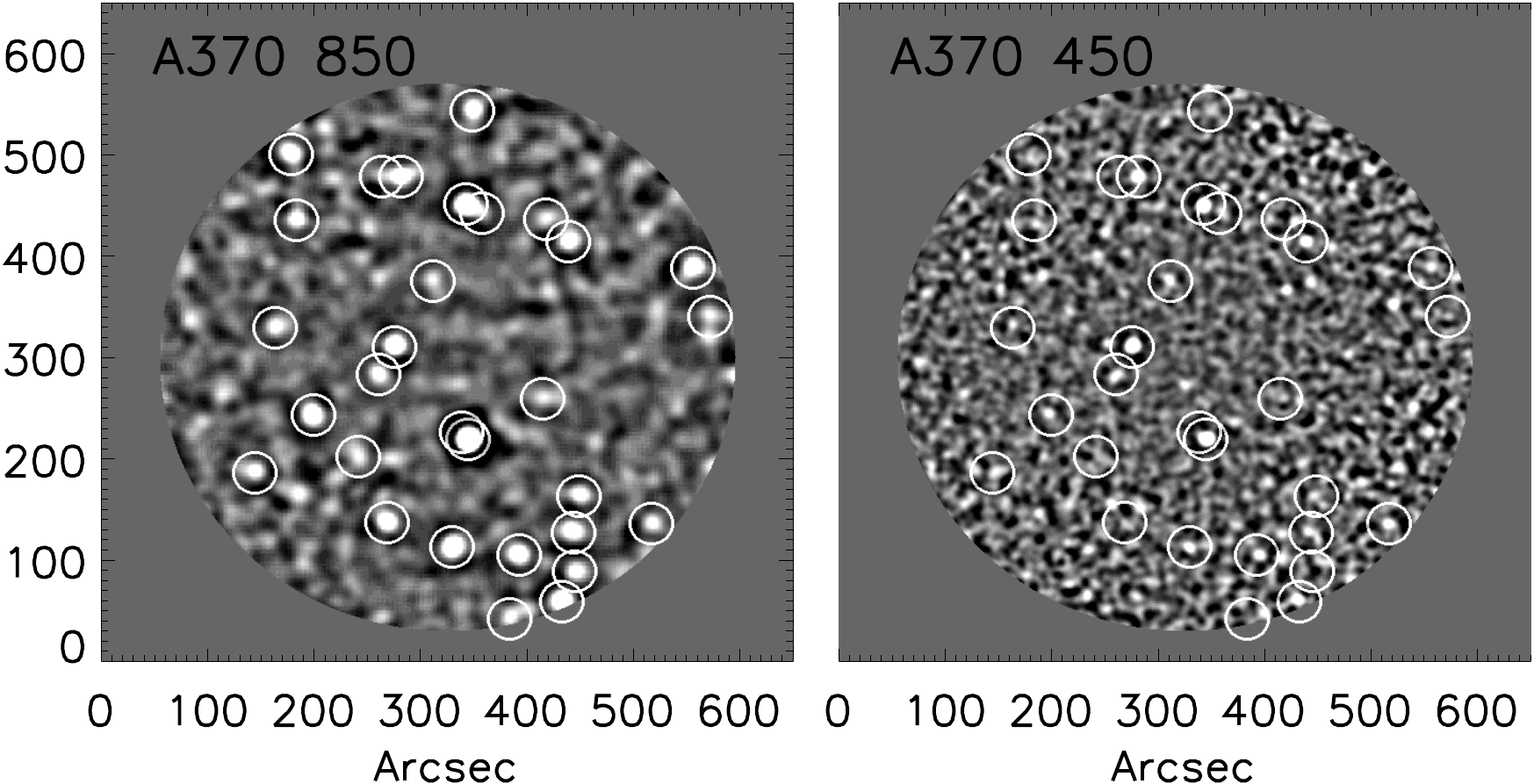}
\centering{\caption{
See caption for Figure~\ref{850_sample}.
\label{850_sample2}}}
\vskip 2cm
\end{figure*}
%---------------------------------------------------------------------

%---------------------------------------------------------------------
% FIGURE A1 Images 
%---------------------------------------------------------------------
\begin{figure*}
\includegraphics[width=7.2in,angle=0]{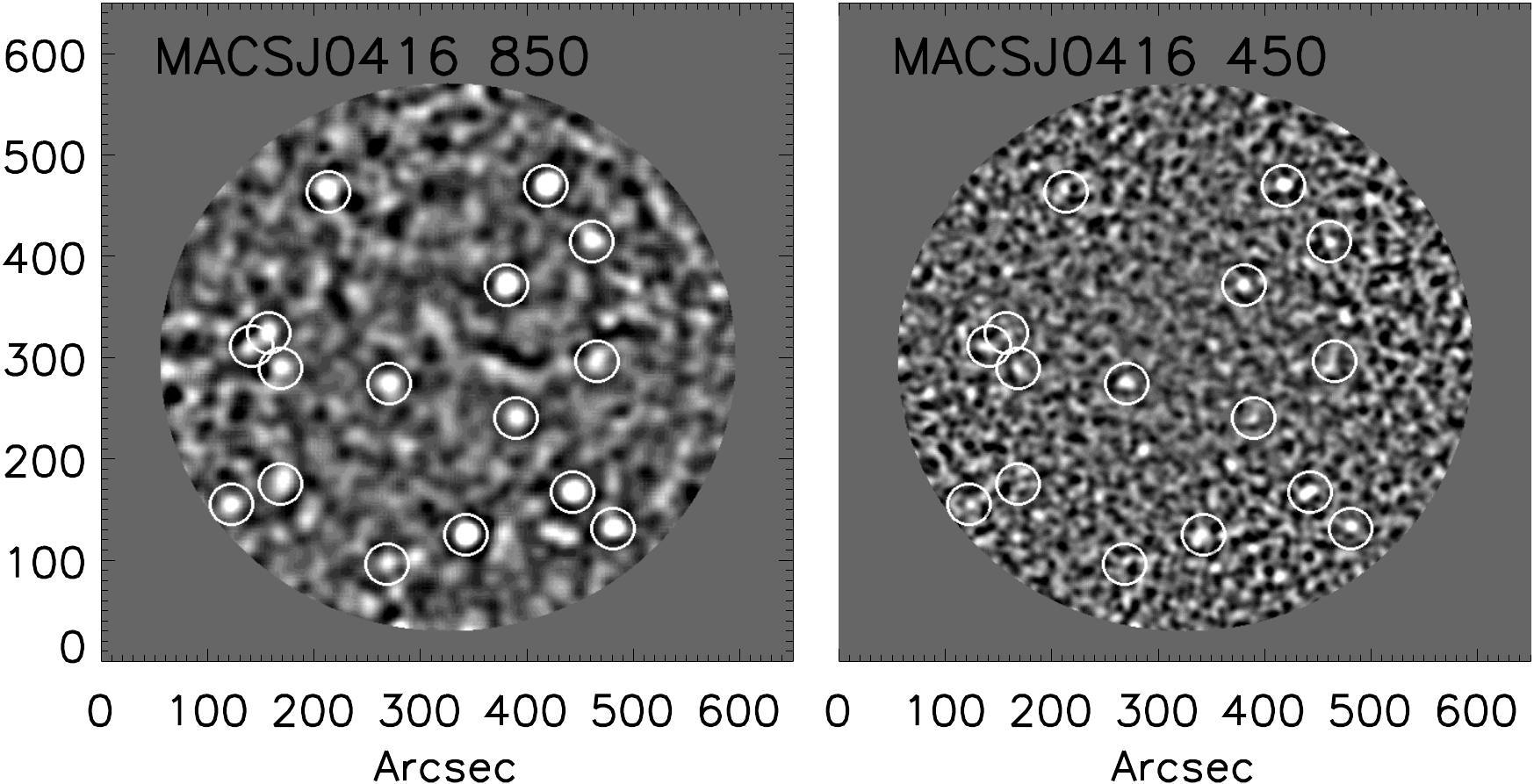}
\vskip 0.4cm
\includegraphics[width=7.2in,angle=0]{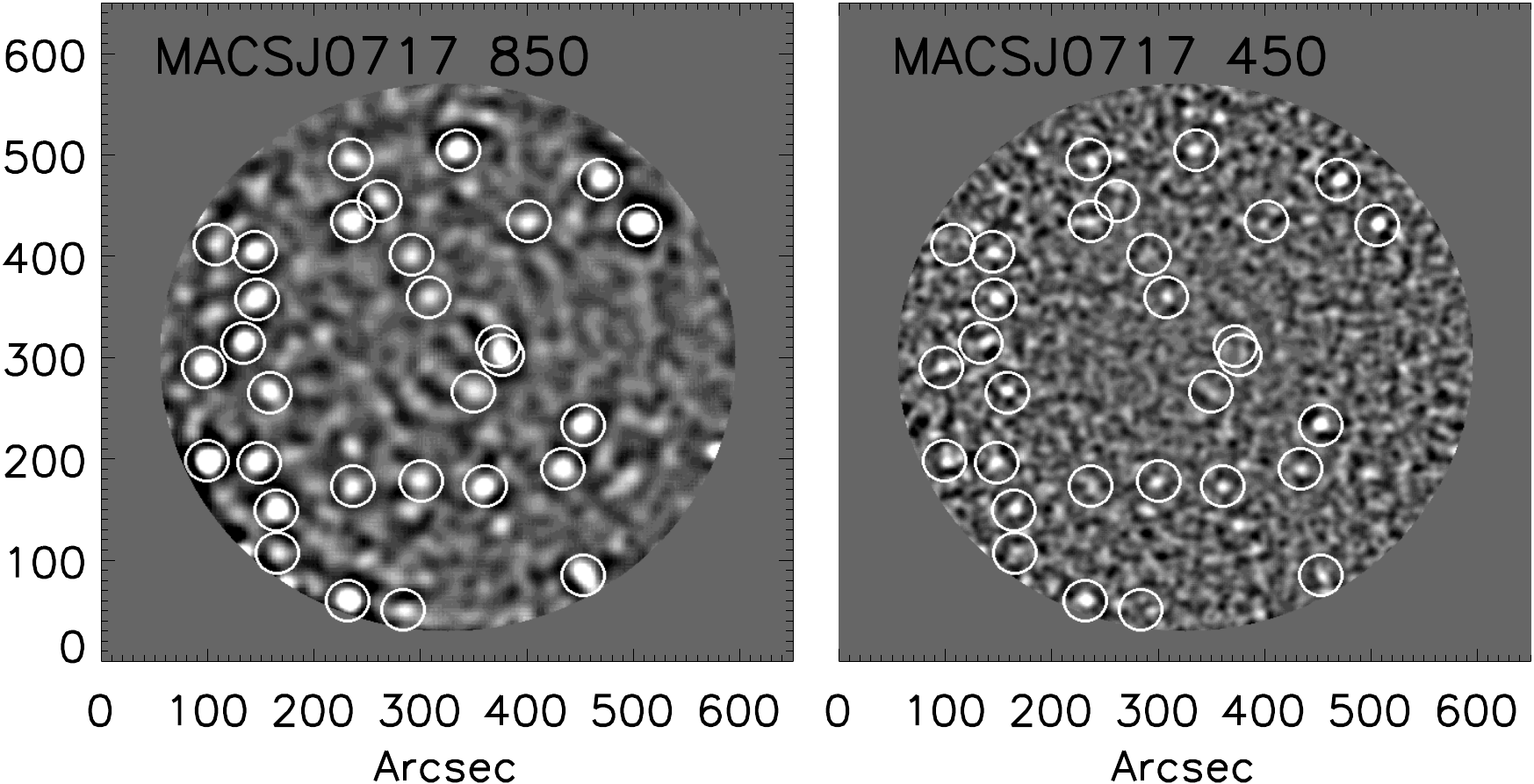}
\centering{\caption{
See caption for Figure~\ref{850_sample}.
\label{850_sample3}}}
\vskip 2cm
\end{figure*}
%---------------------------------------------------------------------

%---------------------------------------------------------------------
% FIGURE A1 Images 
%---------------------------------------------------------------------
\begin{figure*}
\includegraphics[width=7.2in,angle=0]{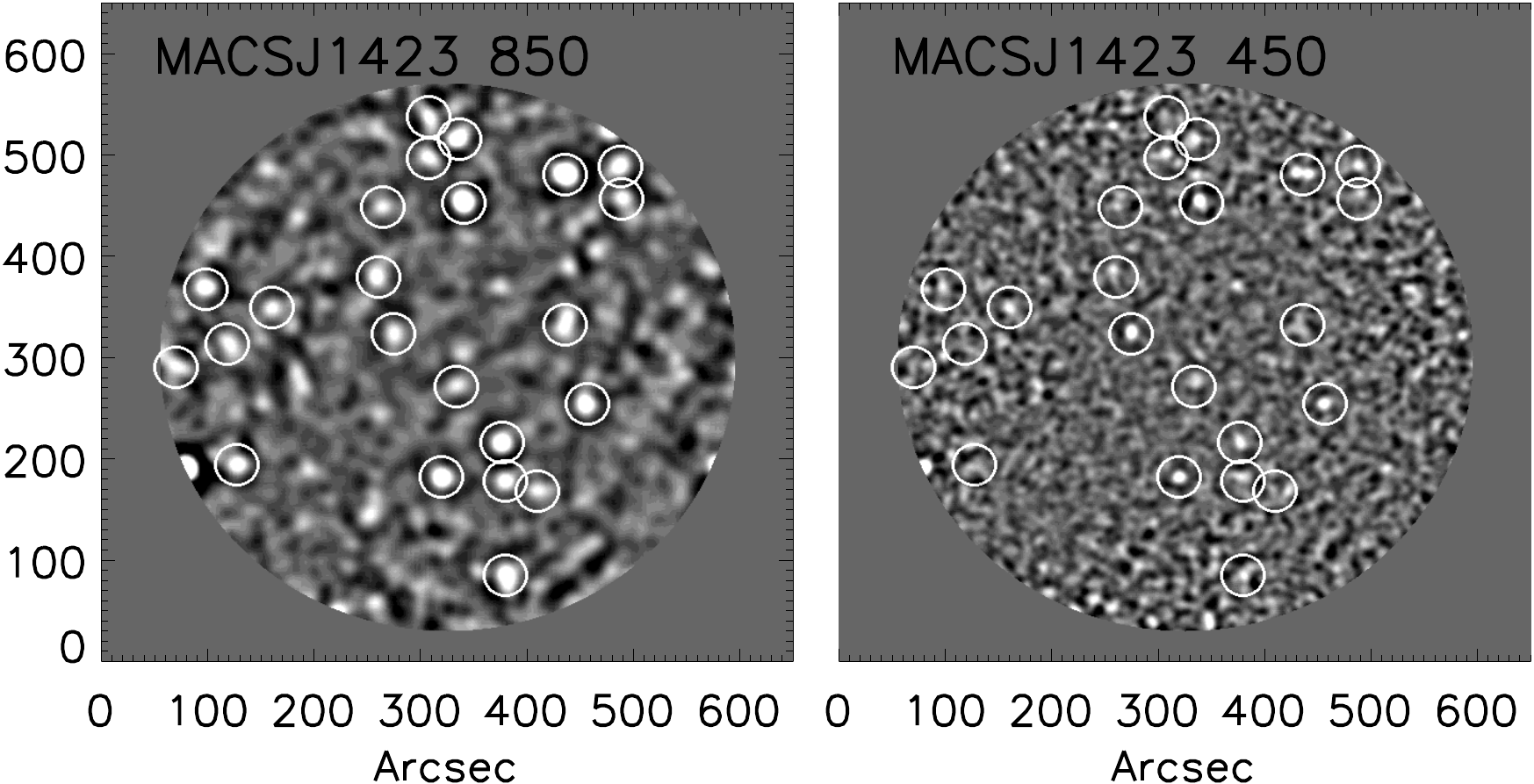}
\vskip 0.4cm
\includegraphics[width=7.2in,angle=0]{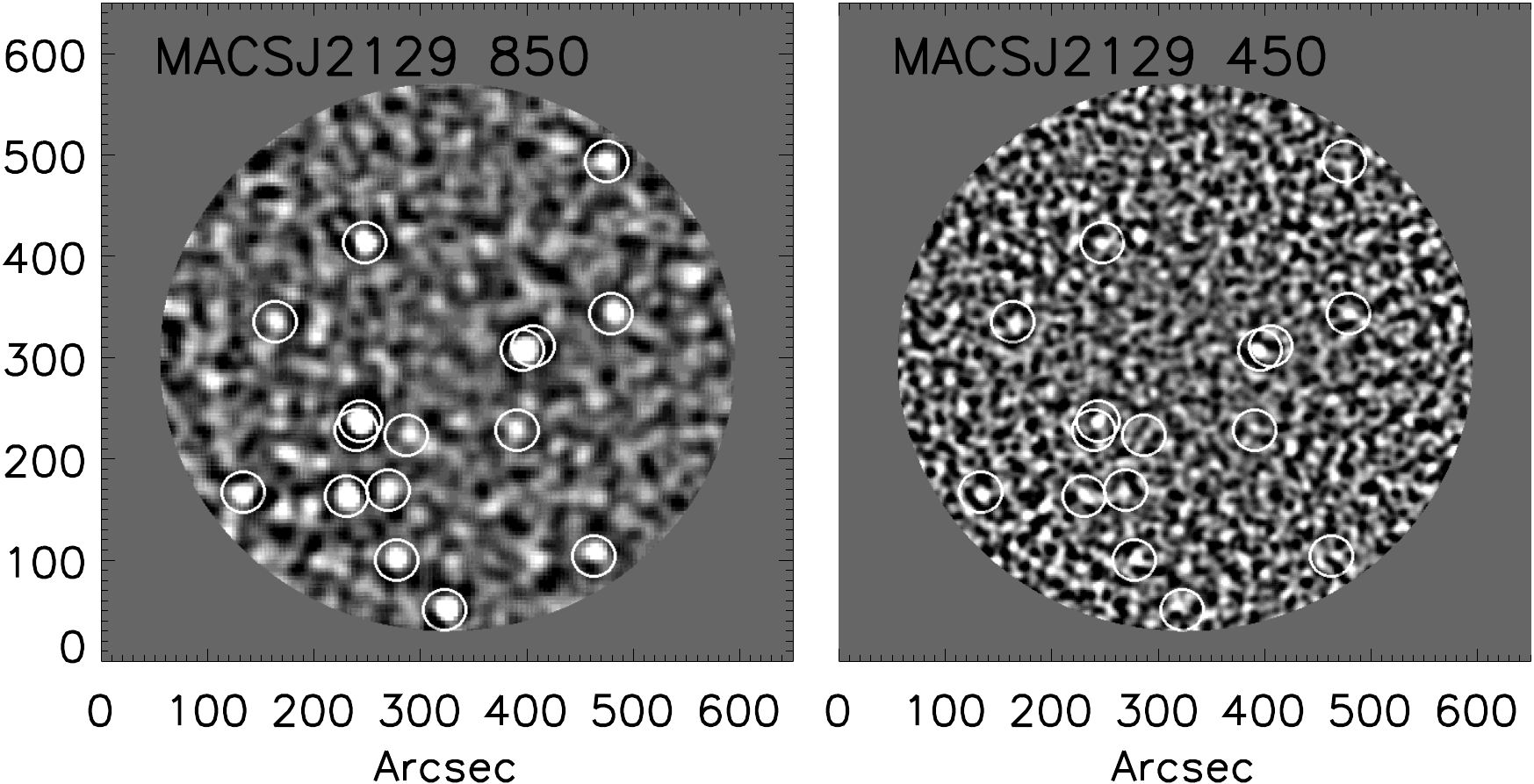}
\centering{\caption{
See caption for Figure~\ref{850_sample}.
\label{850_sample4}}}
\vskip 2cm
\end{figure*}
%---------------------------------------------------------------------

%---------------------------------------------------------------------
% FIGURE A1 Images 
%---------------------------------------------------------------------
\begin{figure*}
\includegraphics[width=7.2in,angle=0]{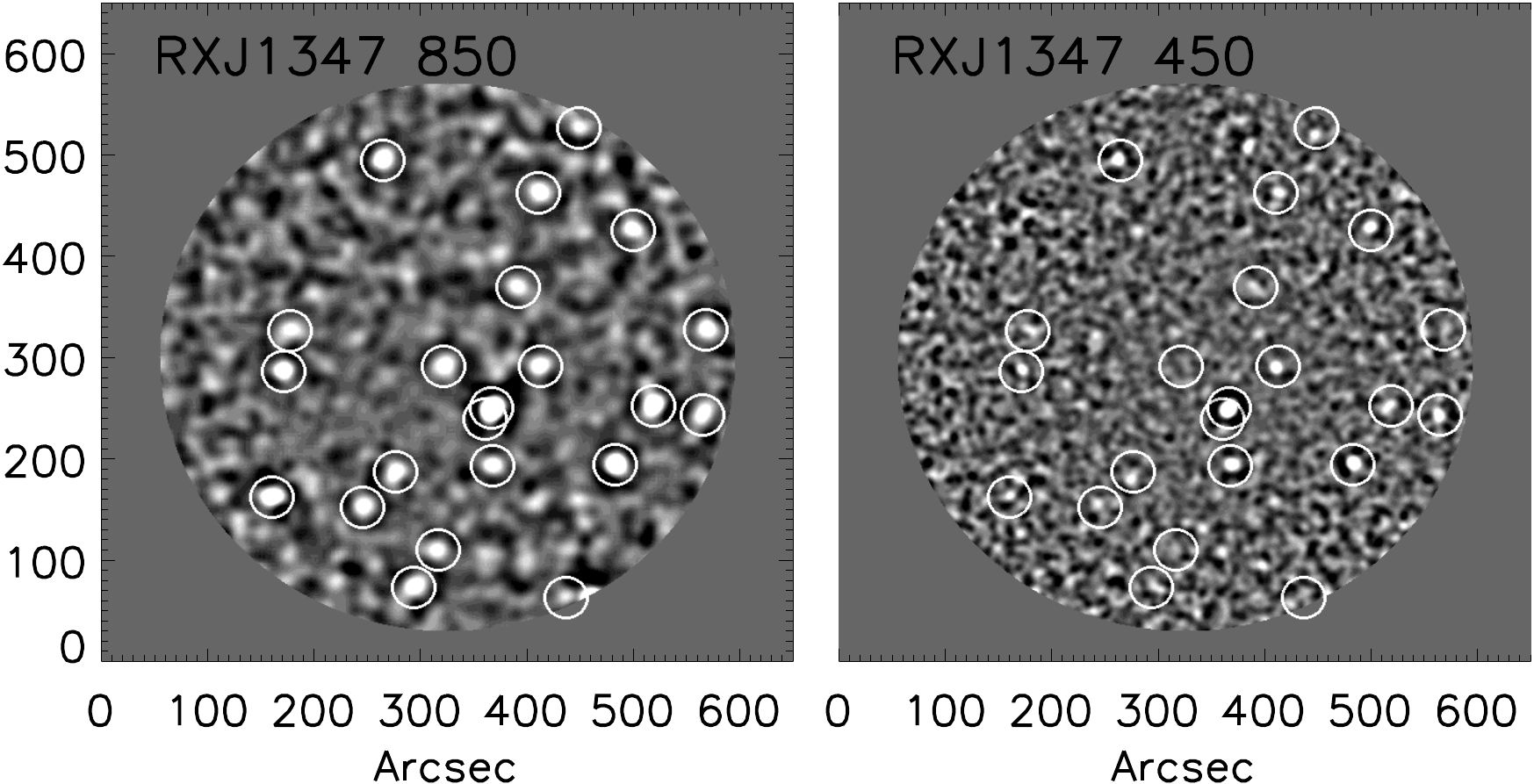}
\centering{\caption{
See caption for Figure~\ref{850_sample}.
\label{850_sample5}}}
\vskip 2cm
\end{figure*}
%---------------------------------------------------------------------


\begin{references}

\reference{aravena20}
Aravena, M., Boogaard, L., Gonz{\'a}lez-L{\'o}pez, J., et al.\ 2020, \apj, 901, 79

\reference{aravena16}
Aravena, M., Decarli, R., Walter, F., et al.\ 2016, \apj, 833, 68

\reference{bakx20}
Bakx, T. J. L. C., Tamura, Y., Hashimoto, T., et al.\ 2020, \mnras, 493, 4294

\reference{barger22}
Barger, A. J., Cowie, L. L., Blair, A. H., \& Jones, L. H..\ 2022, \apj, 934, 56

\reference{barger14}
Barger, A. J., Cowie, L. L., Chen, C.-C., et al.\ 2014, \apj, 784, 9

\reference{barger98}
Barger, A. J., Cowie, L. L., Sanders, D. B., et al.\ 1998, \nat, 394, 248

\reference{barger99}
Barger, A. J., Cowie, L. L., Smail, I., et al.\ 1999, 117, 2656

\reference{barger12}
Barger, A. J., Wang, W.-H., Cowie, L. L., et al.\ 2012, \apj, 761, 89

\reference{bethermin15}
B{\'e}thermin, M., De Breuck, C., Sargent, M., \& Daddi, E.\ 2015, A\&A, 576, L9

\reference{bradac19}
Brada\v{c}, M., Huang, K.-H., Fontana, A., et al.\ 2019, \mnras, 489, 99

\reference{bradac08}
Brada\v{c}, M., Schrabback, T., Erben, T., et al.\ 2008, \apj, 681, 187

\reference{casey13}
Casey, C. M., Chen, C.-C., Cowie, L. L., et al.\ 2013, \mnras, 436, 1919

\reference{casey14}
Casey, C. M., Narayanan, D., \& Cooray, A., 2014, PhR, 541, 45

\reference{castellano16}
Castellano, M., Amor{\'i}n, R., Merlin, E., et al.\ 2016, A\&A, 590, 31

\reference{chapin13}
Chapin, E. L., Berry, D. S., Gibb, A. G., et al.\ 2013, MNRAS, 430, 2545

\reference{chary01}
Chary, R., \& Elbaz, D.\ 2001, \apj, 556, 562

\reference{chen13a}
Chen, C.-C., Cowie, L. L., Barger, A. J., et al. 2013a, ApJ, 762, 81
%Faint Submillimeter Galaxy Counts at 450 μm

\reference{chen13b}
Chen, C.-C., Cowie, L. L., Barger, A. J., et al.\ 2013b, \apj, 776, 131
%Resolving the FIR Background at 450 and 850

\reference{chen14}
Chen, C.-C., Cowie, L. L., Barger, A. J., et al.\ 2014, \apj, 789, 12
%SMA Oservations on Faint SMGs with S850<2 mJy

\reference{colless03}
Colless, M., Peterson, B. A., Jackson, C., et al.\ 2003, arXiv:astro-ph/0306581

\reference{cowie17}
Cowie, L. L., Barger, A. J., Hsu, L.-Y., et al.\ 2017, \apj, 837, 139

\reference{cowie18}
Cowie, L. L., Gonz{\'a}lez-L{\'o}pez, J., Barger, A. J., et al.\ 2018, \apj, 865, 106

\reference{dicriscienzo17}
Di Criscienzo, M., Merlin, E., Castellano, M., et al.\ 2017, A\&A, 607, A30

\reference{dole06}
Dole, H., Lagache, G., Puget, J.-L., et al.\ 2006, A\&A, 451, 417

\reference{eales10}
Eales, S., Dunne, L., Clements, D., et al.\ 2010, PASP, 122, 499
%Herschel ATLAS

\reference{eales99}
Eales, S., Lilly, S., Gear, W., et al.\ 1999, \apj, 515, 518

\reference{edge10}
Edge, A. C., Oonk, J. B. R., Mittal, R., et al.\ 2010, A\&A, 518, L47

\reference{egami10}
Egami, E., Rex, M., Rawle, T D., et al.\ 2010, A\&A, 518, L12
%HLS

\reference{fixsen98}
Fixsen, D. J., Dwek, E., Mather, J. C., et al.\ 1998, \apj, 508, 123

\reference{gomez22}
G{\'o}mez-Guijarro, C., Elbaz, D., Xiao, M., et al.\ 2022, A\&A, 658, 43

\reference{gonzalez17}
Gonz\'alez-L\'opez, J., Bauer, F. E., Romero-Ca\~nizales, C., al.\ 2017, 597, A41

\reference{hoag16}
Hoag, A., Huang, K.-H., Treu, T., et al.\ 2016, \apj, 831, 182

\reference{hodge13}
Hodge, J. A., Karim, A., Smail, I., et al.\ 2013, \apj, 768, 91

\reference{holland13}
Holland, W. S., Bintley, D., Chapin, E. L., et al.\ 2013, \mnras, 430, 2513

\reference{holland99}
Holland, W. S., Robson, E. I., Gear, W. K., et al.\ 1999, \mnras, 303, 659

\reference{hsu17}
Hsu, L.-Y., Cowie, L. L., Barger, A. J., \& Wang, W.-H.\ 2017, \apj, 850, 189
%dichotomy

\reference{hsu16}
Hsu, L.-Y., Cowie, L. L., Chen, C.-C., Barger, A. J., \& Wang, W.-H.\ 2016,
\apj, 829, 25
%number counts and submm flux ratios

\reference{hughes98}
Hughes, D. H., Serjeant, S., Dunlop, J., et al.\ 1998, \nat, 394, 241

\reference{ivison98}
Ivison, R. J., Smail, I., Le Borgne, J.-F., et al.\ 1998, \mnras, 298, 583

\reference{jenness11}
Jenness, T., Berry, D., Chapin, E., et al.\ 2011, in ASP Conf. Ser. 442,
Astronomical Data Analysis Software and Systems XX, ed. I. N. Evans,
A. Accomazzi, D. J. Mink, \& A. H. Rots (San Francisco, CA: ASP), 281

\reference{kroupa01}
Kroupa, P.\ 2001, \mnras, 322, 231

\reference{laporte17}
Laporte, N., Bauer, F. E., Troncoso-Iribarren, P., et al.\ 2017, A\&A, 604, A132

\reference{lefloch05}
Le Floc'h, E., Papovich, C., Dole, H., et al.\ 2005, \apj, 632, 169

\reference{lim20}
Lim, C.-F., Wang, W.-H., Smail, I., et al.\ 2020, \apj, 889, 80
%STUIDES III

\reference{lin18}
Lin, Y.-T., Huang, H.-J., \& Chen, Y.-C. 2018, \aj, 155, 188

\reference{lotz17}
Lotz, J. M., Koekemoer, A., Coe, D., et al.\ 2017, \apj, 837, 97

\reference{maglio11}
Magliocchetti, M., Santini, P., Rodighiero, G.,, et al.\ 2011, \mnras, 416, 1105

\reference{meneghetti17}
Meneghetti, M., Natarajan, P., Coe, D., et al.\ 2017, \mnras, 472, 3177

\reference{merlin16}
Merlin, E., Amor{\'i}n, R., Castellano, M., et al.\ 2016, A\&A, 590, 30

\reference{mobasher09}
Mobasher, B., Dahlen, T., Hopkins, A. et al.\ 2009, \apj, 690, 1074

\reference{molino17}
Molino, A., Ben{\'i}tez, N., Ascaso, B., et al.\ 2017, \mnras, 470, 95

\reference{munoz22}
Mu\~{n}oz Arancibia, A. M.,  Gonz\'alez-L\'opez, J.,  Ibar, E., et al.\ 2022, A\&A, submitted
(arXiv:2203.06195)
 
\reference{oliver12}
Oliver, S. J., Bock, J., Altieri, B., et al.\ 2012, \mnras, 424, 1614
%HERMES

\reference{pope17}
Pope, A., Monta\~na, A., Battisti, A., et al.\ 2017, \apj, 838, 137

\reference{postman12}
Postman, M., Coe, D., Ben{\'i}tez, N., et al.\ 2012, \apjs, 199, 25

\reference{priewe17}
Priewe, J., Williams, L. L. R., Lisenborgs, J., Coe, D., \& Rodney, S. A.\ 2017,
\mnras, 465, 1030

\reference{puget96}
Puget, J.-L., Abergel, A., Bernard, J.-P., et al.\ 1996, A\&A, 308, L5

\reference{rawle16}
Rawle, T. D., Altieri, B., Egami, E., et al.\ 2016, \mnras, 459, 1626

\reference{reddy10}
Reddy, N. A., Erb, D. K., Pettini, M., et al.\ 2010, \apj, 712, 1070

\reference{serjeant03}
Serjeant, S., et al.\ 2003, \mnras, 344, 887

\reference{simpson15}
Simpson, J. M., Smail, I., Swinbank, A. M., et al.\ 2015, \apj, 799, 81

\reference{smail97}
Smail, I., Ivison, R. J., Blain, A. W.\ 1997, \apjl, 490, L5

\reference{smith10}
Smith, G. P., Haines, C. P., Pereira, M. J., et al.\ 2010, A\&A, 518, L18
%LoCuSS

\reference{soucail88}
Soucail, G., Mellier, Y., Fort, B., Mathez, G., \& Cailloux, M.\ 1988, A\&A, 191, L19

\reference{soucail99}
Soucail, G., Kneib, J. P., B{\'e}zecourt, J., et al.\ 1999, A\&A, 343, L70

\reference{spilker16}
Spilker, J. S., Marrone, D. P. , Aravena, M., et al.\ 2016, \apj, 826, 112
%ALMA Imaging and Gravitational Lens Models of South Pole Telescope-Selected 
%Dusty, Star-Forming Galaxies at High Redshifts 

\reference{steinhardt20}
Steinhardt, C. L., Jauzac, M., Acebron, A., et al.\ 2020, \apjs, 247, 64 

\reference{tamura19}
Tamura, Y., Mawatari, K., Hashimoto, T., et al.\ 2019, \apj, 874, 27

\reference{treu15}
Treu, T., Schmidt, K. B., Brammer, G. B., et al.\ 2015, \apj, 812, 114

\reference{umetsu14}
Umetsu, K., Medezinski, E., Nonino, M., et al.\ 2019, \apj, 795, 1634

\reference{wang19}
Wang, T., Schreiber, C., Elbaz, D., et al.\ 2019, Natur, 572, 211

\reference{wang17}
Wang, W.-H., Lin, W.-C., Lim, C.-F., et al.\ 2017, \apj, 850, 37
%STUDIES I

\reference{watson15}
Watson, D., Christensen, L., Knudsen, K. K., et al.\ 2015, Natur, 519, 327

\reference{wong22}
Wong, Y. H. V., Wang, P., Hasimoto, T., et al.\ 2022, \apj, 929, 161
%(arXiv:2202.13613)

\reference{zavala21}
Zavala, J. A., Casey, C. M., Manning, S. M., et al.\ 2021, \apj, 909, 165

\reference{zitrin21}
Zitrin, A.\ 2021, \apj, 919, 54
%Lessons from the First Multiply Imaged Supernova: Revised Strong-lensing Models for the Galaxy Cluster MACS J1149.5+2223
%This is a HFF

\reference{zitrin09}
Zitrin, A., Broadhurst, T., Rephaeli, Y., \& Sadeh, S.\ 2009, \apjl, 707, L102
%The Largest Gravitational Lens: MACS J0717.5+3745 (z = 0.546)
%This is a HFF

\reference{zitrin15}
Zitrin, A., Fabris, A., Merten, J., et al.\ 2015, \apj, 801, 44
%Hubble Space Telescope Combined Strong and Weak Lensing Analysis of the 
%CLASH Sample: Mass and Magnification Models and Systematic Uncertainties

\reference{zitrin13}
Zitrin, A., Meneghetti, M., Umetsu, K.\ et al.\ 2013, \apjl, 762, L30
%CLASH: The Enhanced Lensing Efficiency of the Highly Elongated Merging Cluster MACS J0416.1-2403
%This is a HFF

\reference{zitrin14}
Zitrin, A., Zheng, W., Broadhurst, T., et al.\ 2014, \apj, 793, L12
%A Geometrically Supported z ~ 10 Candidate Multiply Imaged by the Hubble Frontier Fields Cluster A2744
%This is a HFF

\end{references}
\end{document}